\documentstyle{amsppt}
\newcount\mgnf\newcount\tipi\newcount\tipoformule\newcount\greco 
\tipi=2          
\tipoformule=0   

\global\newcount\numsec\global\newcount\numfor
\global\newcount\numapp\global\newcount\numcap
\global\newcount\numfig\global\newcount\numpag
\global\newcount\numnf

\def\SIA #1,#2,#3 {\senondefinito{#1#2}%
\expandafter\xdef\csname #1#2\endcsname{#3}\else
\write16{???? ma #1,#2 e' gia' stato definito !!!!} \fi}

\def \FU(#1)#2{\SIA fu,#1,#2 }

\def\etichetta(#1){(\veroparagrafo.\veraformula)%
\SIA e,#1,(\veroparagrafo.\veraformula) %
\global\advance\numfor by 1%
\write15{\string\FU (#1){\equ(#1)}}%
\write16{ EQ #1 ==> \equ(#1)  }}
\def\etichettaa(#1){(A\veraappendice.\veraformula)
 \SIA e,#1,(A\veraappendice.\veraformula)
 \global\advance\numfor by 1
 \write15{\string\FU (#1){\equ(#1)}}
 \write16{ EQ #1 ==> \equ(#1) }}
\def\getichetta(#1){Fig. \verafigura
 \SIA g,#1,{\verafigura}
 \global\advance\numfig by 1
 \write15{\string\FU (#1){\graf(#1)}}
 \write16{ Fig. #1 ==> \graf(#1) }}
\def\retichetta(#1){\numpag=\pgn\SIA r,#1,{\verapagina}
 \write15{\string\FU (#1){\rif(#1)}}
 \write16{\rif(#1) ha simbolo  #1  }}
\def\etichettan(#1){(n\verocapitolo.\veranformula)
 \SIA e,#1,(n\verocapitolo.\veranformula)
 \global\advance\numnf by 1
\write16{\equ(#1) <= #1  }}

\newdimen\gwidth
\gdef\profonditastruttura{\dp\strutbox}
\def\senondefinito#1{\expandafter\ifx\csname#1\endcsname\relax}
\def\BOZZA{
\def\alato(##1){
 {\vtop to \profonditastruttura{\baselineskip
 \profonditastruttura\vss
 \rlap{\kern-\hsize\kern-1.2truecm{$\scriptstyle##1$}}}}}
\def\galato(##1){ \gwidth=\hsize \divide\gwidth by 2
 {\vtop to \profonditastruttura{\baselineskip
 \profonditastruttura\vss
 \rlap{\kern-\gwidth\kern-1.2truecm{$\scriptstyle##1$}}}}}
\def\verapagina{
{\romannumeral\number\numcap}.\number\numsec.\number\numpag}}

\def\alato(#1){}
\def\galato(#1){}
\def\veroparagrafo{\number\numsec}\def\veraformula{\number\numfor}
\def\veraappendice{\number\numapp}
\def\verapagina{\number\pageno}\def\veranformula{\number\numnf}
\def\verafigura{{\romannumeral\number\numcap}.\number\numfig}
\def\verocapitolo{\number\numcap}\def\veranformula{\number\numnf}
\def\Eqn(#1){\eqno{\etichettan(#1)\alato(#1)}}
\def\eqn(#1){\etichettan(#1)\alato(#1)}
\def\ver{\veroparagrafo}
\def\Eq(#1){\eqno{\etichetta(#1)\alato(#1)}}
\def\eq(#1){\etichetta(#1)\alato(#1)}
\def\Eqa(#1){\eqno{\etichettaa(#1)\alato(#1)}}
\def\eqa(#1){\etichettaa(#1)\alato(#1)}
\def\dgraf(#1){\getichetta(#1)\galato(#1)}
\def\drif(#1){\retichetta(#1)}

\def\eqv(#1){\senondefinito{fu#1}$\clubsuit$#1\else\csname fu#1\endcsname\fi}
\def\equ(#1){\senondefinito{e#1}\eqv(#1)\else\csname e#1\endcsname\fi}
\def\graf(#1){\senondefinito{g#1}\eqv(#1)\else\csname g#1\endcsname\fi}
\def\rif(#1){\senondefinito{r#1}\eqv(#1)\else\csname r#1\endcsname\fi}
\def\bib[#1]{[#1]\numpag=\pgn
\write13{\string[#1],\verapagina}}

\def\include#1{
\openin13=#1.aux \ifeof13 \relax \else
\input #1.aux \closein13 \fi}

\openin14=\jobname.aux \ifeof14 \relax \else
\input \jobname.aux \closein14 \fi
\openout15=\jobname.aux
\openout13=\jobname.bib


\ifnum\tipoformule=1\let\Eq=\eqno\def\eq{}\let\Eqa=\eqno\def\eqa{}
\def\equ{}\fi


{\count255=\time\divide\count255 by 60 \xdef\hourmin{\number\count255}
        \multiply\count255 by-60\advance\count255 by\time
   \xdef\hourmin{\hourmin:\ifnum\count255<10 0\fi\the\count255}}

\def\oramin{\hourmin }

\def\data{\number\day/\ifcase\month\or january \or february \or march \or
april \or may \or june \or july \or august \or september
\or october \or november \or december \fi/\number\year;\ \oramin}

\setbox200\hbox{$\scriptscriptstyle \data $}

\newcount\pgn \pgn=1
\def\foglio{\number\numsec:\number\pgn
\global\advance\pgn by 1}
\def\foglioa{A\number\numsec:\number\pgn
\global\advance\pgn by 1}

\footline={\rlap{\hbox{\copy200}}\hss\tenrm\folio\hss}


\global\newcount\numpunt

\magnification=\magstephalf
\baselineskip=16pt
\parskip=8pt

\voffset=2.5truepc
\hoffset=0.5truepc
\hsize=6.1truein
\vsize=8.4truein 
{\headline={\ifodd\pageno\rightheadline \else \leftheadline \fi}}
\def\rightheadline{\it  {tralala}\hfil\tenrm\folio}
\def\leftheadline{\tenrm \folio \hfil\it  {Section $\ver$}}

\def\a{\alpha}
\def\b{\beta}
\def\d{\delta}
\def\e{\epsilon}

\def\f{\phi}
\def\g{\gamma}
\def\k{\kappa}
\def\l{\lambda}

\def\s{\sigma}
\def\t{\tau}
\def\th{\theta}

\def\z{\zeta}
\def\o{\omega}
\def\D{\Delta}
\def\L{\Lambda}
\def\G{\Gamma}
\def\O{\Omega}
\def\S{\Sigma}

\def\del #1{\frac{\partial^{#1}}{\partial\l^{#1}}}

\def\1{{1\kern-.25em\roman{I}}}
\def\eu{{1\kern-.25em\roman{I}}}
\def\f1{{1\kern-.25em\roman{I}}}

\def\R{{\Bbb R}}  
\def\N{{\Bbb N}}  
\def\P{{\Bbb P}}  
\def\Z{{\Bbb Z}}  
\def\Q{{\Bbb Q}}  
\def\C{{\Bbb C}}  
\def\E{{\Bbb E}}  

\def\la{\langle}
\def\ra{\rangle} 

\def\del{\partial}

\def\dist{\,\roman{dist}}

\let\cal=\Cal

\def\BB{{\cal B}}
\def\CC{{\cal C}}

\def\EE{{\cal E}}
\def\FF{{\cal F}}

\def\HH{{\cal H}}

\def\LL{{\cal L}}
\def\MM{{\cal M}}

\def\OO{{\cal O}}
\def\PP{{\cal P}}
\def\QQ{{\cal Q}}
\def\RR{{\cal R}}
\def\SS{{\cal S}}
\def\TT{{\cal T}}

\def\A{{\cal A}}

\def\chap #1#2{\line{\ch #1\hfill}\numsec=#2\numfor=1}

\def\ba{{\backslash}}

\def\wt{\widetilde}
\def\wh{\widehat}

\def\rar{{\rightarrow}}


\def\note#1{\footnote{#1}}

\def\frac#1#2{{#1\over #2}}
\def\sfrac#1#2{{\textstyle{#1\over #2}}}
\def\ssfrac#1#2{{\scriptstyle{#1\over#2}}}
\def\text#1{\quad{\hbox{#1}}\quad}
\def\newpage{\vfill\eject}
\def\proposition #1{\noindent{\thbf Proposition #1:}}

\def\theo #1{\noindent{\thbf Theorem #1: }}

\def\lemma #1{\noindent{\thbf Lemma #1: }}
\def\definition #1{\noindent{\thbf Definition #1: }}

\def\corollary #1{\noindent{\thbf Corollary #1: }}
\def\proof{{\noindent\pr Proof: }}
\def\proofof #1{{\noindent\pr Proof of #1: }}
\def\endproof{$\diamondsuit$}
\def\remark{\noindent{\bf Remark: }}
\def\thanks{\noindent{\bf Acknowledgements: }}

\font\pr=cmbxsl10
\font\thbf=cmbxsl10 scaled\magstephalf

\font\ch=cmbx12

\font\it=cmti10
\font\bf=cmbx10


\font\tit=cmbx12
\font\aut=cmbx12

\def\s{\char'31}
\centerline{\tit METASTABILITY IN STOCHASTIC DYNAMICS}
\vskip.2truecm
\centerline{\tit OF DISORDERED MEAN-FIELD MODELS}
\vskip.2truecm 
\vskip1.5truecm

\centerline{\aut Anton Bovier 
\note{ Weierstrass-Institut f\"ur Angewandte Analysis und Stochastik,
Mohrenstrasse 39, D-10117 Berlin,\hfill\break Germany.
 e-mail: bovier\@wias-berlin.de},
Michael Eckhoff\note{Institut f\"ur Mathematik, Universit\"at Potsdam,
 Am Neuen Palais 10, D-14469 Potsdam, Germany.\hfill\break
e-mail: eckhoff\@rz.uni-potsdam.de},}
\centerline{\aut V\'eronique Gayrard\note{
Centre de Physique Th\'eorique, CNRS,
Luminy, Case 907, F-13288 Marseille, Cedex 9, France.\hfill\break
email: Veronique.Gayrard\@cpt.univ-mrs.fr},
Markus Klein\note {Institut f\"ur Mathematik, Universit\"at Potsdam,
Am Neuen Palais 10, D-14469 Potsdam, Germany.\hfill\break
e-mail: mklein\@felix.math.uni-potsdam.de}}
\vskip1cm

\vskip2truecm\rm
\def\s{\sigma}
\noindent {\bf Abstract:} We study a class of Markov chains 
that describe  reversible stochastic dynamics of a large class of
disordered mean field models at low temperatures. Our main purpose is
to give a precise relation between the metastable time scales
in the problem to the properties of the rate functions of the 
corresponding Gibbs measures. We derive the analog of the
Wentzell-Freidlin theory in this case, showing that any transition 
can be decomposed, with probability exponentially close to one, into
a deterministic sequence of ``admissible transitions''. For these
admissible transitions we give upper and lower bounds on the 
expected transition times that differ only by a constant factor. 
The distribution rescaled transition times are shown 
to converge to the exponential distribution.
We exemplify our
results in the context of the random field Curie-Weiss model.

\noindent {\it Keywords:} Metastability, stochastic dynamics, Markov chains,
Wentzell-Freidlin theory, disordered systems, mean field models, 
random field Curie-Weiss model.

\noindent {\it AMS Subject  Classification:}  82C44,  60K35 \vfill
$ {} $

\def\cl{\centerline}
\newpage

\vskip 1cm
\cl{\bf Table of contents}
\bigskip

\line{1.\hskip 0.5em Introduction\dotfill 3}
\line{\hskip0.4cm 1.1.  \hskip0.5em General introduction\dotfill 3}
\line{\hskip0.4cm 1.2.  \hskip0.5em General set-up\dotfill 6}
\line{\hskip0.4cm 1.3.  \hskip0.5em Outline of the general strategy\dotfill 
15}
\line{2.  \hskip0.5em Precise estimates on transition probabilities\dotfill 18}
\line{3.\hskip 0.5em Laplace transforms of the transition times in the
elementary situation \dotfill 24}
\line{4.  \hskip0.5em Valleys, trees, and graphs\dotfill 30}
\line{\hskip0.4cm 4.1.  \hskip0.5em Valleystructure and its 
tree-representation\dotfill 30}
\line{\hskip0.4cm 4.2.  \hskip0.5em Construction of the transition process\dotfill 32}
\line{5.  \hskip0.5em Transition times of admissible transitions\dotfill 38}
\line{\hskip0.4cm 5.1.  \hskip0.5em Expected times of admissible transitions\dotfill 38}
\line{\hskip0.4cm 5.2.  \hskip0.5em Laplace transforms of transition times of admissible 
transitions\dotfill 49}
\line{\hskip0.4cm 5.3.  \hskip0.5em The distribution 
 of transition times\dotfill 55}
\line{6.  \hskip0.5em Miscellaneous consequences for the process\dotfill  57}
\line{7.  \hskip0.5em Mean field models and mean field dynamics\dotfill 62} 
\line{8.  \hskip0.5em The random field Curie Weiss model\dotfill 65}
\line{  \hskip0.5em References\dotfill 72}

{\headline={\ifodd\pageno\rightheadline \else \leftheadline \fi}}
\def\rightheadline{\it  {Metastability}\hfil\tenrm\folio}
\def\leftheadline{\tenrm \folio \hfil\it  {Section $\ver$}}
\newpage


\chap{1. Introduction}1
\medskip
\line{\bf \ver.1 General introduction\hfill}
\medskip
This paper is devoted to developing a systematic approach to the
analysis of the long time behaviour of the dynamics of certain
mean field spin systems, where by dynamics we understand of course a 
stochastic dynamics of Glauber type. 
For the purposes of this paper, we will always choose this as reversible 
with respect to the Gibbs measure of the model.
By long time behaviour we mean
that we are interested in time scales on which the phenomena of 
``meta-stability'' occur, i.e. time scales that increase with the 
volume of the system exponentially fast. Our primary motivation
comes from the study of {\it disordered} spin systems, and most particularly 
the so called Hopfield model [Ho,BG1], although in the present paper
we only illustrate our results in a much simpler setting, that of the 
random field Curie-Weiss (RFCWM) model (see e.g. [K1]). Our chief objective
is to be able to control in a precise manner the effect of 
the randomness on the metastable phenomena.

On a heuristic level, metastable phenomena in mean field models are
well understood. The main idea is to consider the dynamics induced 
 on the order parameters
by the Glauber dynamics on the spin space, 
i.e. the macroscopic variables that characterize the model.  
A first issue that arises here, and that we will discuss at length 
below, is that this induced dynamics is in general not Markovian. However,
one may always define a new Markovian dynamics that ``mimics'' the old one 
and that is reversible with respect to the measures induced on order 
parameters by the Gibbs measures.
This dynamics on the order parameters is essentially a random walk in 
a landscape given by the ``rate function'' associated to the distribution of
the order parameters. The accepted picture of the resulting motion is that 
this walk will spend most of its time in the ``most profound valleys''
of the rate function and stay in a given valley for an exponentially long time
of order $\exp(N \D F)$ where $\D F$ is the difference between the minimal
 value of 
the rate function in the valley and its value at the lowest ``saddle point''
over which the process may exit the valley.
An excellent survey on this type of processes is given in 
van Kampen's textbook [vK], although most of the results
presented there, and in particular all those 
 related to the long time behaviour,
 concern the one-dimensional case.
 Rather surprisingly,
one finds very few papers in the literature that really treat this
problem with any degree of mathematical rigour.  One exception is
the classical paper by Cassandro, Galves, Olivieri, and Vares [CGOV] (see
also [Va] for a broader review on metastability)
who consider (amongst others) 
the case of the Curie-Weiss model in which there is 
only a single order parameter and thus the resulting dynamics is that
of a one-dimensional random walk.   More recently, a particular version
of the RFCWM that leads to a two-dimensional problem was treated by 
Mathieu and Picco [MP].  However, there is an abundant literature on
two types of related problems. One of these concerns Markov chains
with finite state space and exponentially small transition probabilities.
They are treated in the work of Freidlin and Wentzell (but see below for a 
discussion) and have since then been investigated intensely (for a small 
selection of recent references see [Sc,OS1,OS2,CC,GT]. In the context of 
stochastic dynamics of spin systems, they occur if finite systems
are considered in the limit of zero temperature. 
\note{Let us mention, however,  that there has been condiderable work done 
on the dynamics of spin systems on infinite lattices; see inparticular
the recent paper by Schonmann and Shlosman [SS] on the metastable behaviour in 
the two-dimensional Ising model in infinite volume, and references therein.}
A second class 
of problems, 
that is in a sense closer to our situation, 
and that can be 
obtained from it  formally by passing to the limit of continuous space 
and time, is that of ``small random perturbations of dynamical systems''
i.e. a 
stochastic differential equation of the form 
$$
dx^\e(t)= b(x^\e(t))dt +\sqrt \e a(x^\e(t)) dW(t)
\Eq(1.1)
$$
where $x^\e(t)\in \R^d$, and $W(t)$ is a $d$-dimensional Wiener process,
and in the case of a reversible dynamics the  drift term $b(x^\e(t),\e)$ is 
given by $b(x,\e)=\nabla F_\e(x)$, $F_e(x)$ being the rate function. 

The basic reference on the problem \eqv(1.1) is the seminal book by Wentzell
and Freidlin [FW] which discusses this  problem (as well as a number of
related ones)   in great detail. Many further references can be found in the
forthcoming second edition of this book.
One of the important aspects of this work is that is devises a scheme that
allows to control the long-time dynamics of the problem through an associated 
Markov chain with finite state space and exponentially small transition 
probabilities. The basic input here are large deviation estimates on the
short time behaviour of the associated processes.
 This 
treatment has inspired a lot of consecutive works which it is impossible to 
summarize to any degree of completeness. For our purposes, an important 
development is a refinement of the estimates which in [FW] are given only 
to the leading exponential order in $\e$ to a full asymptotic expansion.
Relevant references are [Ki1-4,Az,FJ]. The work of [FJ] in particular is
very interesting
 in that it develops
 full asymptotic expansions to all
orders  for certain exit probabilities using purely analytic techniques
based on WKB methods. Very similar results are obtained in [Az] using
refined large deviation techniques. To our knowledge all the
refined treatments that have appeared in the literature treat
only specific ``local'' questions, and there seems to be no coherent 
treatment of the global problem in a complicated (multi-valley) situation
that takes into account sub-leading terms.

The problems we will study require essentially to redo  the work
of Freidlin and Wentzell in the setting of our Markov chains. Moreover,
for the problems we are interested in, it will be important to have a 
more precise control, beyond the leading exponential asymptotics, for
 the global problem, if we want to be able to exhibit the influence of the 
residual randomness. 
The point is that in many disordered mean field models very precise 
estimates of the large deviation properties of the Gibbs measures are 
available. Typically, the rate function is deterministic to leading order
(although {\it not}  equal to the rate function of the averaged system
\note{It is important to keep in mind that the main effect of the disorder 
manifests itself in a deterministic modification of the rate function.
This effect is somewhat reminiscent to the phenomenon of homogenization.}!)
while the next order corrections (typically, but not always, of order 
$N^{-1/2}$) are random. To capture this effect, some degree of precision in the
estimates is thus needed.
On the other hand, we will not really need a full 
asymptotic 
expansion\note{We believe that it is possible to obtain such an expansion for 
the global problem. However, this will require a much more
elaborate analysis which we postpone to future publications.}
of our quantities, and we will put more effort on the control of the 
``global'' behaviour than on the overly precise treatment of ``local'' 
problems.  A main difference is of course that we do not have 
a stochastic differential equation but a Markov chain on a discrete
state space \note{The state space is even finite for any $N$, but its size 
increases with $N$, which renders this fact rather useless.}. Therefore 
one may draw intuition from the proofs given in the continuous case
without being able to use any result proved in that context directly.
Finally, our goal is to give a treatment that is as simple and transparent as 
possible. This is the main reason to concentrate on the reversible case,
and one of our strategies is to use reversibility to as large an extent as 
possible. This allows to replace refined large deviation estimates by 
simple reversibility arguments. Large deviation estimates are then only used 
in a less delicate situations.
In the same spirit, we will take advantage of the 
discrete nature of the problem whenever this is possible (just to compensate
for all the disadvantages we encounter elsewhere). This will surprise 
the reader familiar with the continuous case, but we hope she will be 
convinced at the end that this was a pleasant surprise.

Let us say a final word concerning our preoccupations with the dependence on 
dimensionality.  One of our ultimate goals is to be able to treat, e.g.,
 the Hopfield 
model in the case where the number of order parameters grows with the volume 
of the system. On the level of the mean field dynamics, this requires 
us to be able to treat a system where the dimension of the space grows with the
large parameter. Although we will not consider this situation in this first 
paper, we will achieve a precise control of the dimension dependence 
of sub-leading corrections.


\medskip
\line{\bf \ver.2. The general set-up.\hfill}
\medskip
We will now describe the general class of Markov chains we will consider.
Their relation to disordered spin systems will be explained in Section 7
and a specific example will be discussed in Section 8. Section 7
can be read now, if desired; on the other hand, the bulk of the paper can also
be read without reference to this motivation.

We consider canonical 
Markov chains on a state space $\G_N$ where $\G_N$ is  the 
intersection of some lattice\note{The requirement that $\G_N$ is a lattice 
is made for convenience  and can be weakened considerably, if desired.
What is needed are some homogeneity and rather minimal isotropy assumptions.} 
 (of spacing $O(1/N))$ in $\R^d$ with some connected 
$\L\subset\R^d$ which is either open or the closure of an open set. 
To avoid some irrelevant issues, we will    
assume that $\L$ is either $\R^d$ or a bounded and convex subset of $\R^d$.
$\G_N$ is assumed to
have spacing of order $1/N$, i.e. the cardinality of the state space is
of order $N^d$. Moreover, we identify $\G_N$ with  a graph with finite 
(d-dependent) coordination number respecting the Euclidean structure in the 
sense that a vertex $x\in \G_N$ is connected only to vertices at Euclidean 
distances 
less than $c/N$ from $x$. 
The main example the reader should have in mind is $\G_N= \Z^d/N\cap \L$,
with edges only between nearest neighbors. We denote the set of edges of 
$\G_N$ by $E(\G_N)$.

Let $\Q_N$ be a probability measure on $(\G_N, \BB(\G_N))$. We will set, for
$x\in \G_N$,
$$
F_N(x)\equiv -\frac 1N\ln \Q_N(x)
\Eq(1.2)
$$
We will assume the following properties of $F_N(x)$.

\noindent {\bf Assumptions:}

\item{{\bf R1}}
  $F\equiv \lim_{N\uparrow \infty} F_N$ exists and is a
smooth function $\L\rightarrow\R$; the convergence  is uniform in 
compact subsets of $\hbox{int} \L$.
\item{\bf R2}
$F_N$ can be represented as $F_N=F_{N,0}+\frac 1N F_{N,1}$
where $F_{N,0}$ is twice Lipshitz, i.e. 
$|F_{N,0}(x)-F_{N,0}(y)|\leq C\|x-y\|$ and for any generator of the lattice, 
$k$,

$N|F_{N,0}(x)-F_{N,0}(x+k/N)-(F_{N,0}(y)-F_{N,0}(y+k/N)|\leq C\|x-y\|$,
with $C$ uniform on compact subsets of the interior of $\L$. 
$F_{N,1}$ is only required to be Lipshitz, 
i.e. $|F_{N,1}(x)-F_{N,1}(y)|\leq C\|x-y\|$. 

For the purposes of the present paper we will  make a number of 
assumptions concerning the functions $F_N$ which we will consider as
``generic''. 
An important assumption concerns the structure of the set of minima of the 
functions $F_N$. We will assume 
 that the set $\MM_N\subset\G_N$, 
of {\it local} minima of $F_N$ is finite and of constant cardinality for
all $N$ large enough, and that the sets $\MM_N$ converge, as $N$ tends to 
infinity, to the set $\MM$ of local minima of the function $F$\note{This 
assumption can easily be relaxed somewhat. For example, it would be no 
problem if the function $F_N$ is degenerate on a small set of 
points in the very close (order $N^{-1/2}$) neighborhood of a minimum. One then would just 
choose one of them to represent this cluster. Other situations, e.g. 
when the function  $F$  has local minima on large sets and would lead 
to new effects would require special treatments.}. 

Another set of points that will be important is the set, $\EE_N$,
of  ``essential'' saddle points (i.e. the lowest saddle points one has to 
cross to go from one minimum to another). 
Formally, we define the essential saddle, $z^*(x,y)$, between two minima
$x,y\in\MM_N$ as 
$$
z^*(y,x)\equiv \arg\inf_{\g:\g(0)=y, \g(1)=x} \left(\sup_{t\in [0,1]}
\left[F_N(\g(t))
\right]\right)
\Eq(T.2)
$$
where the infimum is over all paths $\g:[0,1]\rightarrow \G_N$ 
going from $y$ to $x$. 
\note{Note that here  we think of a path as a discontinuous (c\`adl\`ag) function 
that stay at a site in $\G_N$ for some time interval $\d t$ and then jumps
to a neighboring site along an edge  of $\G_N$. This parametrization 
will however be of no importance and allows just some convenient notation.
}
with jumps along the edges of $\G_N$ only\note{We will extend the
definition \eqv(T.2) also to general points $x,y\in\G_N$. In that case it 
may happen that $z^*(x,y)$ is not a saddle point, but one of the endpoints 
$x$ or $y$ itself.}  

A point $z$ is called an essential saddle point if there exist minima
$x,y\in\MM_N$ such that $z^*(x,y)=z$.
The set of all essential saddle point will be denoted by $\EE_N$. 
Our assumptions on the $N$ dependence of the set $\MM_N$ apply in the 
same way to $\EE_N$.

\item{\bf G1} We will assume that there exists $\a>0$ such that 
$\min_{x\neq y\in\MM_N\cup \EE_N} |F_N(x)-F_N(y)|=K_N\geq N^{\a-1}$. 
\item{\bf G2}  We assume that at each minimum the eigenvalues of
the Hessian of $F$ are strictly positive and at each  essential saddle there
is one strictly negative eigenvalue while all others are strictly 
positive.  
\item{\bf G3} All minima and saddles are well in the interior of $\L$, i.e. 
there exists a $\d>0$ such that for any $x\in\MM_N\cup\EE_N$,
$\dist(x,\L^c)>\d$.

\remark We make the rather strong assumptions above in order to be able 
to formulate very general theorems that do not depend on specific properties 
of the model. They can certainly be relaxed. The regularity conditions {\bf R2}
are necessary only for the application of certain large deviation results
in Section 4 and are otherwise not needed.


We recall that in our main applications, $\Q_N$ will be random measures,
but we will forget this fact for the time being and think of $\Q_N$ as some 
particular realization. 

We can now construct a Markov chain $X_N(t)$ with state space given by the 
set of vertices of 
$\G_N$ and time parameter  set either $\N$ or $\R_+$. For this we first
define for any $x,y$ such that $(x,y)\in E(\G_N)$   transition rates
$$
p_N(x,y)\equiv \sqrt {\frac {\Q_N(y)}{\Q_N(x)}} f_N(x,y)
\Eq(S.1)
$$
for some non-negative, symmetric   function  $f_N$.
We will assume that $f_N$ does not introduce too much anisotropy. This can be
expressed by demanding that
\item{\bf R3} There exists $c>0$ such that if $(x,y)\in E(\G_N)$, 
and $\dist (x,\L^c)>\d/2$, (where $\d$ is the same as in assumption {\bf G3}) 
$p_N(x,y)\geq c$. 

Moreover, for applications of large deviation results  we need 
stronger regularity properties analogous to {\bf R2}. 

{\bf R4}  $ \ln f_N(x,y)$ as a function of 
any of its arguments is uniformly Lipshitz   on compact subsets of the 
interior 
of $\L$.  


For the case of  discrete time, i.e. $t\in \N$, we then define the transition 
matrix 
$$
P_N(x,y) \equiv \cases  p_N(x,y), &\text{if} (x,y)\in E(\G_N)\cr
  1-\sum_{z\in \G_N: (x,z) \in E(\G_N)} p_N(x,z), &\text{if} x=y\cr
  0,&\text{else}\endcases
\Eq(S.2)
$$
choosing $f$ such that 
$\sup_{x\in \G_N}\sum_{z\in \G_N: (x,z) \in E(\G_N)} p_N(x,y)
\leq 1$. 

Similarly, in the continuous time case, 
we can use the rates to define the generator 
$$
A_N(x,y) \equiv \cases  p_N(x,y), &\text{if} 
(x,y) \in E(\G_N)\cr
  -\sum_{z\in \G_N: (x,z) \in E(\G_N)} p_N(x,y), &\text{if} x=y\cr
  0,&\text{else}\endcases
\Eq(S.3)
$$

Our basic approach to the analysis of these Markov chains is to observe the
process when it is visiting the positions of the minima of the function $F_N$, 
i.e. the points of the set $\MM_N$, and to record the elapsed time. The 
ideology behind this is that we suspect the process to show the following 
typical behaviour: starting at any given point, it will rather quickly 
(i.e. in some time of order $N^k$) visit a nearby minimum, and then 
visit this same minimum at similar time interval an exponentially large number
of times without visiting any other minimum between successive returns.
Then, at some random moment it will go, quickly again, to some other minimum
which will then be visited regularly a large number of times, and so on.
Moreover, between successive visits of a minimum the process will typically 
not only avoid visits at other minima, but will actually stay very close
to the given minimum. Thus, recording the visits at the minima will be 
sufficient information on the behaviour of the process. These expectations will
be shown to be justified (see in particular Section 7). Incidentally, we 
mention that the
``quick'' processes of transitions can be analysed in detail using large 
deviation methods [WF1-4]. In [BG2] a large deviation principle is proven
for a class of Markov chains including those considered here
that shows that 
the ``paths'' of such quick processes concentrate asymptotically 
near the classical trajectories of some (relativistic) Hamiltonian system.
More precisely, the transitions between minima can be identified as
instanton solutions of the corresponding Hamiltonian system.

Let us mention that the strategy to record visits at single points
is specific to the discrete state space. In the diffusion setting, visits at  
single points do not happen with sufficient probability to contain  pertinent 
information on the process. Indeed, the crucial fact we use is that 
in the discrete case it is excessively difficult for the process to stay 
for a time of order $N^k$ (we will discuss the values of $k$ later) 
in the vicinity of a minimum without visiting it\note{The reader may wonder 
at this point why the minima are so special compared e.g. with their 
neighboring points. In fact they are not, and nothing would change if we 
chose some other  point close to the 
minimum rather than the exact minimum. 
But of course the minima themselves are the
optimal choice, and also the most natural ones.} which in the 
continuum is not the case.
For this reason Freidlin and Wentzell 
record visits not at single points but at certain neighborhoods of minima and
critical points which has the disadvantage that such visits do not exactly 
allow a splitting of the process and this introduces some error terms in 
estimates which in our setting can easily be avoided. This is the main 
advantage we 
draw from working in a discrete space.

The informal discussion above will be made precise in the sequel. 
We place ourselves in the discrete time setting
  throughout this paper, but everything can be transferred to the 
continuous time setup with mild modifications, if desired.
Let us first 
introduce some notation. We will use the symbol $\P$ for the 
law of our Markov chain, omitting the explicit mention 
of the index $N$, and denote by $X_t$ the coordinate variables. 
We will write $\t^y_x$ for the first time the process
conditioned to starting at $y$ hits the point $x$, i.e. we write
$$
\P\left[\t^y_x=t\right]
\equiv \P\left[X_t=x, \forall_{0<s<t}X_s\neq x|X_0=y\right]
\Eq(S.4)
$$
for $t>0$. In the case $x=y$, we will insist that $\t^y_y$ is the time of
the first visit to $y$ {\it after } $t=0$, i.e. $\P[\t^y_y=0]=0$. 
This notation may look unusual at first sight, but we are convinced that the
reader will come to appreciate its convenience.

One of the most useful basic identities which follows directly from the 
strong Markov property and the fact that $\t^y_x$ is a stopping time is the 
following:

\lemma {\ver.1} {\it Let $x,y,z$ be arbitrary points in $\G_N$. Then
$$
\eqalign{
\P\left[\t^y_x=t\right]&=\P\left[\t^y_x=t, \t^y_x<\t^y_z\right]\cr
&+\sum_{0<s<t}\P\left[\t^y_z=s,\t^y_z<\t^y_x\right]\P\left[\t^z_x=t-s\right]
}
\Eq(S.5)
$$
}
\proof Just note that the process either arrives at $x$ before visiting $z$,
or it visits $z$ a first time before $x$. \endproof

A simple consequence is the following basic renewal equation.

\lemma {\ver.2} {\it Let  $x,y\in \G_N$. Then
$$
\eqalign{
\P\left[\t^y_x=t\right]&=\sum_{n=0}^\infty \sum_{{t_1,\dots t_{n+1}}\atop
{\sum_it_i=t}} \prod_{i=1}^n\P\left[\t^y_y=t_i,\t^y_y<\t^y_x\right]
\P\left[\t^y_x=t_{n+1},\t^y_x<\t^y_y\right]
}
\Eq(S.6)
$$
}

The fundamental importance in the decomposition of Lemma \ver.2 lies in the 
fact that objects like the last factor in \eqv(S.6) are ``reversible'', i.e.
they can be  compared to their time-reversed counterpart. 
To formulate a general principle, let us define the time-reversed chain
corresponding to a transition from $y$ to $x$ via
$X^r_t\equiv X_{\t^y_x-t}$. For an event $\A$
 that is measurable with respect to the  sigma algebra 
$\FF\left(X_s, 0\leq s\leq \t^y_x\right)$ we then define the time reversed
event $\A^r$ as the event that takes place for the chain $X_t$ if and only if 
the event $\A$ takes place for the chain $X^r_t$. This allows us to formulate
the next lemma:

\lemma{\ver.3} {\it  Let $x,y\in \G_N$, and let $\A$ be any event
measurable with respect to the sigma algebra 
$\FF\left(X_s, 0\leq s\leq \t^y_x\right)$. Let $\A^r$ denote the time
reversion of the event $\A$.  Then
$$
\Q_N(y)\P\left[\A,\t^y_x<\t^y_y\right]=\Q_N(x)\P\left[\A^r,\t^x_y<\t^x_x\right]
\Eq(S.7)
$$
}

For example, we have
$$
\Q_N(y)\P\left[\t^y_x=t,\t^y_x<\t^y_y\right]
=\Q_N(x)\P\left[\t^x_y=t,\t^x_y<\t^x_x\right]
\Eq(S.8)
$$
Of course the power of Lemma \ver.3 comes to bear when $x$ and $y$ are such 
that  the ratio between $\Q_N(x)$ and $\Q_N(y)$ is very large or very small. 

Formulas like \eqv(S.6) invite the use of Laplace transforms. Let us first 
generalize the notion of  stopping times to arrival times in sets. I.e.
for any set $I\subset \G_N$ we will set $\t^x_I$ to be the time of the first 
visit of the process, starting at $x$, to the set $I$. With this notion we 
define
the corresponding Laplace transforms
$$
G^y_{x,I}(u)\equiv \sum_{t\geq 0} e^{ut} 
\P\left[\t^y_x=t,\t^y_x\leq\t^y_{I}\right]\equiv \E\left[e^{u\t^y_x}
\1_{\t^y_x\leq\t^y_{I}}\right],\quad u\in\C
\Eq(S.9)
$$
(We want to include the possibility that $I$ contains $x$ and/or $y$ for
later convenience).
As we will see it is important to understand what the domains of these 
functions are. 
Since the  Laplace transforms defined in 
\eqv(S.9) are Laplace transforms of the distributions of positive random 
variables, all these functions exist and are analytic at least for 
all $u\in \C$ with $Re(u)\leq 0$. 
Moreover, if $ G^y_{x,I}(u_0)$ is finite for some $u_0\in \R_+$, then it is 
analytic in the half-space $Re(u)\leq u_0$.  As we will see later, each of
 the functions introduced in \eqv(S.9) will actually exist and be finite 
 for some $u_0>0$ 
(depending on $N$). 

Note that in particular
$$
G^y_{x,I}(0)=\P\left[\t^y_x\leq\t^y_{I}\right]
\Eq(S.10)
$$
and
$$
\frac d{du}G^y_{x,I}(u=0)\equiv \dot G^y_{x,I}(0) = 
\E\left[\t^y_x \1_{\t^y_x\leq\t^y_{I}}\right]
\Eq(S.11)
$$
The expected time of reaching $x$ from $y$ conditioned on the event not to 
visit $I$ in the meantime is expressed in terms of these functions as 
$$
\frac {\dot G^y_{x,I}(0)}{G^y_{x,I}(0)}=\E\left[\t^y_x|{\t^y_x\leq\t^y_{I}}
\right]
\Eq(S.12)
$$

An important consequence of Lemma \ver.3 is

\lemma{\ver.4} {\it Assume that $I$ is any subset of $\G_N$ containing 
$x$ and $y$. Then
 $$
\Q_N(y)G^y_{x,I}(u) =\Q_N(x) G^x_{y,I}(u)
\Eq(S.13)
$$
}

\proof Immediate from Lemma \ver.3. \endproof

Lemma \ver.4 implies in particular  that the Laplace transforms of the 
 {\it conditional} times are invariant under reversal, i.e.
$$
\frac {G^y_{x,I}(u)}{G^y_{x,I}(0)}=\frac { G^x_{y,I}(u)}{G^x_{y,I}(0)}
\Eq(S.14)
$$
and in particular
$$
\E\left[\t^y_x|{\t^y_x\leq\t^y_{I}}\right]=
\E\left[\t^x_y|{\t^x_y\leq\t^x_{I}}\right]
\Eq(S.15)
$$

\eqv(S.14) expresses the well-known but remarkable fact that in a reversible
process the conditional times to reach a point $x$ from $y$ without return 
to $y$ are equal to those to reach $y$ from $x$ without return to $x$. 

A special r\^ole will be played by the Laplace transforms
for which the exclusion set are all the minima. We will denote these by
  $g^y_x(u)\equiv G^y_{x,\MM_N}(u)$.
Indeed, 
we  think  of the events $\{\t^y_x\leq\t^y_{\MM_N}\}$, for $x,y\in \MM_N$, 
as elementary transitions and decompose any process going from one minimum 
to another into such elementary transitions. This gives for 
$G^y_x(u)\equiv G^y_{x,x}(u)$:

\lemma {\ver.5} {\it Let $x,y\in \MM_N$. Denote by $\o$ an arbitrary sequence
\break
$\o=\o_0,\o_1,\o_2,\o_3,\dots,\o_{|\tilde\o|}$ of elements 
$\o_i\in \MM_N$. Then we have
$$
G^y_x(u)=\sum_{\o:x\rightarrow y} p(\o) \prod_{i=1}^{|\o|}
\frac {g^{\o_{i-1}}_{\o_{i}}(u)}{g^{\o_{i-1}}_{\o_{i}}(0)}
\Eq(S.16)
$$
where
$$
 p(\o)\equiv \prod_{i=1}^{|\o|} 
\P\left[\t^{\o_{i-1}}_{\o_{i}}\leq\t^{\o_{i-1}}_{\MM_N}\right]
\Eq(S.17)
$$
and $\o:y\rightarrow x$ indicates that the sum is over such walks for which
$\o_0=y$ and $\o_{|\o|}=x$, and $\o_i\neq x$ for all $0<i<|\o|$.
}

Lemma \ver.5 can be thought of as a random walk representation of our process
as observed on the minima only. As we will show soon, the quantities 
$\frac {g^{\o_{i-1}}_{\o_{i}}(u)}{g^{\o_{i-1}}_{\o_{i}}(0)}$ are rather 
harmless, i.e. they do not explode in a small neighborhood of zero, and e.g.
their derivative at zero is at most polynomially large in $N$. 
On the other hand, we will also see that the ``transition probabilities''
$\P\left[\t^{\o_{i-1}}_{\o_{i}}<\t^{\o_{i-1}}_{\MM_N}\right]$
are all exponentially small provided that $\o_{i-1}\neq \o_i$. This means that
a typical walk will contain enormously long ``boring'' chains of repeated 
returns to the same point. It is instructive to observe that these repeated 
returns to a given 
minimum can be re-summed, to obtain a representation in terms
of walks that do not contain zero steps:

\lemma {\ver.6} {\it Let $x,y\in \MM_N$. denote by $\tilde\o$ a sequence
$\tilde\o=\o_0,\o_1,\o_2,\o_3,\dots,\o_{|\o|}$ of elements 
$\o_i\in \MM_N$ such that for all $i$, $\o_i\neq \o_{i+1}$. Then we have
$$
G^y_x(u)=\sum_{\tilde\o:x\rightarrow y} \tilde p(\tilde\o)
 \prod_{i=1}^{|\o|}\frac {1-g^{\o_{i-1}}_{\o_{i-1}}(0)}
{1-g^{\o_{i-1}}_{\o_{i-1}}(u)}
\frac {g^{\o_{i-1}}_{\o_{i}}(u)}{g^{\o_{i-1}}_{\o_{i}}(0)}  
\Eq(S.18)
$$
where
$$
\tilde p(\tilde\o)\equiv \prod_{i=1}^{|\o|}\frac 
{\P\left[\t^{\o_{i-1}}_{\o_{i}}\leq\t^{\o_{i-1}}_{\MM_N}\right]}
{\P\left[\t^{\o_{i-1}}_{\MM_N\ba\o_{i-1}}<\t^{\o_{i-1}}_{\o_{i-1}}\right]}
\Eq(S.19)
$$
}
The reason for writing Lemma \ver.6 in the above form is  that it 
entails as a corollary the following expression for the expected transition 
time:
$${
\E \t^y_x =\sum_{\tilde\o:x\rightarrow y} \tilde p(\tilde\o)
\sum_{i=1}^{|\o|} \left(\frac {\dot g^{\o_{i-1}}_{\o_{i-1}}(0)}
{1-g^{\o_{i-1}}_{\o_{i-1}}(0)}+\frac { \dot g^{\o_{i-1}}_{\o_{i}}(0)}
{g^{\o_{i-1}}_{\o_{i}}(0)}
\right)  
}
\Eq(S.20)
$$
Note that $\tilde p(\tilde \o)$ has indeed 
a natural interpretation as the probability of the sequence of steps 
$\tilde \o$, while each term in the sum is the expected time such a step takes.
Moreover, this time consists of two pieces: the first is a waiting time
which in fact arises from the re-summation of the many returns before a 
transition takes place while the second is the time of the actual transition, 
once it really happens. Note that the first term is enormous since the
denominator, $1-g^{\o_{i-1}}_{\o_{i-1}}(0)=
\P\left[\t^{\o_{i-1}}_{\MM_N\ba\o_{i-1}}<\t^{\o_{i-1}}_{\o_{i-1}}\right]$, 
is, as we will see, exponentially small.

\remark Lemma \ver.6 does provide a representation of the
process on the minima in terms of an embedded Markov chain with exponentially
small transition probabilities. Moreover, we expect that for 
$N$ large, the waiting times will be almost exponentially distributed
(but with very different rates!), while transitions happen essentially 
instantaneously on the scale of even the fastest waiting time. This is 
the analogue of the controlling Markov processes constructed in Freidlin and
Wentzell (see in particular Chap. 6.2 of [FW]).

In the case where $\MM_N$ consists of only two points, Lemma \ver.6
already provides the full solution to the problem since the only walk 
left is the single step $(y,x)$. 

\corollary {\ver.7} {\it Assume that $\MM_N=\{x,y\}$. Then
$$
G^y_x(u)=
\frac {g^{y}_{x}(u)} {1-g^{y}_{y}(u)}
\Eq(S.21)
$$
and 
$$
\E \t^y_x =
\frac {\dot g^{y}_{y}(0)}
{1-g^{y}_{y}(0)}+\frac { \dot g^{y}_{x}(0)}
{g^{y}_{x}(0)}
\Eq(S.22)
$$
}
\proof Just use that in this particular setting, $1-g^y_y(0)=g^y_x(0)$.
\endproof

\remark \eqv(S.22) can be written in the maybe more instructive form
$$
\E \t^y_x =
\frac {1-g^y_x(0)}{g^y_x(0)}\frac {\dot g^{y}_{y}(0)}
{g^{y}_{y}(0)}+\frac { \dot g^{y}_{x}(0)}
{g^{y}_{x}(0)}
\Eq(S.23)
$$
As we will see, all the ratios of the type $\dot g(0)/g(0)$
represent the expected times of a transition conditioned on the event that
this transition happens and should be thought of as ``small''. On the
other hand, the probability $g^y_x(0)$ will be shown to be exponentially small
so that the first factor in the  first term in \eqv(S.23) is extremely large.
Thus, to get a precise estimate on the expected transition time in this case,
it suffices to compute precisely the two quantities
$\dot g^{y}_{y}(0)$  and $g^y_x(0)$ only (the second term in \eqv(S.23)
being negligible in comparison). One might be tempted to think that in the
general case the random walk representation given through Lemma \ver.6
would similarly lead to a  reduction to the problem to that 
of computing the corresponding
quantities at and between all minima. This however is not so. The reason is
that the walks $\tilde\o$ still can perform more complicated multiple loops
and these loops will introduce new and more singular terms 
when appear explicitly in \eqv(S.18) and \eqv(S.20). 
This renders this representation much less useful
than it appears at first sight. On the other hand, the structure of the
representation of Corollary \ver.7 will be rather universal. Indeed, it is
easy to see that with our notations we have the following 

\lemma {\ver.8} {\it Let $I\subset \G_N$. Then for all $y\not\in I\cup x$,
$$
G^y_{x,I}(u)= \frac{G^y_{x,\{I\cup y\}}(u)}{1-G^y_{y,\{I\cup x\}}(u)}
\Eq(S.24)
$$
holds for all $u$ for which  the left-hand side exists. 
}

\proof Separating paths that reach $x$ from $y$ without return to 
$y$ from those that do return, and splitting the latter at the first return 
time, using the strong Markov inequality, we get that
$$
G^y_{x,I}(u) =G^y_{x,\{I\cup y\}}(u) +G^y_{y,\{I\cup x\}}(u) G^y_{x,I}(u)
\Eq(S.24bis)
$$
By construction, if $G^y_{x,I}(u)$ is finite, the second summand being 
 less than
the left-hand side, we have that $G^y_{y,\{I\cup x\}}(u)<1$
 and so \eqv(S.24) follows.\endproof

Lemma \ver.8 will be one of our crucial tools. In particular, since it relates
functions with exclusion sets $I$ to functions with larger exclusion sets, 
it suggests control over the Laplace transforms via induction over the 
size of the exclusion sets.

Lemma \ver.8 has two important consequences that are obtained by 
setting $u=0$ in \eqv(S.24) and by taking the derivative of \eqv(S.24)
with respect to $u$ and evaluating the result at $u=0$: 

\corollary {\ver.9} {\it  Let $I\subset \G_N$. Then for all $y\not\in I\cup x$,
$$
\P\left[\t^y_x<\t^y_I\right]= \frac{\P\left[\t^y_x<\t^y_{I\cup y}\right]}
{\P\left[\t^y_{I\cup x} <\t^y_y\right]}
\Eq(S.25ter)
$$
and
$$
\eqalign{
\E\left[\t^x_y|\t^x_y< \t^x_J\right]=&
\E\left[\t^x_y|{\t^x_y< \t^x_{J\cup x}}\right]
+\frac{\E\left[\t^x_x |{\t^x_x<\t^x_{J\cup y}}\right]}
{\P\left[\t^x_{J\cup y}<\t^x_x\right]}
\P\left[\t^x_x<\t^x_{J\cup y}\right]
}
\Eq(S.25quater)
$$
}


\medskip
\line{\bf \ver.3. Outline of the general strategy.\hfill}
\medskip

As indicated above, an important tool in our analysis will be the use
of induction over the size of exclusion sets by the help of Lemmata \ver.1 and
\ver.8. One of the basic inputs for this will be a priori estimates on the  
quantities $g^x_y(u)$. These will be based on the representation of these 
functions as solutions of certain Dirichlet problems associated to the operator
$(1-e^uP_N)$ with Dirichlet boundary conditions in set containing $\MM_N$.

The crucial point here is to have Dirichlet  boundary conditions
at all the minima of $F_N$ and at $y$. 
Without these boundary conditions, the stochastic 
matrix $P$ is symmetric in the space $\ell_2(\G_N,\Q_N)$ and 
has a maximal eigenvalue $1$ with corresponding (right)
 eigenvector $1$;
since this eigenvector does not satisfy the Dirichlet boundary conditions
at the minima, the spectrum of the Dirichlet operator lies strictly below $1$, 
so that for sufficiently small values of $u$, $1-e^u P$ is invertible. It is
essential  to know by how much the Dirichlet conditions
push the spectrum down. It turns out  that Dirichlet boundary conditions 
at all the minima push
the spectrum by an amount of at least  $CN^{-d-1}$ below one, 
and this will allow
us not only to construct the solution but to get very good control on its
behaviour. If, on the other hand, not all the minima had received 
Dirichlet conditions, we must expect that the spectrum is only pushed down 
by an exponentially small amount, and we will have to devise different 
techniques to deal with these quantities. 

As a matter of fact, while the spectral properties discussed above follow
from our estimates, we will not use these to derive them. The point is that
what we really need are pointwise estimates on our functions, rather than 
$\ell_2$ estimates, and we will actually use more probabilistic techniques
to prove $\ell_\infty$ estimates as key inputs. 
The main result, proven in Section 3, will be the following theorem:

\theo{\ver.10} {\it  
There exists a constant $c>0$ such that 
for all  $x\in \G_N$, $y\in \MM_N$
the
 functions $g^{x}_{y}(u)$ are analytic in the half-plane 
$Re(u)<cN^{-d-3/2}$. Moreover, for such $u$, 
for any non-negative integer $k$ there exists a constant $C_k$ such that
$$
\left|\frac {d^k}{du^k} g^{x}_{y}(u) \right|
\leq C_k
 N^{k(d+3/2)+d/2} 
e^{N [F_N(x)-F_N(z^*(x,y))]}  
\Eq(E.5)
$$
where $z^*(y,x)$ is defined in \eqv(T.2).}

These estimates  are not overly sharp, and there are no
corresponding lower bounds. 
Therefore, our strategy will be to use these estimates only to 
control sub-leading expressions and to use different methods to control the 
leading quantities which will be seen to be certain of the 
 expected return times, like
$\frac {\dot g^{x}_{x}(0)}{g^{x}_{x}(0)}$ and the transition probabilities
$\P\left[\t^x_y<\t^x_x\right]$. The latter quantities will be estimated 
up to a multiplicative error of order $N^{1/2}$ in Section 2. 
In fact we will prove there the following theorem: 

\theo {\ver.11}    {\it With the notation of Theorem \ver.10 
there exists finite positive  constants $c,C$ such that 
if $x\neq y\in \MM_N$, then 
$$
\P\left[\t^y_x<\t_y^y\right]\leq c N^{\frac{d-2} 2} 
 e^{-N[F_N(z^*(y,x))-F_N(y)]}
\Eq(T.6bis)
$$
and
$$
\P\left[\t^y_x<\t_y^y\right]\geq C N^{\frac{d-2} 2} 
 e^{-N[F_N(z^*(y,x))-F_N(y)]}
\Eq(T.6)
$$
}

The estimates for the return times require some more 
preparation and will be stated only  in Section 5, 
but let us mention that the main idea in getting sharp estimates
for them is the
use of the ergodic theorem .


Equipped with these inputs we will, in Section 4, proceed to the analysis
of general transition processes. We will introduce a natural tree structure 
on the set of minima and show that any transition between two minima 
can be uniquely decomposed into a sequence of so-called ``admissible 
transitions'' in such a way that with probability rapidly tending to one (as 
$N\uparrow\infty$), the process will consist of this precise
sequence of transitions. 
This will require large deviation estimates in path space that are special 
cases of more general results that have recently been proven in   [BG2]. 

In Section 5 we will investigate the 
transition times of admissible transitions. In the first sub-section 
 we will prove sharp bounds on the 
expected times of such admissible transitions with upper and lower bounds
differing only be a factor of $N^{1/2}$. This will be based on 
more general upper bounds on expected times of general types of transitions
that will be proven by induction. In the second sub-section  we show that the 
rescaled transition times converge (along subsequences) 
 to exponentially distributed random 
variables. This result again is based on an inductive proof establishing 
control on the rather complicated analytic structure of the Laplace 
transforms of a general class of transition times.  
In Section 6 we use these results to derive some consequences:
We show that during an admissible transition, at any given time, 
the process is
close to the starting point with probability close to 1, that it converges
exponentially to equilibrium, etc. Section 7 motivates the 
connection between our Markov chains and Glauber dynamics 
of disordered mean field models, and in Section 8
we discuss a specific example, the random field Curie-Weiss model.

\noindent{\bf Notation:}  We have made an effort to use a notation 
that is at the same time concise and unambiguous. This has required some 
compromise and it may be useful to outline our policy here. First,
all objects associated with our Markov chains depend on $N$. We make this
evident in some cases by a subscript $N$. However, we have 
omitted this subscript in other cases, 
in particular when there is already a number of other 
indices that are more important (as in $G^x_y(u)$), or in ever recurring 
objects like $\P$ and $\E$, 
and which sometimes will have to be distinguished from
the laws of modified Markov chains by other  subscript. 
Constants $c,C,k$ etc. will always be understood to depend on the details of
the Markov chain, but to be independent of $N$ for $N$ large. There will 
appear constants $K_N>0$  that will depend on $N$ in a way depending on the 
details of the chain, but such that for some $\a>0$, 
$N^{1-\a}K_N\uparrow \infty$ (this can be seen
as a requirement on the chain). Specific letters are reserved for a particular 
meaning only locally in the text. 

\thanks A.B. would like to thank Enzo Olivieri for an inspiring discussion on
reversible dynamics 
that has laid the foundation of this work. M. K. thanks Johannes 
Sj\"ostrand and B. Helffer for helpful discussions. 
The final draft of the paper has benefited from  
comments by and discussions with Enzo Olivieri, Elisabetta Scoppla, 
and Francesca Nardi. 
Finally,  V.G. and A.B.
 thank the Weierstrass-Institute,
Berlin, 
and the Centre de Physique Th\'eorique, Marseille, for hospitality and 
financial support that has made this collaboration possible.

\newpage

\chap{2. Precise estimates on transition probabilities}2

In this section we proof Theorem 1.11. A key ingredient is the following 
variational representation of the probabilities
$$
G^y_x(0)\equiv \P\left[\t^y_x<\t^y_y\right],\quad x,y\in \MM_N
\Eq(T.1)
$$
that can be found in Ligget's book ([Li], p. 99, Theorem 6.1).

\theo {\ver.1}[Li]  {\it Let $\HH^y_x$ denote the space of functions
$$
\HH^y_x\equiv \left\{ h:\G_N\rightarrow [0,1]: h(y)=0, h(x)=1\right\}
\Eq(T.9)
$$
and define the Dirichlet form
$$
\Phi_N(h)\equiv \sum_{x',x''\in\G_N} \Q_N(x')p_N(x',x'')[h(x')-h(x'')]^2
\Eq(T.10)
$$
Then 
$$
 \P\left[\t^y_x<\t_y^y\right]=\frac 1{2\Q_N(y)}\inf_{h\in \HH^y_x}\Phi_N(h)
\Eq(T.11)
$$
} 

\proof See Ligget [Li], Chapter II.6. Note that the set $R$ in Liggett's book  
(page 98) 
will be $\G_N\ba \{x\}$, and our $\HH^y_x$ would be $H_{\G_N\ba \{x\}}$
in his notation.\endproof

\proofof {Theorem 1.11.}  
The proof of the upper bound \eqv(T.6bis) is very easy.
We will just construct a suitable trial-function $h\in \HH^x_y$ and 
bound the infimum in \eqv(T.11) by the value of the Dirchlet form 
$\Phi_N$ at this function. 

To this end 
we construct a `hyper-surface'\note{We actually require no analytic 
properties for the set $\SS_N$ and the term hyper-surface 
should not be taken very seriously.}
  $\SS_N\subset \G_N$ separating $x$ and $y$ such that
\item{i)} $z^*(y,x)\in \SS_N$.
\item{ii)} $\forall z\in \SS_N$, $F_N(z)\geq F_N(z^*(y,x))$.

$\SS_N$ splits $\G_N$ into two components $\G_x$ and $\G_y$
which contain $x$ and $y$, respectively. Let $\chi\in\CC^\infty_0\left(
[0,\infty),[0,1]\right)$ with $\chi(s)=1$ for $s\in [0,1)$ and 
 $\chi(s)=0$ for $s>2$. Then put
$$
h_N(x')=\cases
\chi(N^{1/2}\dist(x',S_N)),&\,\hbox{for}\, x'\in \G_x\cr
    1 ,&\,\hbox{for}\, x'\in \G_y\cr
\endcases
\Eq(T.3)
$$
Clearly, $h_N$ is constant outside a layer $\LL_N$ of width $N^{-1/2}$
around $\SS_N$. Since the transition matrix $p_N(x',x'')$ 
vanishes for $|x'-x''|\geq CN^{-1}$, for some constant $C<\infty$
and since by construction
$$
|h_N(x')-h_N(x'')|\leq cN^{1/2} |x'-x''|
\Eq(T.3a)
$$
we obtain
$$
\eqalign{
\P\left[\t^x_y<\t^x_x\right]&\leq \frac 1{2\Q_N(y)}\inf_{h\in \HH^y_x}
\Phi_N(h_N)\cr
&\leq const. N^{-1}   \sum_{x',x''\in\LL_N} \frac {\Q_N(x')}{\Q_N(x)}
p_N(x',x'')\cr 
&\leq \frac CN \frac{\Q_N(z^*(x,y))}{\Q_N(x)}
 \sum_{x'\in\LL_N} \frac {\Q_N(x')}{\Q_N(z^*(x,y))}
}
\Eq(T.3b)
$$
Since $F_N$ is assumed to have a quadratic saddle point at $z^*(x,y)$, 
using a standard Gaussian approximation the final sum is readily seen to be 
$O(N^{d/2})$ which gives the upper bound \eqv(T.6bis).  
 
The main task of this section will be to establish the corresponding lower 
bound \eqv(T.6).
The main idea of the proof of the lower bound is to reduce the 
problem to a sum of essentially one-dimensional ones which can be solved 
explicitly. The key observation is the following monotonicity 
property of the transition probabilities.

\lemma{\ver.2} {\it Let $\D\subset\G_N$ be a subgraph of $\G_N$ and let
$\wt\P_\D$ denote the law of the Markov chain with transition rates
$$
\wt p_\D(x',x'') =\cases p_N(x',x'') ,&\text{if} x'\neq x'' 
\,,\hbox{and}\,\, (x',x'')\in E(\D)\cr
                      1-\sum_{y':(x',y')\in E(\D) }p_N(x',y'),&\text{if} 
x'=x''\cr
                     0,&\text{else}\endcases
\Eq(T.7)
$$
Assume that $y,x\in\D$. Then 
$$
\P\left[\t^y_x<\t_y^y\right]\geq\wt\P_\D\left[\t^y_x<\t_y^y\right]
\Eq(T.8)
$$
}

\proof 
To prove Lemma \ver.2 from here, just note that for 
any $h\in \HH^y_x$ 
$$
\eqalign{
\Phi_N(h) &\geq \sum_{(x',x'')\in\D} \Q_N(x')p_N(x',x'')[h(x')-h(x'')]^2
\cr
&= \Q_N(\D)\sum_{(x',x'')\in\D} \wt \Q_\D(x')\wt p_{\D}(x',x'')[h(x')-h(x'')]^2
}
\Eq(T.12)
$$
where $ \wt \Q_\D(x)\equiv \Q_N(x)/\Q_N(\D)$.
This implies immediately that 
$$
\inf_{h\in \HH^y_x}\Phi_N(h) \geq \Q_N(\D) \inf_{h\in \HH^y_x}\Phi_\D(h)
= \Q_N(\D) \inf_{h\in \HH^y_x(\D)}\Phi_\D(h)
\Eq(T.13)
$$
where 
$
\HH^y_x(\D)\equiv \left\{ h:\D\rightarrow [0,1]: h(y)=0, h(x)=1\right\}
$. Thus, using Theorem \ver.1 for the process $\wt P_\D$, we see that
$$
 \P\left[\t^y_x<\t_y^y\right]=\frac 1{2\Q_N(y)}\inf_{h\in \HH^y_x}\Phi_N(h)
\geq 
\frac 1{2\wt \Q_\D(y)}\inf_{h\in \HH^y_x(\D)}\Phi_\D(h)=
\wt\P_\D\left[\t^y_x<\t_y^y\right]
\Eq(T.14)
$$
which proves the lemma. \endproof

To make use of this lemma, we will choose $\D$ in a special way. Note that
the simplest choice would be to choose $\D$ as one single  path connecting 
$y$ and $x$ over the saddle point $z^*(y,x)$ in an optimal way. However, 
such a choice would produce a bound of the form $C N^{-1/2}
\exp\left(-NF_N(z^*(y,x))-F_N(y)\right)$ which differs from the 
upper bound by a factor
$N^{d/2}$. It seems clear that in order to improve this bound we must 
choose $\D$ in such a way that it still provides ``many'' paths connecting
$y$ and $x$. To do this we proceed as follows. 
Let $E$ be any number s.t. $F_N(z^*(y,x))>E>\max\left(F_N(y),F_N(x)\right)$
(e.g. choose 
$E=F_N(z^*(y,x))-\frac 12\left(F_N(z^*(y,x))-\max\left(F_N(y),F_N(x)\right)\right)$. 
Denote 
by $D_y$, $D_x$ the connected components of the level set
$\{x'\in\G_N:F_N(x')\leq E\}$ that contain the points $y$, resp. $x$.  

Note that of course we cannot, due to the discrete nature of the set 
$\G_N$, achieve that the function $F_N$ is constant on the actual discrete
boundary of the sets $D_y$, $D_x$. The discrete boundary $\del D$ of any set
$D\subset\G_N$, will be defined as
$$
\del D\equiv \left\{ x\in D|\exists y\in \G_N\ba D \,,\hbox{s.t.}\, (x,y)\in
E(\G_N)\right\}
\Eq(T.141)
$$
  We have, however, that 
$$
\sup_{x' \in \del D_y, x''\in \del D_x} |F_N(x')-F_N(x'')|\leq  C N^{-1}
\Eq(T.15)
$$

Next we choose a family of paths $\g_z:[0,1]\rightarrow \G_N$, indexed by 
$z\in B\subset \SS_N$  with the 
following properties:
\item{i)} $\g_z(0)\in \del D_y$, $\g_z(1)\in \del D_x$
\item{ii)} For $z\neq z'$, $\g_z$ and $\g_{z'}$ are disjoint (i.e. they 
do not have common sites or common edges. 
\item{iii)} $F_N$ restricted to $\g_z$  attains  its maximum at $z$.

Of course we will choose the set $B\subset \SS_N$
 to be a small relative neighborhood in 
$\SS_N$ of 
the saddle $z^*(y,x)$. In fact it will turn out to be enough to take $B$ a
disc of diameter $C N^{-1/2}$ so that its cardinality is bounded by $|B|\leq 
CN^{(d-1)/2}$.

For such a collection, we will set 
$$
\D\equiv D_x\cup D_y\cup\bigcup_{z\in B} V(\g_z)
\Eq(T.14bis)
$$
where $V(\g_z)$ denotes the graph composed of the vertices
that $\g_z$ visits and the edges along which it jumps;  
 the unions are to be understood 
in the sense of the union of the corresponding subgraphs 
of $\G_N$.

\lemma {\ver.3}{
With $\D$  defined above  
we have
$$
\eqalign{
&\wt\P_{\D}\left[\t^{y}_{x}<\t^{y}_{y}\right]\geq
\left(1- CN^{d/2} e^{-N[F_N(z^*(y,x))-E]}\right) \sum_{z\in B}
\frac{\Q_N(\g_z(1))}{\Q_N(y)}  \wt\P_{\g_z}\left[\t^{\g_z(1)}_{\g_z(0)}<
\t^{\g_z(1)}_{\g_z(1)}\right] 
}
\Eq(T.15bis)
$$
}

\proof All paths on $\D$  contributing to the event
$ \left\{\t^{y}_{x}<\t^{y}_{y}\right\}$ must now pass along 
one of the paths $\g_z$. 
    Using the strong Markov property, we split the paths at the first
arrival point in $D_x$ which gives the equality 
$$
 \wt\P_{\D}\left[\t^{y}_{x}<\t^{y}_{y}\right] =\sum_{z\in B}
 \wt\P_{\D}\left[\t^{y}_{\g_z(1)}\leq\t^{y}_{D_x\cup y}\right]
 \wt\P_{\D}\left[\t^{\g_z(1)}_{x}<\t^{\g_z(1)}_{y}\right]
\Eq(T.16)
$$
By reversibility,
$$
\eqalign{
 \wt\P_{\D}\left[\t^{y}_{\g_z(1)}\leq\t^{y}_{D_x\cup y}\right]
&=\frac{\Q_N(\g_z(1))}{\Q_N(y)} 
\wt\P_{\D}\left[\t^{\g_z(1)}_y\leq\t^{\g_z(1)}_{D_x}\right]\cr
&=\frac{\Q_N(\g_z(1))}{\Q_N(y)} 
\wt\P_{\D}\left[\t^{\g_z(1)}_{\g_z(0)}\leq\t^{\g_z(1)}_{D_x}\right]
\wt\P_{\D}\left[\t^{\g_z(0)}_y\leq\t^{\g_z(0)}_{D_x}\right]\cr
}
\Eq(T.17)
$$
where in the last line we used that the path going from $\g_z(1)$ to 
$y$ without further visits to $D_x$ must follow $\g_z$. 
Note further that we have the equality 
$$
\wt\P_{\D}\left[\t^{\g_z(1)}_{\g_z(0)}\leq\t^{\g_z(1)}_{D_x}\right]
=\wt\P_{\g_z}\left[\t^{\g_z(1)}_{\g_z(0)}\leq\t^{\g_z(1)}_{\g_z(1)}\right]
\Eq(T.18)
$$
where the right hand side  is a purely one-dimensional object. 
We will now show that 
the probabilities 
$\wt\P_{\D}\left[\t^{\g_z(1)}_{x}<\t^{\g_z(1)}_{y}\right]$
and $\wt\P_{\D}\left[\t^{\g_z(0)}_y\leq\t^{\g_z(0)}_{D_x}\right]$
are exponentially close to 1. To see this, write
$$
\eqalign{
1-\wt\P_{\D}\left[\t^{\g_z(1)}_{x}<\t^{\g_z(1)}_{y}\right]&=
\wt\P_{\D}\left[\t^{\g_z(1)}_{y}<\t^{\g_z(1)}_{x}\right]
= \frac {\wt\P_{\D}\left[\t^{\g_z(1)}_{y}<\t^{\g_z(1)}_{x\cup\g_z(1) }\right]}
{1-\wt\P_{\D}\left[\t^{\g_z(1)}_{\g_z(1)}<\t^{\g_z(1)}_{x\cup y}\right]}
}
\Eq(T.19)
$$
where the second equality follows from Corollary 1.9. 
Now by reversibility, the numerator in \eqv(T.19) satisfies the
bound
$$
\eqalign{
\wt\P_{\D}\left[\t^{\g_z(1)}_{y}<\t^{\g_z(1)}_{x\cup\g_z(1) }\right]
&\leq \sum_{x'\in B} \wt\P_\D\left[\t^{\g_z(1)}_{x'}<\t^{\g_z(1)}_{x\cup\g_z(1)}\right]
\cr
&\leq |B| \frac{\Q_\D(z^*(y,x))}{\Q_\D(\g_z(1))} = |B| \frac{\Q_N(z^*(y,x))}{\Q_N(\g_z(1))}
}
\Eq(T.20)
$$
On the other hand,
$$
\eqalign{
1-\wt\P_{\D}\left[\t^{\g_z(1)}_{\g_z(1)}<\t^{\g_z(1)}_{x\cup y}\right]
&=\wt\P_{\D}\left[\t^{\g_z(1)}_{x\cup y}<\t^{\g_z(1)}_{\g_z(1)}\right]\cr
&
\geq \wt\P_{\D}\left[\t^{\g_z(1)}_{x}<\t^{\g_z(1)}_{\g_z(1)}\right]
\cr
&\geq \wt\P_{\g}\left[\t^{\g_z(1)}_{x}<\t^{\g_z(1)}_{\g_z(1)}\right]
}
\Eq(T.21)
$$
where $\g$ is a a one dimensional  path going from $\g_z(1)$ to $x$. We will 
show later that
$$ \wt\P_{\g}\left[\t^{\g_z(1)}_{x}<\t^{\g_z(1)}_{\g_z(1)}\right]
  \geq C N^{-1/2}
\Eq(T.22)
$$
Thus we get that 
$$
\wt\P_{\D}\left[\t^{\g_z(1)}_x<\t^{\g_z(1)}_{ y}\right]
\geq 1-CN^{d/2} e^{-N [F_N(z^*(y,x))-E]}
\Eq(T.23)
$$
By the same procedure, we get also that 
$$
\wt\P_{\D}\left[\t^{\g_z(0)}_y\leq\t^{\g_z(0)}_{D_x}\right]
\geq 1-CN^{d/2} e^{-N [F_N(z^*(y,x))-E]}
\Eq(T.24)
$$
Putting all these estimates together, we arrive at the affirmation of the 
lemma.\endproof


We are left to prove the lower bounds for the purely one-dimensional
problems whose treatment is explained for instance in [vK]. 
In fact, we will show that 

\proposition {\ver.4} {Let $\g_z$ be a one dimensional path such that
$F_N$ attains its maximum on $\g_z$ at $z$. Then there is a constant 
$0<C<\infty$ such that
$$
\wt\P_{\g_z}\left[\t^{\g_z(1)}_{\g_z(0)}<\t^{\g_z(1)}_{\g_z(1)}\right]
 \geq CN^{-1/2} e^{-N[F_N(z)-F_N(\g_z(1))]}
\Eq(T.25)
$$
}

\proof Let $K\equiv |\g_z|$ denote the number of edges in the path $\g_z$.
Let us fix the notation $\o_0,\o_1,\dots,\o_{K}$,  for the 
ordered sites of the path $\g_z$, with $\g_z(1)=\o_0, \g_z(0)=\o_{|\g_z|}$.
 
For any site $\o_n$ we introduce the probabilities to jump to the right, resp.
the left 
$$
p(n)=p_N(\o_n,\o_{n+1}),\quad
q(n)=p_N(\o_n,\o_{n-1})
\Eq(T.26)
$$
We will first show that

\lemma{\ver.5} {\it With the notation introduced above,
$$
\wt\P_{\g_z}\left[\t^{\g_z(1)}_{\g_z(0)}<\t^{\g_z(1)}_{\g_z(1)}\right]
=
\left[\sum_{n=1}^{K} \frac{\Q_N(\o_0)}{\Q_N(\o_n)}\frac{1}{q(n)}
\right]^{-1}
\Eq(T.27)
$$
}

\proof  Let us denote by $r(n)$ the solution of the boundary value problem
$$
\eqalign{
&r(n)(p(n)+q(n))=p(n)r(n+1)+q(n)r(n-1),\text {for $0<n<K$}\cr
&r(0)=0,\quad r(K)=1
}
\Eq(T.28)
$$
Obviously we have that
$$
\wt\P_{\g_z}\left[\t^{\g_z(1)}_{\g_z(0)}<\t^{\g_z(1)}_{\g_z(1)}\right]
=p(0)r(1)
\Eq(T.29)
$$
\eqv(T.28) has the following well know  unique  solution
$$
r(n)=\frac{\sum_{k=1}^{n} \prod_{\ell=k}^{K-1}
\frac{p(\ell)}{q(\ell)}}{\sum_{k=1}^{K} \prod_{\ell=k}^{K-1}
\frac{p(\ell)}{q(\ell)}}
\Eq(T.30)
$$
hence,  
$$
\wt\P_{\g_z}\left[\t^{\g_z(1)}_{\g_z(0)}<\t^{\g_z(1)}_{\g_z(1)}\right]
=\frac {p(0)\prod_{\ell=1}^{K-1}\frac{p(\ell)}{q(\ell)}}
{\sum_{k=1}^{K} \prod_{\ell=k}^{K-1}
\frac{p(\ell)}{q(\ell)}}=\frac {p(0)}
{\sum_{k=1}^{K} \prod_{\ell=1}^{k-1}\frac{q(\ell)}{p(\ell)}}
\Eq(T.31)
$$
Now reversibility reads  
$
\Q_N(\o_\ell) p(\ell)=\Q_N(\o_{\ell+1}) q(\ell+1)
$, and this allows to simplify 
$$ 
\prod_{\ell=1}^{k-1}\frac{p(\ell)}{q(\ell)}=
\frac{q(k) \Q_N(\o_k)}{q(1)\Q_N(\o_1)}
\Eq(T.32)
$$
and finally 
$$ 
\wt\P_{\g_z}\left[\t^{\g_z(1)}_{\g_z(0)}<\t^{\g_z(1)}_{\g_z(1)}\right]
=\frac 1{\Q_N(\o_0)\sum_{k=1}^{K} \left[{q(k)\Q_N(\o_k)}\right]^{-1}}
\Eq(T.33)
$$
which is the assertion of the lemma. \endproof

We are left to estimate the sum 
$\Q_N(\o_0)\sum_{k=1}^{K} \frac 1{q(k)\Q_N(\o_k)}$ uniformly in $K$. 
Since $q(k)\geq c>0$ for 
all $1\leq k\leq K$, for an upper bound on this sum it is enough to consider
$$
\Q_N(\o_0)\sum_{k=1}^{K} \frac 1{\Q_N(\o_k)}=
\frac {\Q_N(\o_0)}{\Q_N(z)} \sum_{k=1}^{K} e^{-N[F_N(z)-F_N(\o_k)]}
\Eq(T.34)
$$
Now in the neighborhood of $z$, we can certainly bound 
$$
F_N(z)-F_N(\o_k)\geq c \left(\frac kN\right)^2
\Eq(T.35)
$$
while elsewhere $F_N(z)-F_N(\o_k)>\e>0$ (of course nothing changes 
if the paths have to pass over finitely many saddle points of equal height), 
and from this it follows immediately by elementary estimates that uniformly in 
$K$
$$
\sum_{k=1}^{K} e^{-N[F_N(z)-F_N(\o_k)]}\leq CN^{1/2} 
\Eq(T.36)
$$
which in turn concludes the proof of Proposition \ver.4.\note{Of course we 
could easily be more precise and identify the constant in 
\eqv(T.36) to leading order 
 with the second derivative of $F(z)$ in the direction of $\g$
(see e.g. [vK] where this computation is given in the case of
the continuum setting, and [KMST] where a formal asymptotic expansion 
is derived in the discrete case).}
 \endproof\endproof

Combining Proposition \ver.4 with Lemma \ver.3, we get that
$$
\eqalign{
&\wt\P_{\D}\left[\t^{y}_{x}<\t^{y}_{y}\right]\geq
\left(1- CN^{d/2} e^{-N[F_N(z^*(y,x))-E]}\right) \sum_{z\in B}
\frac{\Q_N(\g_z(1))}{\Q_N(y)} \frac {\Q_N(z)}{\Q_N(\g_z(1))} CN^{-1/2}
\cr&=e^{-N[F_N(z^*(y,x))-F_N(y)]}\left(1- CN^{d/2} 
e^{-N[F_N(z^*(y,x))-E]}\right)
 CN^{-1/2}  \sum_{z\in B} e^{-N[F_N(z)-F_N(z^*(y,x))]}\cr
}
\Eq(T.37)
$$
By our assumptions $F_N(z)-F_N(z^*(y,x))$ restricted to the surface
$\SS_N$ is bounded from above by a quadratic function in a small neighborhood 
of $z^*(y,x)$ and so, if $B$ is chosen to be such a neighborhood,
the lower 
bound claimed in Theorem 1.11 follows immediately by a standard Gaussian 
approximation of the last sum. \endproof\endproof

\vskip1cm

\def\ra{{\rangle}}
\def\la{{\langle}} 

\chap{3. Laplace transforms
of transition times in the elementary situation}3

In this section we shall prove Theorem 1.10, which is our basic estimate
for the Laplace transforms of elementary transition times. We shall need
the sharp estimates on the transition probabilities which we obtained in the
previous section based on Lemma 2.2. 
Combined with reversibility they lead 
to an estimate on the hitting time $\t^x_{\MM_N}$. This is the basic
analytic result needed to estimate the Laplace transforms, using their usual
representation as solutions of an appropriate boundary value problem.
Let us recall the notation
$$  
G^x_{y,\S}(u) = \E\left[e^{u \t^x_y} \1_{\t^x_y \leq \t^x_\S} \right],
\quad g^x_y(u)= G^x_{y,\MM_N}(u)
$$
In this section $\S$ will always 
denote a proper nonempty
subset of $\G_N$ that contains $\MM_N$. Moreover, we will assume that $y$ is 
not in the interior of $\S$, i.e. it is not impossible that $y$ is reached 
before $\S\ba y$ from $x$, since otherwise $G^x_{y,\S}(u)=0$
trivially.

%
%

To prove Theorem 1.10, it is enough to show that
$$
 g^{x}_{y}(u) 
\leq C_0
 N^{d/2} 
e^{N [F_N(z^*(x,y))-F_N(x)]}  
\Eq(3.1bis)
$$
for real and positive $u\leq cN^{-d-3/2}$.
Note that $z^*(x,y)$ is defined in \eqv(T.2) in such a way that 
$z^*(x,y)$ equals to $x$ if $y$ can be reached from $x$ without passing a
point at which $F_N$ is larger than $F_N(x)$.
 Analyticity then follows since
$g^x_y(u) $ is a Laplace transform of the distribution of 
a positive random variable, and the estimates for $k\geq 1$ follow using  
Cauchy's inequality.


In the sequel we will fix $y\in\S$ and $\MM_N\subset\S\subset\G_N$.
It will be useful to define the function
$$
v_u(x)=\left\{
\eqalign{  &G^x_{y,\S}(u) \text{for} x\notin\S 
\cr                       
&1              \text{for} x=y 
\cr                       
&0              \text{for} x\in \S\ba y.
}                      \right.
\Eq(3.3)
$$
As explained in the introduction, $v_u(x)$ 
is analytic near $u=0$ 
(so far without any control in $N$ on the region of analyticity).  

Similarly, we define the function
$$
w_{0}(x)=\left\{\eqalign {
&\E[\t^x_{\MM_N}] \text{for} x\notin\MM_N 
\cr
&0                  \text{for} x\in \MM_N 
}                       
\right.  
\Eq(3.5)
$$
Observe that as a consequence of Lemma 1.1 of 
the introduction we get (for  any $x,y \in \G_N, \S \subset \G_N$) that
$$
G^x_{y,\S}(u)=e^uP_N(x,y)+e^u\sum_{z\notin\S}P_N(x,z)G^z_{y,\S}(u).
\Eq(3.9)
$$
Using this  identity one readily deduces that $v_u$ is the unique solution of 
the boundary value problem
$$
(1-e^u P_N)v_u(x)=0 \quad(x \notin\S),\qquad 
v_u(y)=1,\quad v_u(x)=0 \quad(x\in\S\ba y).
\Eq(3.4)
$$
and, in the same way, $w_u$ is the unique solution of 
$$
(1-P_N)w_{0}(x)=1  \quad(x\notin\MM_N),\qquad  w_{0}(x)=0 \quad(x\in\MM_N).
\Eq(3.6)
$$

We shall use these auxiliary functions to prove the crucial

\lemma{\ver.1} {\it
There is a constant $C\in\R$ such that for all $N$ large
enough
$$
T_N:= \max_{y\in\G_N}\E[\t_{\MM_N}^y] \leq CN^{d+1}
\Eq(3.7)
$$ 
}

\proof
In view of the Kolmogorov forward equations it suffices to consider
the case $y \notin \MM_N$.  
We set $\S=\MM_N \cup y,$ where $y \notin \MM_N$. 
Then $v_0(x)$ defined in \eqv(3.3) solves the Dirichlet problem
$$
\eqalign{
&(1-P_N)v_0(x)=0 \quad(x\notin\S)\cr
&
v_0(y)=1,\qquad v_0(x)=0 \quad(x\in\MM_N)
}
\Eq(3.8)
$$
Moreover, \eqv(3.9) with $u=0$ and $x=y$  reads  
(since  $G^y_{y,\S}(0)= \P[\t^y_y \leq \t^y_\S]$)
$$
1 - \P\left[\t^y_\S <\t^y_y \right] = \sum_{z \in \G_N} p_N(y,z) v_0(z)
\Eq(3.10)
$$
which can be written as
$$
(1-P_N)v_0(y)=\P[\t^y_{\MM_N}<\t^y_y]
\Eq(3.81)
$$
We shall use $v_0(x)$ as a fundamental solution for $1-P_N$ and, using the
 symmetry of $P_N$ in $\ell^2(\G_N,\Q_N)$, we get
$$
\eqalign{
\Q_N(y)\P[\t_{\MM_N}^{y}<\t_{y}^{y}]\E[\t_{\MM_N}^{y}]
&=\la (1-P_N)v_0,w_{0} \ra_{\Q}
\cr
&=\la v_0,(1-P_N)w_{0} \ra_{\Q}
\cr
&=\Q_N(y)+\sum_{x \notin\S}
\Q_N(x)\P[\t^{x}_y < \t^{x}_{\MM_N}],
}
\Eq(3.11)
$$
where in the last step we have used equation \eqv(3.6)
and the fact that $y \notin \MM_N.$
This gives the crucial formula for the expected hitting time
in terms of the invariant measure $\Q_N$ 
 and transition probabilities, namely
$$
\E[\t_{\MM_N}^{y}]=
\sum_{x \notin \S}
{\Q_N(x) \over \Q_N(y)}
{\P[\t^{x}_{y}<\t^{x}_{\MM_N}] \over \P[\t^{y}_{\MM_N}<\t^{y}_{y}]}
+{1 \over \P[\t^{y}_{\MM_N}<\t^{y}_{y}]}.
\Eq(3.12)
$$
We remark that in this sum only those values of $x$ with 
$\Q_N(x) \geq \Q_N(y)$ can give a large contribution. 
To estimate the probabilities in
equation \eqv(3.12) we choose, given the  starting point 
$y \notin \MM_N$,  an appropriate minimum 
$z \in \MM_N$ near $y$ such that there is  a path $\g:y \rar z$
(of moderate cardinality) so that $F_N$ attains its maximum
on $\g$ at $y$ (note that such a $z$ exists trivially always).  
Then the variational principle in equation \eqv(T.14)
(with $\g$ as the subgraph $\Delta$) gives
$$
\P[\t_{\MM_N}^y<\t_y^y] \geq \P[\t_{z}^y<\t_y^y] 
\geq  \wt{\P}_\g [\t_{z}^y<\t_y^y],
\Eq(3.13)
$$
where the first inequality is a trivial consequence of $z \in \MM_N.$
But then Proposition 2.5 can be applied to get the lower bound
$$
\P[\t_{\MM_N}^y<\t_y^y] \geq  C N^{-1/2}
\Eq(3.15)
$$
for some constant $C$.

To estimate the other probability in \eqv(3.12) we use Corollary 1.9 to 
write, for $x\not\in\S$, 
$$
\P[\t^{x}_{y}<\t^{x}_{\MM_N}]
={\P[\t^{x}_{y}<\t^{x}_{\MM_N\cup x}] \over
\P[\t^{x}_{\S}<\t^{x}_{x}]}
\Eq(3.17)
$$
Since $\MM_N \subset \S$, we obtain from \eqv(3.15) that for $x\not\in\S$,
$$
\P[\t^{x}_{\S}<\t^{x}_{x}] \geq \P[\t^{x}_{\MM_N}<\t^{x}_{x}] \geq  C N^{-1/2}
\Eq(3.18)
$$
Reversibility then gives the upper bound
$$
\P[\t^{x}_{y}<\t^{x}_{\MM_N\cup x}]
={\Q_N(y) \over \Q_N(x)} \P[\t^{y}_{x}<\t^{y}_{\MM_N\cup y}]
\leq \min \left(1,{\Q_N(y) \over \Q_N(x)} \right).
\Eq(3.19)
$$
Thus, inserting \eqv(3.18) and \eqv(3.19) into \eqv(3.17)  we
obtain from the representation \eqv(3.12) that
$$
\E[\t^y_{\MM_N}]\leq  C N (1 + \sum_{x \notin \S} 1)
\leq CN^{d+1} .
\Eq(3.20)
$$
for some constant $C$. This proves the lemma.
\endproof

Next we need an estimate on the Laplace transform $G^x_{y,\S}(u)$.
This will be obtained from an integral representation of our auxiliary
function $v_u(x)$, choosing $u$ smaller than the estimate on the inverse
of the maximal expected time $T_N$ obtained in Lemma \ver.1.
More precisely, we shall prove

\lemma{\ver.2}  {\it
Assume that $\MM_N \subset \S \subset \G_N.$  Then there is a constant
$c>0$ such that for all $u\leq cN^{-d-1}$ and all $ x,y \in \G_N $,
$$
G^x_{y,\S}(u)\leq 2  
\Eq(3.21)
$$
Furthermore, there are constants $b,c >0$ such that for all 
$u\leq cN^{-d-3/2}$ and $ y \in \G_N\ba \MM_N$,
$$
1-G^y_{y,\S}(u)\geq bN^{-1/2}
\Eq(3.22)
$$
}

\proof
As mentioned
in the beginning of this chapter we can assume without loss of generality that
$y \in \partial\S$. 
Then it follows
from equation \eqv(3.4) that the function 
$w_u(x):= v_u(x) - v_0(x)
$
solves the Dirichlet problem
$$
\eqalign{
&(1-P_N) w_u(x) = (1-P_N) v_u(x) =(1-e^{-u}) v_u(x), \quad (x \notin \S)\cr
&\qquad w_u(x) =0 \quad (x \in \S)
}
\Eq(3.22b)
$$
The relation between resolvent and semi-group gives  
the following representation for $x \notin \S$
$$
w_u(x) = \E \left[ \sum _{t=0}^{\t^x_\S -1} f(X_t) \right],
\quad f(x):=(1-e^{-u})v_u(x)
\Eq(3.23)
$$
that in turn yields  the integral equation 
$$
v_u(x)= \P[\t^x_y = \t^x_\S] +  (1-e^{-u})\E 
\left[\sum _{t=0}^{\t^x_\S -1} v_u(X_t)
\right].
\Eq(3.24)
$$
for the function $v_u$. We can now use our a priori bounds from Lemma \ver.1 
 on the expectation 
of the stopping time $\t^x_\S$ to extract an upper bound for the 
sup-norm of this function. Namely, 
setting $M(u):= \sup_{x \notin \S} v_u(x)$ we obtain 
the estimate
$$
M(u) \leq 1 + |1-e^{-u}|  \max_{x \in \G_N} \E[\t^x_\S] M(u)
\leq 1+  {1 \over 3} M(u),
\Eq(3.25)
$$
where we have used that  $|u| < c N^{-d -1}$
with $c$ sufficiently small. 
This gives for $x\not\in\S$,
$$
G^x_{y,\S}(u)\leq 3/2 
\Eq(3.25b)
$$
The estimate of the Laplace transform
$G^x_{y,\S}(u)$ 
is trivial for negative $u$ or for $x \in \S \ba \partial\S$. 
In the case $x \in \partial \S$, \eqv(3.21) follows from 
\eqv(3.9), using \eqv(3.25b). 

To prove the  estimate \eqv(3.22) on the Laplace transform $G^y_{y,\S}$
of the recurrence time to the boundary point $y \in \partial \S,$
(in particular $y \in \S \ba \MM_N$ under our assumptions), observe
that for any $\d>0$, there exists $c>$ such that for $|u| < c N^{-d-3/2}$,
using  Lemma \ver.2 to estimate
$\E[\t^x_\S] \leq \E[\t^x_{\MM_N}]$ from above, it follows that
$$
\E\left[ \sum _{t=0}^{\t^x_\S -1} (1-e^{-u}) \right] \leq \d N^{-1/2}
\Eq(3.26)
$$
Inserting this estimate and the a priori bound \eqv(3.21)  into
 \eqv(3.24) together with the a priori bound gives then that
$$
G^x_{y,\S}(u) \leq  \P[\t^x_y = \t^x_\S] + 2\d  N^{-1/2},
\Eq(3.26b)
$$

Inserting \eqv(3.26b) into \eqv(3.9), which represents $G^y_{y,\S}(u)$ via
$G^x_{y,\S}(u)$ for $x \notin \S,$ it follows that modulo 
$\d N^{-1/2}$ one has, for $|u| < c N^{-d-3/2}$,
$$
\eqalign{
1-G^y_{y,\S}(u) 
&\geq 1 -e^u P_N(y,y) - e^u \sum_{x \notin \S} p_N(y,x) \P[\t^x_y = \t^x_\S]
\cr
&=    1- \P[\t^y_\S=\t^y_y]
\cr
&= \P[\t^y_\S < \t^y_y].
}
\Eq(3.27)
$$
Since $\MM_N \subset \S$ and $y \in \S \ba \MM_N$ one obtains from
\eqv(3.18)
that 
$$
1-G^y_{y,\S}(u) 
\geq \P[\t^y_{\MM_N} < \t^y_y] - 2\d N^{-1/2} 
\geq  b  N^{-1/2}
\Eq(3.27b)
$$
for some $b>0,$ choosing $\d$ sufficiently small in equation \eqv(3.26). This proves Lemma \ver.2.
\endproof

We are now ready to give the

\proofof {Theorem 1.10}
Note that  when  $F_N(x)=F_N(z^*(x,y))$, Lemma \ver.2 already provides 
the desired (actually a sharper) estimate. 
It remains to consider the case   
$z^*(x,y)\neq x$.


Here we can,  as in the proof of Theorem 1.11 in Section 2,
construct a discrete
separating hyper-surface $\SS_N$ containing the minimal saddle
$z^*(x,y)$  and  separating $y$ and $x.$
Since the process starting at $x$ must hit $\SS_N$ before hitting $y,$
path splitting at $\SS_N$ gives
$$ 
g^x_y(u)=\sum_{z\in\SS_N}G_{z,\O}^x(u)g_y^z(u), \qquad \O= \MM_N \cup \SS_N.
\Eq(3.28)
$$
We treat the cases $x \in \MM_N$ and $ x \notin \MM_N$ separately.
In the latter case we need an additional renewal argument, while in the former
all loops are suppressed since 
the process is killed upon arrival at $x \in \MM_N.$
For $x \notin \MM_N$ the renewal equation \eqv(S.24) reads
$$
G_{z,\O}^x(u)=
(1-G_{x,\O}^x(u))^{-1}
G_{z,\O\cup x}^x(u) 
\Eq(3.29)
$$
By Lemma \ver.2 and reversibility we have
$$
G_{z,\O\cup x}^x(u)
={\Q_N(z) \over \Q_N(x)}G_{x,\O}^z(u)
\leq 2 {\Q_N(z) \over \Q_N(x)},
\Eq(3.30)
$$
using Lemma \ver.2. Combining \eqv(3.30) and \eqv(3.22) of  Lemma \ver.2
we get from the renewal equation \eqv(3.29)
$$
G_{z,\O}^x(u) \leq CN^{1/2}{\Q_N(z) \over \Q_N(x)},
\quad (z\in\SS_N, u \leq c N^{-d-3/2})
\Eq(3.31)
$$
for $c>0$ sufficiently small.

If $x \in \MM_N,$ we directly apply the reversibility argument to
$G_{z,\O}^x(u)$ (without renewal) and obtain a sharper estimate, i.e.
\eqv(3.31) with $N^{1/2}$ deleted on the right hand side.

Inserting \eqv(3.31) into \eqv(3.28)  and using   \eqv(3.21)
to estimate the Laplace transform $g^z_y(u) = G^z_{y,\MM_N}(u)$
we finally get, for $ u \leq c N^{-d-3/2}$,
$$
g^x_y(u)\leq CN^{1/2}  \Q_N(x)^{-1} \sum_{z \in \SS_N} \Q_N(z) = 
{\cal O}(N^{d/2}) e^{-N (F_N(z^*(x,y)) -F_N(x))},
\Eq(3.32)
$$
where the last equality is obtained by a 
standard gaussian approximation as \eqv(T.37). 
All estimates on the derivatives $k \geq 1$ now follow from
Cauchy's inequality and the obvious extension of our estimates
to complex values of $u.$
This completes the proof of Theorem 1.10.\endproof\endproof


\vskip1cm

\chap{4. Valleys, trees and graphs}4

In this chapter we provide the setup for the inductive treatment of the 
global problem. Although this description is not particularly original, and 
is essentially equivalent to the approach of   Freidlin and Wentzell [WF], 
we give a 
self-contained exposition of our version that we find particularly suitable 
for the specific problem at hand. 
To keep the description as simple as possible, we  make the assumption 
that $F_N$ is ``generic'' in the sense that no accidental 
symmetries or other ``unusual'' structures occur. This will be made more 
precise below. 
For the case of a random system, this appears a natural assumption.

\medskip
\line{\bf \ver.1. The valley structure and its tree-representation\hfill}
\medskip
We recall from Section 1 the definition \eqv(T.2) of essential saddle points. 
Under our general assumptions (G1), any esssential saddle has
 the property that 
 the  connected (according to the graph structure on $\G_N$)
component of the level set $\L_z \equiv \{x\in\G_N :F_N(x)\leq F_N(z)\}$ 
that contains $z$ falls into two disconnected components when
$z$ is removed from it. 

These two components are called ``valleys'' and denoted
by $V^\pm(z)$, with the understanding that
$$
\inf_{x\in V^+(z)}F_N(x)<\inf_{x\in V^-(z)}F_N(x)
\Eq(V.1)
$$
holds. We denote by $\EE_N$ the set of all essential saddle points.

With any valley we associated two characteristics: its ``height'',
$$
h(V^i(z))\equiv F_N(z)
\Eq(V.2)
$$
and its  ``depth''
$$
d(V^i(z))\equiv F_N(z)-\inf_{x\in V^i(z)}F_N(x)
\Eq(V.3)
$$

The essential topological structure of the landscape $F_N$ is encoded in a tree
structure that we now define  on the set $\MM_N\cup\EE_N$. To construct this, 
we define, for any essential saddle $z\in\EE_N$, the two points
$$
z^\pm_z=\cases \arg\max_{z_i\in \EE_N\cap V^\pm(z)} F_N(z_i),&\text {if}
 \EE_N\cap V^\pm(z)\neq \emptyset\cr
\MM_N\cap V^\pm(z) ,& \text {else}
\endcases
\Eq(V.4)
$$
(note that necessarily the set $\MM_N\cap V^\pm(z)$ consists of a single point
if $\EE_N\cap V^\pm(z)= \emptyset$). Now draw a link from any
essential saddle to the 
two points $z^\pm_z$. This produces a connected tree, $\TT_N$, with vertex set 
$\EE_N\cup\MM_N$ having the property that all the vertices with coordination 
number $1$ (endpoints) correspond to local minima, while all other vertices 
are essential saddle points.  An alternative equivalent way to construct this
tree is by starting from below: Form each local minimum, draw a link 
to the lowest essential saddle connecting it to other minima. Then from each 
saddle point that was reached before, draw a line to the lowest saddle
above it that connects it to further minima. Continue until exhaustion.
We see that under our assumption of non-degeneracy, both procedures 
give a unique answer. (But note that in a random system the answer can 
depend on the value of  
$N$!)  

The tree $\TT_N$ induces a natural hierarchical distance   between two 
points in $\EE_N\cup \MM_N$, given by the length of the shortest path 
on $\TT_N$ needed to join them. 
We will also call  the ``level'' of a vertex its distance to the root,
$z_0$. 

The properties of the long-time behaviour of the process will be mainly 
read-off from the structure of the tree $\TT_N$ and the values of $F_N$ on 
the vertices of $\TT_N$. However, this information 
will not be quite sufficient. In fact, we will see that the information encoded
in the tree contains all information on the time-scales of ``exits'' from 
valleys; what is still missing is how the process descends into a neighboring 
valley after such an exit. It turns out that all we need to know in addition 
is which minimum the process visits {\it first} after crossing a saddle 
point. This point deserves some discussion. First, we note that the techniques 
we have employed so far in this paper are insufficient to answer such a 
question. Second, it is clear that without further assumptions,
there will not be a deterministic answer to this question; that is,
in general it is possible that the process has the option to visit
various minima first with certain probabilities. If this situation 
occurs, one should compute these probabilities; this appears, however, 
an exceedingly difficult task that is beyond the scope of the present paper.
We will therefore restrict our attention to the situation where $F_N$ is such 
that there is always one minimum that is visited first with overwhelming 
probability. To analyse this problem, we need to discuss an issue that we
have so far avoided, that of sample path large deviations for the 
(relatively) short time behaviour of our processes.  A detailed treatment of
this problem is given in [BG2] and, as this issues concerns the present paper 
only marginally, we will refer the interested reader to that paper 
and keep the discussion here to a minimum. What we will need here is that for 
``short'' times, i.e. for times $t=TN$, $T<\infty$, the process starting
at  any point $x_0$ at time $0$ 
will remain (arbitrarily) close (on the macroscopic scale) to certain
deterministic trajectories $x(t, x_0)$ with probability exponentially close to 
one\note{Convergence of this type of processes to 
deterministic trajectories was first proved on the level of the 
law of large numbers by Kurtz [Ku].}. 
These trajectories are solutions of certain differential equations 
involving the function $F$. In the continuum approximation they are just the 
gradient flow of $F$, i.e. $\frac d{dt}x(t)=-\nabla F(x(t)),\quad x(0)=x_0$, 
and    while the equations are more complicated in the discrete 
case they are essentially of similar nature. In particular, all critical 
points are always fixpoints. 
We will assume that the probability to reach a 
$\d$-neighborhood of the boundary of $\L$  in finite time 
$T$ will be exponentially small for all fixed $T$.
We will assume further  that  at each essential saddle 
the deterministic paths starting in a neighborhood of $z$ lead into 
uniquely specified minima within the two valleys connected through $z$. As 
we will see, these paths will determine the behaviour of the process.

We will incorporate these information in our graphical representation by
decorating the tree by adding   two 
yellow
\note{This color was used in the original drawing on a blackboard in 
the office of V. G.  in the
CPT, Marseille, and is retained here for historical reasons.}  
arrows pointing from each essential saddle to the 
minima in each of the branches of the tree emanating from it into 
which the deterministic paths lead. (These branches 
are essentially  obtained by following the gradient flow 
from the saddle into the next 
minimum on both sides.) 
We denote the tree decorated with the yellow arrows by 
$\wt\TT_N$.

\medskip
\line{\bf \ver.2. Construction of the transition process\hfill }
\medskip 
 
We are in principle interested in questions like ``how long does the
process take to get from one minimum to another?''. 
This question is more subtle 
than one might think. A related question, that should precede the previous one,
is actually ``how does the process get from one minimum to another one?'', 
and we will first make this question precise and provide an answer.

We recall that in \eqv(S.16) we have given a representation of the 
process going from $y$ to $x$ in terms of a random walk on the minima.
As we pointed out there, this representation was not
extremely useful. We will now show that it is possible to give another 
decomposition of the process that is much more useful.

Let us consider the event $\FF(x,y)\equiv
\{\t^y_x<\infty\}$ with $x,y\in \MM_N$. Of course this event has probability
one. We  now describe an algorithm that will allow to decompose this 
event, up to a set of exponentially small measure, into a sequence of 
``elementary'' transitions of the form
$$
\FF(x_i,z_i,x_{i+1})\equiv 
\left\{ \t^{x_i}_{x_{i+1}}\leq\t^{x_i}_{\TT_{z_i,x_i}^c\cap\MM_N}\right\}
\Eq(V.5)
$$
where $x_i,x_{i+1}\in \MM_N$, $z_i$ is the first common ancestor 
of $x_i$ and $x_{i+1}$ in the tree $\TT_N$, and $\TT_{z_i,x_i}$ is the 
branch of $\TT_N$ emanating from $z_i$ that contains  $x_i$, and
$\TT_{z_i,x_i}^c\equiv \TT_N\ba \TT_{z_i,x_i}$. 
We will write $\TT_z$ for the union of all branches emanating from 
$z$.
The motivation for this definition is contained in the following 

\proposition {\ver.1} {\it Let $x,y\in \TT_{z,y}\cap\MM_N$, and $\bar y\in 
\TT_{z,x}^c\cap\MM_N$. Then there is a constant $C<\infty$ such that  
$$
\P
\left[ \t^x_{\bar y}<\t^x_y\right]\leq \inf_{z'\in\TT_{z,y}\ba z}
C
e^{-N [F_N(z)-F_N(z')]}  
\Eq(V.6)
$$
}
\remark Note that by construction we have $F_N(z)-F_N(z')>0$ for all lower 
saddles in the branch $\TT_{z,y}$. Thus the proposition asserts that with 
enormous probability, the process starting from any minimum in a given valley 
visits all other minima in that same valley before visiting any 
minimum outside of this valley. As a matter of fact, the same also holds for 
general points. Thus what the proposition says is that up to the first exit 
from a valley, the process restricted to this valley behaves like 
an ergodic one. 

\proof We use Corollary 1.9 with $I$ consisting of a single point. This gives
$$
\eqalign{
 \P\left[ \t^x_{\bar y}<\t^x_y\right]&=\frac{\P\left[\t^x_{\bar y}<\t^x_{x\cup
\bar y}\right]}{\P\left[\t^x_{y\cup \bar y}<\t^x_{x}\right]}
\leq \frac{\P\left[\t^x_{\bar y}<\t^x_{x}\right]}
{\P\left[\t^x_{y}<\t^x_{x}\right]}
}
\Eq(V.7)
$$
Using the upper and lower bounds from Theorem 1.11 for the numerator and 
denominator, resp., we get 
$$
\P\left[ \t^x_{\bar y}<\t^x_y\right]\leq C e^{-N[F_N(z)-F_N(z')]}
\Eq(V.11)
$$
where $z'$ is the lowest saddle connecting $x$ and $y$. \eqv(V.11) yields the 
proposition.\endproof

Proposition \ver.1 implies in particular that the process will visit
the lowest minimum in a given valley before exiting from it, with enormous
probability. This holds true on any level
of the hierarchy of valleys. These visits at the lowest minima thus serve as a 
convenient breakpoint to organize any transition into elementary steps
that start at a lowest minimum of a given valley and exit just into the 
next hierarchy. This leads to the following definition.

\definition {\ver.2} {\it A transition $\FF(x,z,y)$ is called 
admissible, if
\item{i)} $x$ is the deepest minimum in the branch $\TT_{z,x}$,
i.e. $F_N(x)=\inf_{x'\in \TT_{z,x}}F_N(x)$.
\item{ii)}    $z$ and $y$ are connected by a yellow arrow in
 $\wt\TT_N$.
}
  
\remark We already understand why an admissible transition should start 
at deepest minimum: if it would not, we would know that the process would 
first go 
there, and we could decompose it into a first transition to this lowest 
minimum, and then an admissible transition to $y$. What we do not see yet,
is where the condition on the endpoint (the yellow arrow) comes from.
The point here is that upon exiting the branch $\TT_{z,x}$, the process 
has to arrive somewhere in the other branch emanating from $z$.
We will show later that with exponentially large probability this is the 
first  minimum  which the deterministic path staring  from $z$ leads to.

\proposition {\ver.3} {\it If $\FF(x,z,y)$ is an admissible 
transition, then there exists $K_N>0$, satisfying 
$N^{1-\a}K_N\uparrow\infty$ such that 
$$
\P\left[ \FF(x,z,y) \right]\geq 1-e^{-NK_N} 
\Eq(V.12)
$$
}

\remark To proof this proposition, we will use the large deviation estimates 
that require the stronger regularity assumptions ${\bf R2, R4}$, as well as
the structural assumptions discussed in the beginning of this section. 
These are to some extent technical and clearly not necessary. 
Alternatively, one can replace these by the assumption that 
Proposition \ver.3 holds, i.e. for any $z\in\EE$ and $x\in \TT_{z,x}$
there is a unique $y\in\TT_{z,x}^c\cup\MM_N$ such that \eqv(V.12) holds.

\proof The proof is based on the fact that the process will, with probability 
one, hit the set $\TT_{z,x}^c$ eventually. Thus, if we show that 
given $x$ and $z$, for all   
$\tilde y\in \TT_{z,x}^c$ with $\tilde y\neq y$,
$\P\left[\t^x_{\tilde y}<\t^x_{ \TT_{z,x}^c\ba {\tilde y}}\right]$ is  
exponentially small, 
the proposition follows.  
To simplify the notation, let us set $I= \TT_{z,x}^c\cap\MM_N$. 
Note that the case ${\tilde y}\not\in \TT_z$ is already covered by 
Proposition \ver.1, 
so we assume that ${\tilde y}\in \TT_z$.
Using Corollary 1.9
$$
\eqalign{
\P\left[\t^x_{\tilde y}<\t^x_{I\ba {\tilde y}}\right]&=
\frac{\P\left[\t^x_{\tilde y}<\t^x_{I\ba {\tilde y}\cup x}\right]}
{\P\left[\t^x_I<\t^x_x\right]}
}
\Eq(V.101)
$$
By reversibility, 
$$
\P\left[\t^x_{\tilde y}<\t^x_{I\ba {\tilde y}\cup x}\right]=
e^{N[F_N(x)-F_N({\tilde y})]}
\P\left[\t^{\tilde y}_x<\t^{\tilde y}_{I}\right]
\Eq(V.102)
$$
Now construct the separating hyper-surface $\SS_N$ passing through $z$ as in 
the proof of Theorem 1.11. Then 
$$
\P\left[\t^{\tilde y}_x<\t^{\tilde y}_{I}\right]=
\sum_{z'\in \SS_N}
\P\left[\t^{\tilde y}_{z'}\leq \t^{\tilde y}_{I\cup \SS_N}\right]
\P\left[\t^{z'}_x<\t^{z'}_I\right]
\Eq(V.103)
$$
Putting all things together, and using reversibility once more, 
we see that
$$
\eqalign{
\P\left[\t^x_{\tilde y}<\t^x_{I\ba {\tilde y}}\right]&=\frac {1}
{\P[\t^x_I<\t^x_x]}
\sum_{z'\in \SS_N}e^{-N[F_N(z')-F_N(x)]}
\P\left[\t^{z'}_{\tilde y}\leq \t^{z'}_{I\cup \SS_N}\right]
\P\left[\t^{z'}_x<\t^{z'}_I\right]
}
\Eq(V.104)
$$
Using that $\P[\t^x_I<\t^x_x]\geq \P[\t^x_{y'}<\t^x_x]$ for any $y'\in I$, 
together with  the lower bound of Theorem 1.11  and the trivial bound
$  \P\left[\t^{z'}_x<\t^{z'}_I\right]\leq 1$, we see that
$$
\eqalign{
\P\left[\t^x_{\tilde y}<\t^x_{I\ba {\tilde y}}\right]&\leq C^{-1} N^{-(d-2)/2}\sum_{z'\in \SS_N}e^{-N[F_N(z')-F_N(z)]}
\P\left[\t^{z'}_{\tilde y}\leq \t^{z'}_{I\cup \SS_N}\right]\cr
&\leq C^{-1} N^{-(d-2)/2}\sum_{{z'\in \SS_N}\atop{F_N(z')-F_N(z)\leq K_N}}
 e^{-N[F_N(z')-F_N(z)]}
\P\left[\t^{z'}_{\tilde y}\leq \t^{z'}_{I\cup \SS_N}\right]\cr
&+N^{d/2+1} e^{-NK_N}
}
\Eq(V.104bis)
$$
Under our assumptions the condition $F_N(z')-F_N(z)\leq K_N$ implies that
$|z'-z|\leq C' \sqrt  {K_N}$, i.e. all depends on 
the term 
$\P\left[\t^{z'}_{\tilde y}\leq \t^{z'}_{I\cup \SS_N}\right]$ for $z'$ very close to the 
saddle point $z$. 
Now, heuristically, we must expect that with large 
probability the process will first arrive at the minimum that is reached from 
$z'$ by following the `gradient' of $F_N$. 

Let us now show that this is the case. Let us first remark that using the same arguments as in the proof of Proposition \ver.1, it is clear that the 
probability that the process will hit the set where
$F_N(x')>F_N(z^*(x,{\tilde y}))+\d'$, $\d'>0$, before reaching ${\tilde y}$ is of order 
$\exp(-\d' N)$
so that this possibility is
negligible. Denote the complement of this set by $L_{\d'}$. 
 Now consider the ball $D_\d$ of radius $\d$
centered at
 $z$, where $\d$ should be large enough such that the intersection of 
$L_{\d'}$ with $\SS_N$ is well contained in the interior of $D_\d$. The 
set $L_{\d'}\cap D_\d$  is then separated by $\SS_N$ into two parts, and we call
$C_\d$ the part that is on the side of $I$. According to the previous 
discussion, if the process is to reach $I$, it has to pass through 
the surface $\S\equiv \del C_\d\cap \del D_\d$. Finally, let  
$R_\d$  denote the ball of radius $\d$ centered at $y$. 
Note first that 
$$
\eqalign{
&\P\left[\t^{z'}_{\tilde y}< \t^{z'}_{I\cup \SS_N\cup L_{\d'}^c}\right]\cr
&\leq\P\left[\t^{z'}_{\tilde y}< \t^{z'}_{I\cup \SS_N\cup L_{\d'}^c}, \t^{z'}_{\tilde y}<
\t^{z'}_{R_\d}\right]+
\P\left[\t^{z'}_{\tilde y}< \t^{z'}_{I\cup \SS_N\cup L_{\d'}^c}, \t^{z'}_{\tilde y}>
\t^{z'}_{R_\d}\right]\cr
&\leq \P\left[ \t^{z'}_{\tilde y}<
\t^{z'}_{R_\d \cup \SS_N\cup L_{\d'}^c}\right]
+\sum_{x''\in R_\d}\P\left[\t^{x''}_{\tilde y}< \t^{x''}_{y}\right]
}
\Eq(V.300)
$$
The second term is exponentially small by standard reversibility arguments.
It remains to control the first. 
$$
\eqalign{
& \P\left[ \t^{z'}_{\tilde y}<
\t^{z'}_{R_\d \cup \SS_N\cup L_{\d'}^c}\right]
\cr
&=\sum_{x'\in\S}  \P\left[ \t^{z'}_{x'}\leq \t^{z'}_{\S}\right]
\P\left[\t^{x'}_{\tilde y}<\t^{x'}_{R_\d}\right]\cr
&\leq |\S| \sup_{x'\in\S} \P\left[\t^{x'}_{\tilde y}<\t^{x'}_{R_\d}\right]
}
\Eq(V.301)
$$
Now under the assumptions on $F$, for all $x'\in \S$,
 the deterministic paths $x(t,x')$
reach $R_\d$ in finite time $T$ (i.e. in a microscopic time 
$TN$) without getting close to $\tilde y$. 
Therefore, for some $\rho>0$
$$
 \P\left[\t^{x'}_{\tilde y}<\t^{x'}_{R_\d}\right]
\leq \P\left[\sup_{t\in [0,NT]} |X_t-x(t,x')|>\rho \big |X_0=x'\right]
\Eq(V.302)
$$
But the large deviation theorem of [BG2] implies that 
there exists $\e\equiv \e(\rho,T)>0$, such that 
$$
\limsup_{N\uparrow\infty} 
\frac  1N\ln \P\left[\sup_{t\in [0,NT]} |X_t-x(t,x')|>\rho \big |X_0=x'\right]
\leq -\e(\rho,T)
\Eq(V.303)
$$
so that, e.g., for all large enough $N$,
$$
 \P\left[\sup_{t\in [0,NT]} |X_t-x(t,x')|>\rho \big |X_0=x'\right]
\leq e^{-N\e(\rho,T)/2}
\Eq(V.304)
$$
Tt then suffices to observe that 
$$
\sum_{y'\in I}\P\left[\t^x_{y'}<\t^x_{I\ba y'}\right] =\P\left[\t^x_I<\infty\right]
=1
\Eq(V.105bis)
$$
and so, since $\P\left[\t^x_{\tilde y}<\t^x_{I\ba \tilde y}\right]
\leq \exp(-N K_N)$, for all $\tilde y\neq y$, 
$\P\left[\t^x_{y}<\t^x_{I\ba y}\right]>1-\exp(-NK_N)$.\endproof

 Note that the above argument also shows 
that if $\FF(x,z,y)$ is admissible, and $y'\in I$, then 
$$
\P\left[\t^x_{y'}\leq\t^x_{I\cup x}\right]
\leq 
\P\left[\t^x_y\leq \t^x_{I\cup x}\right]
e^{-NK_N}
\Eq(V.106)
$$

\theo {\ver.4} {\it Let $x,y\in\MM_N$. Then there is a unique sequence of 
admissible events  $\FF(x_i,z_i,x_{i+1})$, $i=1,\dots,k$, such that\note{
We hope the notation used here is self-explanatory: E.g. $ \{\t^x_y<\infty\}$
stands for $\cup_{t<\infty} \{X_0=x,X_1\neq y,\dots,X_{t-1}\neq y,X_t=y\}$.}
$$
\{\t^x_y<\infty\} \supset \{\t^x_y=\sum_{i=1}^k\t^{x_i}_{x_{i+1}}<\infty\} 
\cap\bigcap_{i=1}^k  \FF(x_i,z_i,x_{i+1})
\Eq(V.13)
$$
and such that the sequences are free of cycles, i.e. the points 
$x_i,i=1,\dots k+1$ are all distinct. 
Moreover,  there is a strictly positive constant $K_N$, such that 
$$
\P\left[  \{\t^x_y=\sum_{i=1}^k\t^{x_i}_{x_{i+1}}<\infty\}
\cap\bigcap_{i=1}^k \FF(x_i,z_i,x_{i+1}) \right]\geq 1-e^{-NK_N} 
\Eq(V.14)
$$
}

\proof There is a simple algorithm that allows to construct the sequence
of admissible transitions. Let $z$ be the first common ancestor 
of $x$ and $y$ in $\TT_N$.  First we notice that 
we will `never' (that is to say  with exponentially small probability) 
visit a minimum that is not contained in the 
two branches emanating from $z$ before visiting all of $\TT_z$. 
Given this restriction,
starting from $x$, we make the maximal 
admissible transition, i.e. one traverses the highest possible saddle
for which the starting point is a lowest minimum of its branch. 
This leads to some point $x_2$, from which we continue as before, with the 
restriction that the first common ancestor of $x_2$ and $y$ now determines
the maximal allowed transition. This process is continued until an admissible
transition reaches $y$. 
It is clear that this algorithm determines a  sequence of admissible 
transitions. We have to show that this is the only one containing no 
loops. 

Note first that the condition that no transition leaves the branches of the 
youngest common 
ancestor  follows since Proposition \ver.3 ensures that the target 
point is reached before exit from this valley with probability close to one.
It is easy to see that we should always choose the maximal admissible 
transition. Suppose we start in some point that is the deepest minimum in 
some valley that does not contain the target point, and we perform an 
admissible transition that does not exit from this valley. Then we must
return to this point at least once more before reaching the 
target which means that our sequence of admissible transitions contains a 
loop. Therefore, at each step the choice of the next admissible transition 
is uniquely determined. 

Finally, from Proposition \ver.3 the estimate \eqv(V.14) follows immediately.
\endproof

\remark  We see that the same type of reasoning would also allow us
to deal with degenerate situations where e.g. integral curves of the gradient
bifurcate and transitions to several 
points $y$ may have non-vanishing probabilities. The picture of the
deterministic sequence of admissible transitions should then be replaced  by
a (cycle free) random process of admissible transitions. The precise 
computation 
of the corresponding probabilities would however require more refined 
estimates than those presented here (except if this can be done 
by using exact symmetries).  

\remark Theorem \ver.4 asserts that for fixed large $N$ a transition 
occurs along an essentially deterministic  sequence of admissible transitions.
When dealing with the dynamics of system with quenched disorder, this
deterministic (with respect to the Markov chain) sequence will however 
depend on the realization of the quenched disorder, 
{\it and on the volume $N$}. In a typical situation, this will give 
rise to a manifestation of dynamical  ``chaotic size dependence''
(in the spirit of Newman and Stein (see e.g. [NS] for an overview).

In the sequel we will  always be interested in  computing the times
(expected or distribution) of transitions conditioned on the 
canonical chain of admissible transitions constructed in Theorem \ver.4. 
We mention that in general, these do not coincide with the unconditional 
transition times. Namely, in general, there can occur unlikely  excursions 
(into deeper valleys) that take  extremely long times so that they dominate
e.g. the expected transition times. Physically, this is clearly not the most 
interesting quantity. 

\vskip1cm

\chap{5. Transition times of admissible transitions}5

From the discussion above it is clear that the most basic quantities we need
to control to describe the long time behaviour of our processes are the
times associated with an admissible transition. Note that an admissible 
transition $\FF(x,z,y)$
 can also be considered as a first exit from the valley 
associated with the saddle $z$ and the minimum $x$.
We proceed in three steps, considering first the expectations
of these times, then the Laplace transforms, and finally 
the probability distributions themselves.

\medskip
\line{\bf \ver.1. Expected times of admissible transitions.\hfill}
\medskip

 A first
 main result is       the following theorem.

\theo{\ver.1} {\it Let  $\FF(x,z,y)$ be an admissible transition, and 
assume that $x$ is a generic quadratic minimum. 
Then there exist
finite positive  constants 
$c,C$ such that, for $N$ large enough, 
$$
\eqalign{
&\E\left[\t^x_y|{\FF(x,z,y)}\right] \leq CNe^{N[F_N(z)-F_N(x)]}
\cr
&\E\left[\t^x_y|{\FF(x,z,y)}\right]\geq  cNe^{N[F_N(z)-F_N(x)]}
}
\Eq(V.15)
$$
where $K_N$ satisfies $N^{1-\a}K_N\uparrow\infty$ for some $\a>0$.}

\remark In dimension $d=1$ the upper bound captures the true behaviour
(see e.g. [vK] where the expected transition time in $d=1$ is computed
in the continuous case. Note that the extra factor $N$ in our estimates is
just a trivial scaling factor between the microscopic discrete time and the 
appropriate macroscopic time scale).  We expect that the upper bound
has the correct behaviour in all dimensions.

  Before proving the theorem, we will prove some more crude but
more general estimates. For this we introduce some notation.
Let $I\subset \MM_N$. We define
$$
d_I(x,y) \equiv \inf_{x'\in I\cup y}\left[ F_N(z^*(x,x')) -
F_N(x)\right]
\Eq(V.24)
$$
to be the effective  depth of a valley associated with the minimum $x$ 
 with exclusion at the set $I$. Recall  that  $z^*(x,y)$ denotes 
 the lowest saddle connecting $x$ and $y$, as defined in \eqv(T.2). 
Note  that Theorem 1.11 implies that 
$$
CN^{(d-2)/2} e^{-Nd_I(x,y)} \leq
\P\left[\t^x_{I\cup y}\leq \t^x_x\right]
\leq c(|I|+1) N^{(d-2)/2} e^{-Nd_I(x,y)} 
\Eq(V.24ter)
$$
With these notations we will show the following 

\lemma {\ver.2} {\it Let $I\subset \MM_N$, and 
$\P\left[\t^x_y\leq \t^x_I\right]>0$ (This can of course only fail in a 
one-dimenssional situation).
There exist $C<\infty$ 
 such that  for any $x,y\in \MM_N$,
 we have that for all $N$ large enough,
$$
\eqalign{
&\E \left[\t^x_y|\t^x_y\leq \t^x_{I}\right]\leq 
 C N^{d+3} 
+CN^{d+3}\sup_{x'\in \MM_N\ba \{I\cup y\}}\left(
\P\left[\t^x_{x'}<\t^x_y|\t^x_y\leq\t^x_I\right]
e^{N d_{I}(x',y)}\right)
}
\Eq(V.26)
$$
If the set  $\MM_N\ba \{I\cup y\}$ is empty, we use the convention that
the sup takes the value one. 
}


\remark Note that 
$$
\eqalign{
\P\left[\t^x_{x'}<\t^x_y|\t^x_y<\t^x_I\right] =
\frac{\P\left[\t^x_{x'}<\t^x_y<\t^x_I\right]}{\P\left[\t^x_y<\t^x_I\right]}
=\frac{\P\left[\t^x_{x'}<\t^x_{y\cup I}\right]\P\left[\t^{x'}_y<\t^{x'}_I
\right]}{\P\left[\t^x_y<\t^x_I\right]}
\leq   \frac{\P\left[\t^x_{x'}<\t^x_y\right]}{\P\left[\t^x_y<\t^x_I\right]}
}
\Eq(V.2000)
$$
and thus, using the same arguments as in the proof of Proposition 4.1,
whenever $\P\left[\t^x_y<\t^x_I\right]$ is close to one, 
we have the more explicit bound
$$
\P\left[\t^x_{x'}<\t^x_y|\t^x_y<\t^x_I\right] \leq C
\min\left(e^{-N[F_N(z^*(x,x'))-F_N(z^*(x,y))]},1\right)
\Eq(V.2001)
$$
This will be important in the appliction of this Lemma to the proof of 
Theorem \ver.1.


\proof 
The starting point of the proof of Lemma \ver.2 is the observation that 
it holds for $I=\MM_N$. We formulate this as a distinct lemma.

\lemma{\ver.3} {\it 
 Assume that $\P\left[\t^x_y\leq \t^x_{\MM_N}\right]>0$. 
Then there exists a constant $C<\infty$ such that for all $x,y\in\MM_N$,
and all $N$ large enough,
$$
\E\left[\t^x_y|\t^x_y\leq\t^x_{\MM_N}\right]
\leq CN^{d+3}
\Eq(V.2010)
$$
}

\proof
The first important observation is then that 
$\P\left[\t^x_y\leq \t^x_{\MM_N}\right]$ cannot be too small, i.e.

\lemma{\ver.4}{\it For any $x,y\in \MM_N$, 
there exists $L<\infty$, such that for all $N$ large enough
$$
\P\left[\t^x_y\leq \t^x_{\MM_N}\right]\geq e^{-NL}
\Eq(V.2003)
$$
}

\proof  
Now fix any $T>0$ 
Clearly 
$$
\P\left[\t^x_y\leq \t^x_{\MM_N}\right]
\geq \P\left[X_{[TN]}=y, \forall{0< t> [NT]}, X_t\not\in
\MM_N|X_0=x\right] 
\Eq(V.20000)
$$
So all we have  to show is that the finite-time probability 
in \eqv(V.20000)  is larger than $\exp(-LN)$ for some constant
$L<\infty$. 
But this is obvious by just fixing a trajectory consisting of $[NT]$ steps
and leading from $x$ to $y$ without visiting the set $\MM_N$ on the way 
(making sure that $T$ is chosen large enough to allow such a trajectory)
and observing that the probability that the process is doing just this is
at least $c^{TN}$, with $c$ is the constant from assumption (R3). 
\endproof

Next we will use the fact that in Lemma 3.2 we have shown that
$g^x_y(u)\leq 2$ for $ u<N^{-d-1}$.
Now for all $T<\infty$, 
$$\eqalign{
\E\left[\t^x_y\1_{\{\t^x_y\leq \t^x_{\MM_N}\}}\right] &\leq
T\P\left[\t^x_y\leq \t^x_{\MM_N}\right] +
\sum_{i=T}^\infty \E \left[\t^x_y \1_{\{i\leq \t^x_y\leq i+1\}}
\1_{\{\t^x_y\leq 
\t^x_{\MM_N}\}}\right]\cr
&\leq T\P\left[\t^x_y\leq \t^x_{\MM_N}\right] +
\sum_{i=T}^\infty (i+1) 
\inf_{v\geq 0} e^{-vi}
\E \left[ e^{v\t^x_y}\1_{\t^x_y\leq \t^x_{\MM_N}}\right]\cr
&\leq
  T\P\left[\t^x_y\leq \t^x_{\MM_N}\right] +
\sum_{i=T}^\infty (i+1) 
2e^{-cN^{-(d+1)} i}\cr
&\leq T  \P\left[\t^x_y\leq \t^x_{\MM_N}\right] +
Ce^{-cN^{-(d+1)} T} N^{ d+1 }(T+N^{d+1})
}
$$
Now choose $T=N^{d+3}$. Then 
$$
\E\left[\t^x_y|\t^x_y\leq\t^x_{\MM_N}\right]\leq N^{d+3}
+CN^{2d+4}\exp(-cN^2+KN)
$$
which for $N$ large enough is bounded by $CN^{d+3}$ for some constant $C$, 
as desired.
\endproof

This gives us a
 starting point to 
prove the lemma by downward induction over the size of the set $I$. 
Actually, the structure of the induction is a bit more complicated. We have to
distinguish the cases when the  starting point $x$ is contained in the 
exclusion set $I$ and when it is not. We will then proceed in two steps:

\item{(i)} Show that if  \eqv(V.26) holds for all $J\subset\MM_N$ 
with cardinality  
$|J|=k$ and
all $x,y\in \MM_N$, and if \eqv(V.26) holds 
for all $J$ of cardinality $|J|=k-1$ for  all 
$y\in \MM_N$ and
$x\not\in J\cup y$, then 
\eqv(V.26) holds  for all $I$ with cardinality $|I|=k-1$ 
and all $x,y\in \MM_N$.
\item{(ii)}  Show that  if \eqv(V.26) holds for all 
$J$ with cardinality  $|J|=k$
and all $x,y\in \MM_N$, then 
  \eqv(V.26) holds for all  $J$ of cardinality $|J|=k-1$ for all
$y\in\MM_N$ and $x\not \in J\cup y$.

If we can establish both  steps, we can  conclude that since
 \eqv(V.26) holds for
$I=\MM_N$ and all $x,y\in \MM_N$, it holds for all $I\subset\MM_N$. 
 
We now proof both assertions.  Note that $C$  will denote in the 
course of the proof a generic finite numerical constant. We will not keep
track of the  changes of its value in the course of the induction. 
We will set $\k=d+3$.

\noindent{\bf Step  (i)}: We need only to consider sets $J$ of cardinality
$k-1$ with  $x\in J\cup y$.  We can assume without loss that
$y\notin J$.
$$
\eqalign{
&\E \left[\t^x_y\1_{\t^x_y<\t^x_J}\right]\cr
&= \E \left[\t^x_y \1_{\t^x_y
\leq\t^x_{\MM_N}}\right]
+\sum_{x'\in\MM_N\ba J\ba \{x,y\}} \E\left[(\t^x_{x'}+\t^{x'}_y) \1_{\t^x_{x'}
\leq\t^x_{\MM_N}}
\1_{\t^{x'}_y<\t^{x'}_{J}}\right]
\cr
&=\E \left[\t^x_y \1_{\t^x_y
\leq\t^x_{\MM_N}}\right] +\sum_{x'\in\MM_N\ba J\ba \{x,y\}}
\Biggl(  \E\left[\t^x_{x'} \1_{\t^x_{x'}
\leq\t^x_{\MM_N}}\right]\P\left[\t^{x'}_y<\t^{x'}_{J}\right]\cr
&+
\P\left[\t^x_{x'}
\leq\t^x_{\MM_N}\right] \E\left[\t^{x'}_y\1_{\t^{x'}_y<\t^{x'}_{J}}
\right]\Biggr)
}
\Eq(V.28)
$$
Dividing by $\P\left[\t^x_y<\t^x_J\right]$, 
we get from \eqv(V.28) that
$$
\eqalign{
\E \left[\t^x_y|{\t^x_y<\t^x_J}\right]&\leq  \E \left[\t^x_y|{\t^x_y
\leq\t^x_{\MM_N}}\right]
 \cr
&+\sum_{x'\in\MM_N\ba J\ba \{x,y\}}
\Biggl(\frac{  \E\left[\t^x_{x'}|{\t^x_{x'}
\leq\t^x_{\MM_N}}\right]}
{\P\left[\t^x_{x'}<\t^x_{x}\right]}
\frac{\P\left[\t^x_{x'}
\leq\t^x_{\MM_N}\right]
 \P\left[\t^{x'}_y<\t^{x'}_{J}\right]}{\P\left[{\t^x_y<\t^x_J}\right]}\cr
&+ \E\left[\t^{x'}_y|{\t^{x'}_y<\t^{x'}_{J}}
\right]
\frac{\P\left[{\t^{x'}_y<\t^{x'}_{J}}\right]
\P\left[\t^{x}_{x'}<\t^{x}_{\MM_N}\right]}{\P\left[{\t^x_y<\t^x_J}\right]}
\Biggr)
}
\Eq(V.29)
$$
The first summand in the last line 
in \eqv(V.28)
produces exactly the first term in \eqv(V.26). For the second,
observe that 
$$
\eqalign{
\frac{\P\left[{\t^x_{x'}
\leq\t^x_{\MM_N}}\right]
 \P\left[\t^{x'}_y<\t^{x'}_{J}\right]}{\P\left[{\t^x_y<\t^x_J}\right]}
&\leq 
\frac{\P\left[{\t^x_{x'}
\leq\t^x_{J\cup y}}\right]
 \P\left[\t^{x'}_y<\t^{x'}_{J}\right]}{\P\left[{\t^x_y<\t^x_J}\right]}\cr
&=\P\left[\t^{x}_y<\t^{x}_y|{\t^x_y<\t^x_J}\right]
}
\Eq(V.30)
$$
which makes the entire term smaller than $CN^\k$. For the last term
we may use the induction hypothesis for the conditional expectation time 
appearing in it to get
$$
\eqalign{
&\E\left[\t^{x'}_y|{\t^{x'}_y<\t^{x'}_{J}}
\right]\frac{\P\left[{\t^{x'}_y<\t^{x'}_{J}}\right]
\P\left[\t^{x}_{x'}<\t^{x}_{\MM_N}\right]}{\P\left[{\t^x_y<\t^x_J}\right]}
\cr&\leq CN^\k+
CN^\k\sup_{x''\in\MM_N\ba\{J\cup y\}} \frac
{\P\left[\t^{x'}_{x''}<\t^{x'}_y<\t^{x'}_J\right] 
\P\left[{\t^{x}_{x'}<\t^{x}_{\MM_N}}\right]}
{\P\left[{\t^x_y<\t^x_J}\right]}
e^{Nd_{J}(x'',y)}
\cr&=CN^\k+CN^\k\sup_{x''\in\MM_N\ba\{J\cup y\}} \frac
{\P\left[\t^{x}_{x''}<\t^{x}_y<\t^{x}_J\right]}
{\P\left[\t^{x}_y<\t^{x}_J\right]}e^{Nd_{J}(x'',y)}
\frac {\P\left[\t^{x'}_{x''}<\t^{x'}_y<\t^{x'}_J\right] \
P\left[{\t^{x}_{x'}<\t^{x}_{\MM_N}}\right]}{\P\left[\t^{x}_{x''}<\t^{x}_y<\t^{x}_J\right]}
}
\Eq(V.31)
$$
But the last factor satisfies
$$
\eqalign{
\frac {\P\left[\t^{x'}_{x''}<\t^{x'}_y<\t^{x'}_J\right] \
P\left[{\t^{x}_{x'}<\t^{x}_{\MM_N}}\right]}
{\P\left[\t^{x}_{x''}<\t^{x}_y<\t^{x}_J\right]}
&\leq
\frac {\P\left[\t^{x'}_{x''}<\t^{x'}_y<\t^{x'}_J\right] \
P\left[{\t^{x}_{x'}<\t^{x}_{J\cup y\cup x''}}\right]}{\P\left[\t^{x}_{x''}<\t^{x}_y<\t^{x}_J\right]}
\cr &\leq \frac {\P\left[\t^x_{x'}<\t^{x}_{x''}<\t^{x}_y<\t^{x}_J\right]}
{\P\left[\t^{x}_{x''}<\t^{x}_y<\t^{x}_J\right]}\leq 1
}
\Eq(V.32)
$$
So that \eqv(V.31) actually gives a term of the desired form. This proves the
first inductive step.
 
To complete the proof we need to turn to 

\noindent{\bf Step (ii):} Here we must consider $J$  such that 
$x\not\in J\cup y$.

We will first consider the sub-case when $x$ is such that 
$$
d_J(x,y)=\sup_{x'\in \MM_N\ba\{J\cup y\}} d_J(x',y)
\Eq(V.33)
$$
Note that in this situation
$$
\eqalign{
\sup_{x'\in \MM_N\ba\{J\cup y\}} \P\left[\t^x_{x'}<\t^x_y|\t^x_y<\t^x_J\right]
e^{Nd_J(x',y)} &\geq \frac{ \P\left[\t^x_{x}<\t^x_y<\t^x_J\right]}{
\P\left[\t^x_y<\t^x_J\right]}
e^{Nd_J(x,y)} \cr
&=\P\left[\t^x_{x}<\t^x_{y\cup J}\right]
e^{Nd_J(x,y)} 
}
\Eq(V.34)
$$
But 
$$
 \P\left[\t^x_{x}<\t^x_{y\cup J}\right] =1-
\P\left[\t^x_{J\cup y}<\t^x_{x}\right]\geq 1- CN^{(d-2)/2}e^{-Nd_J(x,y)}
\Eq(V.35)
$$
so that in this case 
it will be enough to prove that 
$$
\E\left[\t^x_y|\t^x_y<\t^x_J\right] \leq C N^\k e^{Nd_J(x,y)} 
\Eq(V.36)
$$
Recall from Corollary 1.9 that
$$
\eqalign{
\E\left[\t^x_y|\t^x_y< \t^x_J\right]=&
\E\left[\t^x_y|{\t^x_y< \t^x_{J\cup x}}\right]
+\frac{\E\left[\t^x_x |{\t^x_x<\t^x_{J\cup y}}\right]}
{\P\left[\t^x_{J\cup y}<\t^x_x\right]}
\P\left[\t^x_x<\t^x_{J\cup y}\right]
}
\Eq(V.32a)
$$
By the  induction hypothesis, the first term in \eqv(V.32a) 
satisfies the bound
$$
\E\left[\t^x_y|{\t^x_y< \t^x_{J\cup x}}\right]
\leq CN^\k\sup_{x'\in\MM_N\ba \{J\cup y\cup x\}}
\P\left[\t^x_{x'}<\t^x_y| \t^x_y<\t^x_{J\cup x}\right]
e^{Nd_{J\cup x}(x',y)}
\leq  C N^\k e^{Nd_{J}(x,y)}
\Eq(V.32b)
$$
as desired (note that $d_{J\cup x}(x',y)\leq d_J(x',y)$ by definition
and $d_J(x',y)\leq d_J(x,y)$ by assumption \eqv(V.33)).
For the second term, we use again the induction hypothesis to get that 
$$
\eqalign{
&\frac{\E\left[\t^x_x |{\t^x_x<\t^x_{J\cup y}}\right]}
{\P\left[\t^x_{J\cup y}<\t^x_x\right]}
\P\left[\t^x_x<\t^x_{J\cup y}\right]
\cr&\leq  CN^\k\sup_{x'\in\MM_N\ba \{J\cup y\cup x\}}
\frac {\P\left[\t^x_{x'}<\t^x_x<\t^x_{J\cup y}\right]}
{\P\left[\t^x_{J\cup y}<\t^x_x\right]}
e^{Nd_{J\cup y}(x',x)}+\frac {CN^\k}{\P\left[\t^x_{J\cup y}<\t^x_x\right]}\cr
&\leq C^{-1} N^{-(d-2)/2}  e^{Nd_J(x,y)} CN^\k\sup_{x'\in\MM_N\ba \{J\cup y\cup x\}}
\Biggl(\P\left[\t^x_{x'}<\t^x_x\right] e^{Nd_{J\cup y}(x',x)}\Biggr)+\frac {CN^\k}{\P\left[\t^x_{J\cup y}<\t^x_x\right]}\cr
&\leq C^{-1}c    e^{Nd_J(x,y)}CN^\k 
\sup_{x'\in\MM_N\ba \{J\cup y\cup x\}}
\Biggl(e^{-N \left(F_N(z^*(x,x'))-F_N(x)\right)}
 e^{Nd_{J\cup y}(x',x)}\Biggr)\cr
 &+N^{-(d-2)/2} c^{-1} CN^\k  e^{Nd_J(x,y)}
}
\Eq(V.32c)
$$
It remains to show that
$$
\eqalign{
&e^{-N \left(F_N(z^*(x,x'))-F_N(x)\right)}
 e^{Nd_{J\cup y}(x',x)}\cr
&=
e^{-N \left(F_N(z^*(x,x'))-F_N(x)\right)}
 e^{\inf_{x''\in J\cup y\cup x}
N\left(F_N(z^*(x'',x'))-F(x')\right)}
}
\Eq(V.32cbis)
$$
is bounded by one. We consider two cases:
\item{(i)} Assume that 
$$
\inf_{x''\in J\cup y\cup x}
F_N(z^*(x'',x'))-F(x')=F_N(z^*(x,x'))-F_N(x')
\Eq(V.32cc)
$$
 Define $z^*(x,A)$ by $F_N(z^*(x,A))=\inf_{ x'\in A} F_N(z^*(x,x'))$. Then in 
this case, $x$ is `closer' to $x'$ then to $J\cup y$, so that
$$
z^*(x,J\cup y)=z^*(x',J\cup y)
\Eq(V.32d)
$$
Thus by assumption \eqv(V.33)
$$
\inf_{x''\in J\cup y}
F_N(z^*(x'',x'))-F(x')=F_N(z^*(x,J\cup y))-F_N(x')
<F_N(z^*(x,J\cup y))-F(x)
\Eq(V.32e)
$$
This implies that $F_N(x)<F_N(x')$, and since 
$$
\eqalign{
&e^{-N \left(F_N(z^*(x,x'))-F_N(x)\right)}
 e^{\inf_{x''\in J\cup y\cup x}
N\left(F_N(z^*(x'',x'))-F(x')\right)}
\leq e^{-N \left(F_N(x')-F_N(x)\right)}
}
\Eq(V.32ebis)
$$
the sup is bounded by 1 as 
desired.
\item{(ii)} We are left with the case   
$$
\inf_{x''\in J\cup y\cup x}
F_N(z^*(x'',x'))-F(x')<F_N(z^*(x,x'))-F_N(x')
\Eq(V.32ff)
$$
 Here $x'$ is closer to $J\cup y$ then to $x$, and so
$$
z^*(x,J\cup y)=z^*(x,x')
\Eq(V.32f)
$$
But then, again by \eqv(V.33)
$$
F_N(z^*(x,x')-F_N(x)=
F_N(z^*(x,J\cup y))-F_N(x)\geq \inf_{x''\in J\cup y\cup x}
 F_N(z^*(x',x''))-F_N(x') 
\Eq(V.32h)
$$
which has the desired implication.  This covers the case when 
\eqv(V.33) holds. 

Let us now turn to the  case when \eqv(V.33) does not hold.
Let $x^*\in \MM_N$ be such that 
$$
d_J(x^*,y)=\sup_{x'\in \MM_N\ba\{J\cup y\}} d_J(x',y)
\Eq(V.33a)
$$
By assumption $x^*\neq x$.
We can  write
$$
\eqalign{
\E \left[\t^x_y|{\t^x_y<\t^x_J}\right]&
=\frac {\E \left[\t^x_y \1_{{\t^x_y<\t^x_J}\1_{\t^x_{x^*}>\t^x_y}}\right]}
{\P\left[{\t^x_y<\t^x_J}\right]}
+\frac {\E \left[\t^x_y \1_{{\t^x_y<\t^x_J}\1_{\t^x_{x^*}<\t^x_y}}\right]}
{\P\left[{\t^x_y<\t^x_J}\right]}
\equiv (I)+(II)
}
\Eq(V.33b)
$$

Now 
$$
(I) =\E \left[\t^x_y |{\t^x_y<\t^x_{J\cup x^*}}\right]
\frac {\P\left[{\t^x_y<\t^x_{J\cup x^*}}\right]}
{\P\left[{\t^x_y<\t^x_{J}}\right]}
\Eq(V.33c)
$$
and using the induction hypothesis,
$$
\eqalign{
(I)&\leq CN^\k + CN^\k\sup_{x'\in\MM_N\ba\{J\cup y\cup x^*\}} \frac {\P\left[
\t^x_{x'}<\t^x_y<\t^x_{J\cup x^*}\right]}
{\P\left[\t^x_y<\t^x_{J}\right]} e^{Nd_{J\cup x^*}(x',y)}
\cr
&\leq CN^\k+CN^\k \sup_{x'\in\MM_N\ba\{J\cup y\}} \frac {\P\left[
\t^x_{x'}<\t^x_y<\t^x_{J}\right]}
{\P\left[\t^x_y<\t^x_{J}\right]} e^{Nd_{J}(x',y)}
}
\Eq(V.33d)
$$
as desired.  On the other hand,
$$
\eqalign{
(II)=&
\frac{\E\left[\t^x_{x^*}\1_{\t^x_{x^*}<\t^x_{J\cup y}}\right]}
{\P\left[\t^x_y<\t^x_J\right]} \P\left[\t^{x^*}_y<\t^{x^*}_J\right]+\frac{\E\left[\t^{x^*}_y\1_{\t^{x^*}_y<\t^{x^*}_{J}}\right]}
{\P\left[\t^x_y<\t^x_J\right]} \P\left[\t^x_{x^*}<\t^x_{J\cup y}\right]\cr
\cr&\equiv (IIa)+(IIb)
}
\Eq(V.33e)
$$
To treat (IIa) we can use the induction hypothesis to get
$$
\eqalign{
(IIa)&=\E\left[\t^x_{x^*}|\t^x_{x^*}<\t^x_{J\cup y}\right]
\P\left[\t^x_{x^*}<\t^x_y|\t^x_y<\t^x_J\right]\cr
&\leq CN^\k+CN^\k \sup_{x'\in\MM_N\ba\{J\cup x^*\cup y\}} \frac
{\P\left[\t^x_{x'}<\t^x_{x^*}<\t^x_{J\cup y}\right]}
{\P\left[\t^x_y<\t^x_J\right]} \P\left[\t^{x^*}_y<\t^{x^*}_{J}\right]
e^{Nd_{J\cup y}(x',x^*)}
\cr&= CN^\k+CN^\k\sup_{x'\in\MM_N\ba\{J\cup x^*\cup y\}} \frac
{\P\left[\t^x_{x'}<\t^x_{x^*}<\t^x_y<\t^x_{J}\right]}
{\P\left[\t^x_y<\t^x_J\right]} 
e^{Nd_{J\cup y}(x',x^*)}
\cr&\leq  CN^\k+CN^\k\sup_{x'\in\MM_N\ba\{J\cup x^*\cup y\}} \frac
{\P\left[\t^x_{x'}<\t^x_y<\t^x_{J}\right]}
{\P\left[\t^x_y<\t^x_J\right]} 
e^{Nd_{J}(x',y)}
}
\Eq(V.33f)
$$
since $d_{J\cup y}(x',x^*)=\inf_{x''\in J\cup y\cup x^*}[F_N(z^*(x',x''))
-F_N(x')]\leq d_J(x',y)$,
and finally
$$
\eqalign{
(IIb)&= \E\left[\t^{x^*}_y |\t^{x^*}_y<\t^{x^*}_J\right]
\frac{\P\left[\t^x_{x^*}<\t^x_{J\cup y}\right]
\P\left[\t^{x^*}_y<\t^{x^*}_J\right]}{\P\left[\t^x_y<\t^x_J\right]}
\cr
&= \E\left[\t^{x^*}_y |\t^{x^*}_y<\t^{x^*}_J\right]
\P\left[\t^x_{x^*}<\t^x_y|\t^x_y<\t^x_{J}\right]
\cr &\leq CN^\k\P\left[\t^x_{x^*}<\t^x_y|\t^x_y<\t^x_{J}\right]
e^{Nd_J(x^*,y)}
}
\Eq(V.33g)
$$
where the last line is obtained by using that the conditional expectation 
in  the one-but-last line the conditional expectation is of the form 
considered just before. 
Putting all together we see that the lemma is proven.\endproof

\proofof{Theorem \ver.1}
Lemma \ver.2 can now be used to prove the theorem. For this, let
$\FF(x,z,y) $ be an admissible transition and fix 
$I=\TT_{z,x}^c\ba y\cap\MM_N$. 

We have already seen in \eqv(V.35) that $\P\left[\t^x_x<\t^x_{I\cup y}\right]$
differs from 1 only by an exponentially small term. Moreover,
using \eqv(V.106),
we see that in the case of an admissible transition,
$$
\eqalign{
\P\left[\t^x_y<\t^x_x\right]\leq 
&\P\left[\t^x_{I\cup y}<\t^x_x\right]
\leq\P\left[\t^x_{y}<\t^x_x\right]+ \sum_{y'\in I}
\P\left[\t^x_{y'}<\t^x_{x\cup I}\right]
\cr&\leq \P\left[\t^x_{y}<\t^x_x\right] (1+|I|e^{-NK_N})
}\Eq(V.52)
$$
Therefore \eqv(V.32a) implies in this case that
$$
\E\left[\t^x_y|\t^x_y<\t^x_I\right]= 
\frac {\E\left[\t^x_x|\t^x_x<\t^x_{I\cup y}\right]}
{\P\left[\t^x_{y}<\t^x_x\right]} \left(1+\OO(e^{-NK_N})\right) 
+\E\left[\t^x_y|\t^x_y<\t^x_{I\cup x}
\right] 
\Eq(V.206)
$$
Now we have already precise bounds on the denominator
 of the first term (see Section 2), 
and using the upper bound from Lemma \ver.2 (taking into account that
we are now in the situation of the remark following that Lemma!)
 we see that, 
under the assumptions of the theorem, the 
second term   is by a factor $N^\k\exp(-K_N N)$  smaller than the 
first. It remains to estimate precisely the numerator in the first term.
The essential idea here is to use the ergodic theorem. It may be useful to 
explain this first in a simpler situation where there is only a single 
minimum 
present and consider the quantity $g^y_y(u)$. Let $D\subset \G_N$ be 
the local valley associated to $y$, that is the connected component of the 
level set of the saddle point that connects $y$ to the rest of the world.
The basic idea is to show that the expected recurrence time at $y$ 
(without visits at other points of $\MM_N$)
is up to exponentially small errors equal to the same time of another
Markov chain  $\wt X_D(t)$ with state space $D$ 
with transition rates 
$
\wt p_D(x,z)$ defined as in \eqv(T.7)
 and whose invariant measure, $\wt \Q_D$, is easily seen to be just
 $\Q_N$ conditioned on $D$, i.e. $\wt \Q_D(x)\equiv \Q_N(x)/\Q_N(D)$
for any $x\in D$. 
Then, by the ergodic theorem, we have that
$$
\wt \E_D\t^y_y =\frac 1{\wt \Q_D(y)}=\frac {\Q_N(D)}{ \Q_N(y)}
\Eq(EE.2)
$$
This quantity can be estimated very precisely via sharp large deviation 
estimates. It will typically exhibit a behaviour of the form $C N^{d/2}$.

To arrive at this comparison, we simply divide the paths in our process 
into those reaching the boundary of $D$ and those who don't, i.e. 
we write
$$
g^y_y(u) =\E \left[ e^{u\t^y_y} \1_{\t^y_y\leq\t^y_{\MM_N}}
\1_{\t^y_{\del D}<\t^y_y}\right]+
\E \left[ e^{u\t^y_y}\1_{\t^y_{\del D}>\t^y_y}\right]
\Eq(EE.3)
$$
Let us denote by $D^+$ and $D^-$ the two sets obtained by adding 
and removing, respectively, one layer of points to, resp. from, $D$. 
Note that on the event $\{\t^y_{\del D}>\t^y_y\}$ the processes
$X(t)$ and $\wt X_{D^+}(t)$ have the same law until time $\t^y_y$,
so that
$$
\E \left[ e^{u\t^y_y}\1_{\t^y_{\del D}>\t^y_y}\right]
=\wt\E_{D^+} \left[ e^{u\t^y_y}\1_{\t^y_{\del D}>\t^y_y}\right]
\leq \wt\E_{D^+} \left[ e^{u\t^y_y}\right]
\Eq(EE.4)
$$
We will show that this is the dominant term in \eqv(EE.3), the first summand 
on the right being exponentially small. Indeed 
$$
\eqalign{
\E \left[ e^{u\t^y_y} \1_{\t^y_y\leq\t^y_{\MM_N}}
\1_{\t^y_{\del D}<\t^y_y}\right]&=
\sum_{z\in \del D}\E \left[ e^{u(\t^y_z+\t^z_y)} 
\1_{\t^y_z\leq\t^y_{y\cup \del D}}
\1_{\t^z_y\leq\t^z_{\MM_N}}
\right]\cr
&=\sum_{z\in \del D}\E \left[ e^{u\t^y_z}
\1_{\t^y_z\leq\t^y_{y\cup \del D}}\right]
\E\left[e^{u\t^z_y}
\1_{\t^z_y\leq\t^z_{\MM_N}}
\right]\cr
}
\Eq(EE.5)
$${
Using Theorem 1.10, for small enough $u$, the first factor is bounded by 
\break $const.N^\k e^{-N [F_N(z)-F_N(y)]}$, while the second is bounded by 
$const. N^\k$. This gives the desired upper bound 
$$
g_y^y(u) \leq 
\wt\E_{D^+} \left[ e^{u\t^y_y}\right]+ C N^\k
e^{ -N [F_N(z^*)-F_N(y)]}
\Eq(EE.6)
$$
where $z^*$ denotes the lowest saddle point in $\del D$. To get the 
corresponding lower bound, just note that 
$$
\wt\E_{D^+} \left[ e^{u\t^y_y}\1_{\t^y_{\del D}>\t^y_y}\right]
=\wt\E_{D^+} \left[ e^{u\t^y_y}\right]-
\wt\E_{D^+} \left[ e^{u\t^y_y}\1_{\t^y_{\del D}\leq\t^y_y}\right]
\Eq(EE.7)
$$
But the last term in \eqv(EE.7) can be treated precisely as in \eqv(EE.5), so 
that we arrive at 
$$
g_y^y(u) \geq 
\wt\E_{D^+} \left[ e^{u\t^y_y}\right]- C N^\k 
e^{ -N [F_N(z^*)-F_N(y)]}
\Eq(EE.6bis)
$$
Differentiating and using reversibility and 
the upper bounds from Theorem 1.10, as well as the 
obvious lower bound
$$
g_y^y(0) = 1-\P\left[\t^y_{\MM_N} <\t^y_y\right]
\geq 1-\sum_{x\in \MM_N\ba y} g^y_x(0) 
\geq 1-|\MM_N|N^{d-1}  e^{-N[F_N(z^*)-F_N(y)]}
\Eq(EE.8)
$$
gives in the same way  that 
$$
\frac {\dot g_y^y(0)}{g_y^y(0)} = \frac {\Q_N(D^+)}{\Q_N(y)} + \OO(N^\k)
 e^{-N[F_N(z^*)-F_N(y)]}
\Eq(EE.7bis)
$$
}

The same ideas can now be carried over to the estimation 
of the return time in an admissible situation, using the estimates from Lemma
\ver.2.

\proposition {\ver.5} {\it Let $\FF(x,z,y)$ be an admissible transition.
Let $D$ denote the level set of the saddle $z$. Then for $K_N>0$ satisfying 
$N^{1-\a}K_N \uparrow \infty$, for some $\a>0$, 
$$
\E\left[\t^x_x| \t^x_x<\t^x_{I\cup y}\right] =\frac{\Q_N(D^+)}{\Q_N(x)}
+\OO\left(e^{-NK_N}\right)
\Eq(V.208)
$$
where $I=\TT^c_{z,x}\ba y\cap\MM_N$.
}

\proof Basically,  the proof goes as outlined above. 
 With $D$ defined as the level set of the saddle 
$z$,  we can decompose
$$
\E\left[\t^x_x \1_{\t^x_x<\t^x_{I\cup y}}\right]
=\E\left[\t^x_x \1_{\t^x_x<\t^x_{I\cup y}}\1_{\t^x_x<\t^x_{\del D}}
\right]+\E\left[\t^x_x \1_{\t^x_x<\t^x_{I\cup y}}\1_{\t^x_x>\t^x_{\del D}}
\right]
\Eq(V.21)
$$
The first summand gives precisely
$$
\E\left[\t^x_x \1_{\t^x_x<\t^x_{I\cup y}}\1_{\t^x_x<\t^x_{\del D}}\right]
=\E\left[\t^x_x \1_{\t^x_x<\t^x_{\del D}}\right]
\Eq(V.22)
$$
so that from this term alone we would get the same estimate as in 
\eqv(EE.7bis). 
We have to show that the second term does not give a relevant 
contribution.
Note that as in \eqv(EE.5) we can split
paths at the first visits to $\del D$. This gives
$$
\eqalign{
\E\left[\t^x_x \1_{\t^x_x<\t^x_{I\cup y}}\1_{\t^x_x>\t^x_{\del D}}
\right]&=\sum_{z'\in\del D} \E\left[\t^x_{z'} \1_{\t^x_{z'}\leq
\t^x_{\del D}}\1_{\t^x_{z'}<\t^x_x}\right]
\P\left[\t^{z'}_x<\t^{z'}_{I\cup y}\right]\cr
&+\P\left[\t^x_{z'}\leq
\t^x_{\del D},{\t^x_{z'}<\t^x_x}\right]\E\left[\t^{z'}_x
\1_{\t^{z'}_x<\t^{z'}_{I\cup y}}\right]
}
\Eq(V.23)
$$
Now $
\P\left[\t^x_{z'}\leq
\t^x_{\del D},{\t^x_{z'}<\t^x_x}\right]$ is bounded by 
$e^{-N[F_N(z)-F_N(x)]}$, and by reversibility\hfill\break 
$\E\left[\t^x_{z'} \1_{\t^x_{z'}\leq
\t^x_{\del D}}\1_{\t^x_{z'}<\t^x_x}\right]
\leq e^{-N[F_N(z)-F_N(x)]} 
\E\left[\t^{z'}_x \1_{\t^{z'}_x<
\t^{z'}_{\del D}}\right]$, so all we have to show is that the two quantities
$\E\left[\t^{z'}_x \1_{\t^{z'}_x<
\t^{z'}_{\del D}}\right]$ and $\E\left[\t^{z'}_x
\1_{\t^{z'}_x<\t^{z'}_{I\cup y}}\right]$ (which are more or less the
same) are not too large. 
But this follows from our previous bounds by splitting the process going 
from $z'$ to $x$ at its first visit to a point in $\TT_{z,x}\cap\MM_N$, e.g.
$$
\eqalign{
&\E\left[\t^{z'}_x
\1_{\t^{z'}_x<\t^{z'}_{I\cup y}}\right]
=\E\left[\t^{z'}_x
\1_{\t^{z'}_x\leq\t^{z'}_{\MM_N}}\right]\cr
&+\sum_{x'\in \MM_N\ba I}\left(
\E\left[\t^{z'}_{x'}\1_{\t^{z'}_{x'}\leq \t^{z'}_{\MM_N}}\right]
\P\left[\t^{x'}_x<\t^{x'}_{I\cup y}\right]
+\P\left[{\t^{z'}_{x'}\leq \t^{z'}_{\MM_N}}\right]
\E\left[\t^{x'}_x\1_{\t^{x'}_x<\t^{x'}_{I\cup y}}\right]\right)\cr
}
\Eq(V.207)
$$
Lemma \ver.2 and Theorem 1.10 can now be used on the expectations in
\eqv(V.207), and this implies the desired result.\endproof

Now if (as we assume) $x$ is a quadratic minimum of $F_N$,
$\frac {\Q_N(D^+)}{\Q_N(x)} =CN^{d/2}$, and using this together with 
Theorem 1.11 we get the estimates of Theorem \ver.1.\endproof\endproof

\remark The reader will have observed that we could also prove 
lower bounds for more general transitions, complementing Lemma \ver.2. 
But the point is that these would depend in a complicated way 
on the global specifics of the function $F_N$, contrary to the situation 
of admissible transitions for which we get the very simple estimates of 
Theorem \ver.1. The beauty of the construction lies in some sense in the fact 
that the general ``worst case'' upper bounds of Lemma \ver.2 suffice to
obtain the precise estimates of the theorem.

\vskip1cm

\medskip
\line{\bf \ver.2 Laplace transforms of transition times of admissible transitions\hfill}
\medskip
Theorem \ver.1 gives precise estimates on the expected transition times
for an elementary transition. We will now show that as expected, the 
distribution of these transition times is asymptotically exponential.
This will be done by controlling the Laplace transforms for small arguments.

\theo {\ver.6} {\it Let $\FF(x,z,y)$ be an admissible transition.  
Set $\bar \t^x_y\equiv \E\left[\t^x_y|\FF(x,z,y)\right]$.
Then
$$
\E\left[e^{v\t^x_z/\bar\t^x_y}|\FF(x,z,y)\right] =
\frac 1{1-v} +e^{-N K_N}f(v) 
\Eq(V.40)
$$
where for any $\d>0$, for  $N$ large enough, $f$ is bounded and 
analytic in the domain
$|Re(v)|<1-\d$
}

\proof The main ingredient of the proof lies in controlling the 
analytic structure of the Laplace transforms.
The procedure 
will be similar to that in the proof of Lemma \ver.2, that is we consider 
the entire family of functions
$G^x_{y,I}(u)$ and establish the corresponding domains by induction, starting
with the case $I=\MM_N$ where the analytic estimates of Theorem 1.10 hold. 
It will be convenient to use functions where the argument $u$ has been 
properly rescaled. The naive expectation might be that the Laplace transform 
will exist for values of $u$ up to the inverse of the 
corresponding expected transition time. However, this is not so. The point 
is that Laplace transforms are much more sensitive to ``deep valleys''
than the expected times for which such valleys contribute less if they
are unlikely to be visited. However the Laplace transform will only partly
 benefit from 
this, but simply explode at a value corresponding to the deepest valley that 
is at all allowed to be visited. 

We introduce some more notation for convenience. Set
$$
t_{I}(x,y)\equiv e^{Nd_I(x,y)}
\Eq(V.60)
$$
and
$$
T_I(y)\equiv \sup_{x\in \MM_N\ba\{I\cup y\}} t_{I}(x,y)
\Eq(V.61)
$$
With the notation of Section 1, we define
$$
\wh G^x_{y,I}(v)\equiv \frac{  G^x_{y,I}\left(v/T_I(y)\right)}
{\P\left[\t^x_y\leq\t^x_I\right]}
\Eq(V.62)
$$
 The following key lemma gives us control over
how this happens. It is the analogue of Lemma \ver.2. 

\lemma{\ver.7} {\it Let $I\subset \MM_N$, and 
let $x,y\in\MM_N$. 
Then $\wh G^x_{y,I}(v)$ can be represented in the form 
$$
\eqalign{
&\wh G^x_{y,I}(v)=a^0_{x,y,I}(v/T_I(y))\cr
&+\sum_{x'\in \MM_N\ba\{I\cup y\}}
\P\left[\t^x_{x'}<\t^x_y|\t^x_y\leq \t^x_I\right]
a^{x'}_{x,y,I}\left(v\frac {t_I(x',y)}{T_I(y)}\right)
}
\Eq(V.63)
$$
where 
$a^0_{x,y,I}$ and
$
a^{x'}_{x,y,I}
$, 
 for any $x'\in \MM_N\ba\{I\cup y\}$ are complex functions 
that have the properties (for a  finite constant $C$, and $\k=d+3$):
\item{(i)} They are bounded by $CN^\k$ and analytic in the 
domain  $|Re(u)| <CN^{-\k}$,
\item{(ii)} They are real and positive  
for real $v$.  
}

\proof  
An important corollary of analyticity are corresponding bounds on the
derivatives. Namely, 
by a standard application of the Cauchy integral formula it follows that
for any function $a$ which is bounded and analytic in 
the domain $|Re(v)| <CN^{-\k}$,
for 
$|Re(v)|<\frac C2 N^{-\k}$ we have, 
$$
\left|\frac {d^n}{dv^n} a(v)\right|
\leq n! C \frac {2^nN^{n\k}}{C^n}\sup_{v:Re (v)< CN^{-\k}}a(v)
\Eq(V.43)
$$
This will be used repeatedly in the sequel.

The first step in the proof is to show that \eqv(V.63)
holds  if $I=\MM_N$. The proof of this fact is completely analogous to that of
Lemma \ver.3 and we will skip the details. Let us just mention that we will
get the boundedness of  $\wh G^x_{y,\MM_N}(v)$ in a domain $Re(u)<c N^{-d-3}$
(while analyticity was established in the larger domain 
$Re(u)<cN^{-d-1}$ in section 3). This provides the starting point for 
our induction like in Lemma \ver.2. 
Again we assume that the Lemma holds for all $I$ of cardinality 
greater than or equal to $\ell$, and we consider sets $J\subset\MM_N$ 
of cardinality $\ell -1$.  
As before, we first show that the case $x\in J$ reduces easily to the case
$x\not \in J$. Without loss of generality we assume $y\notin J$. Namely,
in the former case,
$$
\eqalign{
&\wh G^x_{y,J}(v) =\frac{g^x_{y}(v/T_J(y))}{\P\left[\t^x_y<\t^x_J\right]}+
\sum_{x'\in \MM_N\ba\{J\cup y\} }\frac{g^x_{x'}(v/T_J(y)) 
\P\left[\t^{x'}_y<\t^{x'}_J\right]}{\P\left[\t^x_y<\t^x_J\right]}
 \wh G^{x'}_{y,J}(v)
}
\Eq(V.63a)
 $$
Inserting the induction hypothesis for the $\wh G^{x'}_{y,J}(v)$, we get
$$
\eqalign{
&\wh G^x_{y,J}(v) =\frac{g^x_{y}(v/T_J(y))}{\P\left[\t^x_y<\t^x_J\right]}
+\sum_{x''\in \MM_N\ba\{J\cup y\} }\frac{g^x_{x'}(v/T_J(y)) 
\P\left[\t^{x'}_y<\t^{x'}_J\right]\P\left[\t^x_{x'}\leq\t^x_{\MM_N}\right]}
{g^x_{x'}(0)\P\left[\t^x_y<\t^x_J\right]}
 a^0_{x',y,J}(v/T_J(y))
\cr
&+\sum_{x'\in \MM_N\ba\{J\cup y\}}\sum_{x''\in \MM_N\ba\{J\cup y\}} 
\frac{g^x_{x'}(v/T_J(y))}{ g^x_{x'}(0)}
\frac{\P\left[\t^x_{x'}\leq \t^x_{\MM_N}\right]
\P\left[\t^{x'}_y<\t^{x'}_J\right]}{\P\left[\t^x_y<\t^x_J\right]}
\cr
&\times\P\left[\t^{x'}_{x''}<\t^{x'}_y|\t^{x'}_y<\t^{x'}_J\right]
a^{x''}_{x',y,J}\left(v\frac {t_J(x'',y)}{T_J(y)}\right)
}
\Eq(V.64)
$$
Remember that in the proof of Lemma \ver.2 (\eqv(V.31)-\eqv(V.32))
 we have established that
$$
\eqalign{
\frac{\P\left[\t^x_{x'}\leq \t^x_{\MM_N}\right]
\P\left[\t^{x'}_y<\t^{x'}_J\right]}{\P\left[\t^x_y<\t^x_J\right]}
 \P\left[\t^{x'}_{x''}<\t^{x'}_y|\t^{x'}_y<\t^{x'}_J\right]
\leq  \P\left[\t^{x}_{x''}<\t^{x}_y|\t^{x}_y<\t^{x}_J\right]
}
\Eq(V.65)
$$
which, since the analytic properties of $a^{x''}_{x',y,J}$ and 
$a^{x''}_{x,y,J}$ are the same,
 shows that \eqv(V.64) provides the claimed representation.

The more subtle part of the proof concerns the case $x\not\in J$.
We proceed again as in the proof of Lemma 5.2 and consider first the case
when 
$T_J(y)=d_J(x,y)$. 
Note again that in this case, the representation in the Lemma reduces to 
$$
\wh G^x_{y,J}(v)= a^x_{x,y,J}(v)
\Eq(V.65a)
$$
since all the other terms in the sum \eqv(V.63) are smaller and more regular.

We use of course that 
$$
G^x_{y,J}(u)=\frac {G^x_{y,J\cup x}(u)}{1-G^x_{x, J\cup y}(u)}
\Eq(V.66)
$$
 which implies 
$$
\eqalign{
&\wh G^x_{y,J}(v)=\frac {\P\left[\t^x_y<\t^x_{J\cup x}|\t^x_y<\t^x_J\right]
\wh G^x_{y,J\cup x}\left(v\frac{T_{J\cup x}(y)}{T_J(y)}\right)}
{1-\P\left[\t^x_x<\t^x_{J\cup y}\right]
\wh G^x_{x,J\cup y}\left(v\frac{T_{J\cup y}(x)}{T_J(y)}\right)}
\cr
&=\frac {\wh G^x_{y,J\cup x}\left(v\frac{T_{J\cup x}(y)}{T_J(y)}\right)}
{1-\sfrac{\P\left[\t^x_x<\t^x_{J\cup y}\right]}
{\P\left[\t^x_x>\t^x_{J\cup y}\right]}
v\frac{T_{J\cup y}(x)}{T_J(y)}\int_0^1d\th\wh G'{}^x_{x,J\cup y}
\left(\th v\frac{T_{J\cup y}(x)}{T_J(y)}\right)}
}
\Eq(V.67)
$$
The numerator again poses no problem since it permits to obtain the desired 
representation by the inductive hypothesis. Potential danger comes from the 
denominator. But using the induction hypothesis, we see that
(similar to \eqv(V.32b))
$$
\eqalign{
&\frac{T_{J\cup y}(x)}{T_J(y)} \wh G'{}^x_{x,J\cup y}
\left(\th v\frac{T_{J\cup y}(x)}{T_J(y)}\right)\cr
&=\frac{T_{J\cup y}(x)}{T_J(y)}
\Biggl[\sum_{x'\in \MM_N\ba\{J\cup x\cup y\}} 
\P\left[\t^x_{x'}<\t^x_x\right]\cr 
&\times a'{}^{x'}_{x,x,J\cup y}\left(\th v\frac {t_{J\cup y}(x',x)}{T_J(y)}\right)
\frac {t_{J\cup y}(x',x)}{T_{J\cup y(x)}}
+a^0_{x,x,J\cup y}\left(\frac{\th v}{T_J(y)}\right)
\frac 1{T_{J\cup y}(x)}
\Biggr]\cr
&=\sum_{x'\in \MM_N\ba\{J\cup x\cup y\}} 
\P\left[\t^x_{x'}<\t^x_x\right]  \frac {t_{J\cup y}(x',x)}{T_{J} (y)}
a'{} ^{x'}_{x,x,J\cup y}\left(\th v\frac {t_{J\cup y}(x',x)}{T_J(y)}\right)
\cr
&+a^0_{x,x,J\cup y}\left(\frac{\th v}{T_J(y)}\right)
\frac 1{T_{J}(y)}
}
\Eq(V.68)
$$
All we need is to bound the modulus of this expression from above for $v$ real.
This gives
$$
\eqalign{
&\left|\frac{T_{J\cup y}(x)}{T_J(y)} 
\wh G'{}^x_{x,J\cup y}\left(\th v\frac{T_{J\cup y}(x)}{T_J(y)}\right)
\right|\cr
&\leq \sum_{x'\in \MM_N\ba\{J\cup x\cup y\}} 
\P\left[\t^x_{x'}<\t^x_x\right]\frac {t_{J\cup y}(x',x)}{T_{J} (y)}
CN^\k+\frac {cN^\k}{T_J(y)}
}
\Eq(V.69)
$$
In the second part of the proof of Lemma \ver.2  (\eqv(V.32c) 
we have shown that
$$
\eqalign{
&\P\left[\t^x_{x'}<\t^x_x\right] {t_{J\cup y}(x',x)}\leq  {CN^{(d-2)/2}}  
}
\Eq(V.70)
$$
Thus we deduce from \eqv(V.69) that
$$
\eqalign{
&\left|\frac{T_{J\cup y}(x)}{T_J(y)} \wh G'{}^x_{x,J\cup y}
\left(\th v\frac{T_{J\cup y}(x)}{T_J(y)}\right)\right|\leq
\frac {CN^\k}{T_J(y)}\cr
}
\Eq(V.71)
$$
We see thus that the numerator in 
\eqv(V.67) will not vanish for $v<C^{-1}N^{-\k}$.  
Thus $\wh G^x_{y,J}(v)$  is bounded and
analytic in the strip $|Re(v)|<C^{-1}N^{-\k}$. 

In the general case, we proceed as in the proof of Lemma \ver.2 and 
decompose all paths into those avoiding the deepest minimum $x^*$ and those
visiting it. A simple computation yields then 
$$
\eqalign{
\wh G^x_{y,J}(v)
=&
\wh G^x_{y, J\cup x^*}\left(v \frac {T_{J\cup x^*}(y)}  {T_{J}(y)}\right)
\frac {\P\left[\t^x_{y}<\t^x_{J\cup x^*}\right]}
 {\P\left[\t^x_{y}<\t^x_{J}\right]}\cr
&+\P\left[\t^x_{x^*}<\t^x_y|\t^x_y<\t^x_J\right]
\wh G^x_{x^*, J\cup y}\left(v \frac {T_{J\cup y}(x^*)}  {T_{J}(y)}\right)
\wh G^{x^*}_{y, J}\left(v\right)
}
\Eq(V.72)
$$
Clearly the first term is by the inductive assumption at least as good as 
desired, and since we have just shown that the last factor in the second
term is bounded and analytic in $|Re(v)|<CN^{-\k}$, using the inductive 
assumption for $\wh G^x_{x^*, J\cup y}\left(v \frac {T_{J\cup y}(x^*)} 
 {T_{J}(y)}\right)$, one sees easily that the desired representation 
holds. This concludes the proof of the Lemma.\endproof

We are now ready to prove Theorem \ver.6. For this we have to improve the 
previous analysis in the case where $x$ is the deepest minimum in the allowed 
set  $\MM_N\ba I$. In that case $T_I(y)$ is strictly larger than
any of the terms $t_{I\cup x}(x',y)$ and $t_{I\cup y}(x',x)$
and it will pay to use Taylor expansions to second order.
 Also, we will be more precise in the rescaling of the variables $u$ and define
$$
\wt G^x_{y,I}(v)\equiv \wh G^x_{y,I}(v T_I(y)/\bar T)
\Eq(V.74)
$$
with 
$$
\bar T\equiv \frac{\E\left[\t^x_x\1_{\t^x_x<\t^x_{y\cup I}}\right]}
{\P\left[\t^x_{I\cup y}<\t^x_x\right]}
\Eq(V.49)
$$ 
Note that we use  $\bar T$ instead of $\bar\t^x_y$ in the proof because this
will simplify the following formulas. But recall from the proof of 
Theorem \ver.1 that 
$$
\left|\frac{\E\left[\t^x_y|{\t^x_y<\t^x_{I}}\right]}{\bar T}-1\right|
\leq e^{-NK_N}
\Eq(V.49bis)
$$
Thus in the final results they can be interchanged without harm.
Then 
$$
\wt G^x_{y,I}(v) =
\frac {\P\left[\t^x_y<\t^x_{I\cup x}|\t^x_y<\t^x_I\right]
\wh G^x_{y,I\cup x}\left(v \frac {T_{I\cup x}(y)} {\bar T}\right)}
{1-\P\left[\t^x_x<\t^x_{I\cup y}\right] \wh G^x_{x,I\cup y}
\left(v \frac {T_{I\cup y}(x)}{\bar T}\right)}
\Eq(V.50)
$$
Using the analyticity properties established in the preceding lemma, we 
now proceed to a more careful computation, using second order Taylor 
expansions. This yields
$$
\eqalign{
&\wt G^x_{y,I}(v)\cr
&=\frac{\P[\t^x_y<\t^x_{I\cup x}|\t^x_y<\t^x_{I}]\left(
1+\frac v{\bar T} \E[\t^x_y |{\t^x_y<\t^x_{I\cup x}}]   
+\ssfrac {(vT_{I\cup x}(y))^2}{2 \bar T^2} 
\wh G''{}^x_{y,I\cup x}\left(\tilde \th v \ssfrac {T_{I\cup x}(y)}{\bar T}
\right)\right)}
{\P[\t^x_{y\cup I}<\t^x_{x}]
-\frac v{\bar T}\E[\t^x_x\1_{\t^x_x<\t^x_{I\cup y}}]
-\ssfrac {(vT_{I\cup y}(x))^2}{2 \bar T^2} \P[\t^x_x<\t^x_{I\cup y}]
\wh G''{}^x_{x,I\cup y}\left(\tilde \th v \ssfrac {T_{I\cup y}(x)}{\bar T}
\right)
}
}
\Eq(V.51)
$$
for some $0\leq\tilde\th\leq 1$.
We use Corollary 1.9 to get
$$
\eqalign{
\wt G^x_{y,I}(v)
&=\frac{1+\frac v{\bar T} \E\left[\t^x_y |{\t^x_y<\t^x_{I\cup x}}\right]   
+\frac {(vT_{I\cup x}(y))^2}{2 \bar T^2} 
\wh G''{}^x_{y,I\cup x}\left(\tilde \th v \frac {T_{I\cup x}(y)}{\bar T}
\right)}
{1-\frac{v}{\bar T} \frac{\E[\t^x_x\eu_{\t^x_x<\t^x_{I\cup y}}  ]}
{\P\left[\t^x_{y\cup I}<\t^x_x\right]}
-\frac {(vT_{I\cup y}(x))^2\P[\t^x_x<\t^x_{I\cup y}]}
{2 \bar T^2\P[\t^x_{I\cup y}<\t^x_x]} 
\wh G''{}^x_{x,I\cup y}\left(\tilde\th  v \frac {T_{I\cup y}(x)}{\bar T}
\right)}
\cr
&=\frac{1+\frac v{\bar T} \E\left[\t^x_y |{\t^x_y<\t^x_{I\cup x}}\right]   
+\frac {(vT_{I\cup x}(y))^2}{2 \bar T^2} 
\wh G''{}^x_{y,I\cup x}\left(\tilde \th v \frac {T_{I\cup x}(y)}{\bar T}
\right)}
{1-{v}
-\frac {(vT_{I\cup y}(x))^2}
{2 \bar T\E\left[\t^x_x|\t^x_x<\t^x_{I\cup y}\right]} 
\wh G''{}^x_{x,I\cup y}\left(\tilde\th v \frac {T_{I\cup y}(x)}{\bar T}
\right)} 
}
\Eq(V.54)
$$
The term we must be most concerned with is the second order term in the 
denominator. Here we must use the full analyticity properties proven in Lemma
\ver.7. 
This gives, after computations analogous to those leading to \eqv(V.69)
and using the obvious lower bound on $\bar T$  
$$
\eqalign{
&\left|\frac {(vT_{I\cup y}(x))^2}
{2 \bar T\E\left[\t^x_x|\t^x_x<\t^x_{I\cup y}\right]} 
\wh G''{}^x_{x,I\cup y}\left(\tilde\th  v \frac {T_{I\cup y}(x)}{\bar T}
\right)\right|
\leq C^2N^{2k}\frac {|v|^2}{2\E\left[\t^x_x|\t^x_x<\t^x_{I\cup y}\right]^2} 
 \cr
&\times\sum_{x'\in\MM_N\ba\{J\cup y\cup x\}}
e^{-N[F_N(z^*(x',x))-F_N(x)+F_N(z^*(x,y))-F_N(x)]} 
\left(t_{I\cup y}(x',x)\right)^2
}
\Eq(V.80)
$$
Now according to our hypothesis that $x$ is the lowest minimum in 
$I$, it follows that 
$t_{I\cup y}(x',x)\leq e^{N[F_N(z^*(x,x'))-F_N(x)]}$
so that \eqv(V.80) is finally bounded by 
$$
\eqalign{
& C^2N^{2k} C' N^{-d}\frac {|v|^2}{2}
|I| \sup_{x'\in\MM_N\ba\{I\cup y\cup x\}}
e^{-N[F_N(z^*(x,y)-F_N(z^*(x',x))]} \cr
&= C^2N^{2k}C' N^{-d}\frac {|v|^2}{2}
|I| e^{-N\d}
}
\Eq(V.81)
$$
where $\d$ is strictly positive.

All the other terms in \eqv(V.54) 
except the leading ones are even smaller. Note moreover that 
both $T_{I\cup y}(x) $ and $T_{I\cup x}(y)$ are exponentially small
compared to $\bar T$, so that all these error terms as functions
of $v$ are analytic if   $|Re(v)|<1$.
 This allows to write
$
\wt G^x_{y,I}(v)$ in the form
$$
\wt G^x_{y,I}(v) =\frac 1{1-v} + e^{-NK_N/2}\frac{e_1(v)}
{1-v} + \frac {e_2(v)}{(1-v)(1-v-e^{-NK_N/2} e_3(v))}
\Eq(V.82)
$$
where all $e_i$ are analytic and uniformly (in $N$) bounded in the domain
$|Re(v)|< 1$.
 This concludes the proof of the theorem.\endproof\endproof

\medskip
\line{\bf \ver.3. The distribution of transition times.\hfill}
\medskip

From Theorem \ver.6   
one obtains of course some information on the distribution function. 

\corollary {\ver.8} {\it Under the same assumptions as in Theorem \ver.6,
we have:
\item{(i)} For any $\d>0$, for sufficiently large $N$,
$$
\P\left[\t^x_y>t\bar T|\FF(x,z,y)\right]\leq  e^{-(1-\d)t}/\d
\Eq(V.83)
$$
\item{(ii)} Assume that $N_i$ is a sequence of volumes tending to infinity
such that for all $i$, $\FF(x,z,y)$ is an admissible transition. Then, 
for any $t\geq 0$,
$$
\lim_{i\uparrow \infty}\P_{N_i}\left[\t^x_y>t\bar \t_{N_i}|\FF(x,z,y)\right]= 
e^{-t}
\Eq(V.84)
$$
}

\proof (i) is an immediate consequence of the the Laplace transform is bounded
for real positive $v$ with $v<1$ and the exponential Chebyshev inequality.
(ii) is a standard consequence of the fact that the Laplace transform 
converges pointwise for any 
purely imaginary $v$ to that of the exponential distribution, and is 
analytic in a neighborhood of zero.\endproof

With a little more work we can also complement the upper bound \eqv(V.83)
by a  lower bound on the distribution of the survival time in a valley.

\proposition {\ver.9} {\it Let $\FF(x,z,y)$ be an admissible transition, and
set $I=\MM_N\ba \TT_{z,x}$. Let $h(N)$ be any sequence tending to zero as
$N$ tends to  infinity. Then, for some $\k<\infty$, 
for any
$0<\a_t<1$ we have that 
$$
\P\left[\t^x_{I}>t\right]\geq \cases e^{-t /(\bar T(1-\a_t))} \a_t^2 
\left(1-h(N)\right) &\,\hbox{if}\,\,\, t>h(N)\frac{ CN^{d/2}} {N^\k +T_I(x)}\cr
1-o(1) &\,\hbox{if}\,\,\, 
t\leq h(N)\frac{ CN^{d/2}} {N^\k +T_I(x)}
\endcases
\Eq(V.85)
$$
}  
 
\proof The proof of this lower bound consists essentially in guessing the 
strategy the process will follow in order to realize the event in question
which will be to return a specific number of times to $x$ without visiting the 
set $I$. For, obviously,
$$
\eqalign{
\P\left[\t^x_{I}>t\right]&\geq \sum_{{s_1,\dots,s_n\geq 1}
\atop{s_1+\dots+s_n>t}}\P\left[\forall_{i=1}^n X_{s_i}=x, 
\forall_{s\leq s_1+\dots+s_n} X_s\not\in I\right]\cr
&=\sum_{{s_1,\dots,s_n\geq 1}
\atop{s_1+\dots+s_n>t}}
\prod_{i=1}^n\P\left[\t^x_x=s_i\leq\t^x_{I}\right]\cr
&=\left(\P\left[\t^x_x\leq\t^x_{I}\right]\right)^n
\sum_{{s_1,\dots,s_n\geq 1}
\atop{s_1+\dots+s_n>t}}\prod_{i=1}^n
\P\left[\t^x_x=s_i|\t^x_x\leq\t^x_{I}\right]\cr
}
\Eq(V.86)
$$
We introduce the family of independent, identically distributed variables
$Y_i$ taking values in  the positive integers  such that for  
$\P\left[Y_i=s\right]=\P\left[\t^x_x=s|\t^x_x\leq\t^x_{I}\right]$.
Then \eqv(V.86) can be written as
$$
\P\left[\t^x_{I}>t\right] \geq 
\left(\P\left[\t^x_x\leq\t^x_{I}\right]\right)^n
\P\left[\sum_{i=1}^n Y_i>t\right]
\Eq(V.87)
$$
We have good control on the  first factor in \eqv(V.87). We need a lower 
bound on the second probability. The simplest way to proceed is to use the
inequality, going back to Paley and Zygmund, 
that asserts that for any random variable $X$ 
with finite expectation,
and any $\a>0$,
$\P[X>(1-\a)\E X]\geq \a^2\frac{(\E X)^2}{\E X^2}$.  We will use this
with $X=\sum_i Y_i$. Thus
$$
\eqalign{
\P\left[\sum_{i=1}^n Y_i>t\right] &=\P\left[\sum_{i=1}^n Y_i>\frac t{n\E Y_1} n\E Y_1\right]
\cr
&\geq \left(1-\frac t{n\E Y_1}\right)^2 \frac {n^2(\E Y_1)^2}
{n(n-1) (\E Y_1)^2 +n\E Y_1^2}\cr
&= \left(1-\frac t{n\E Y_1}\right)^2 \frac 1{1+\frac 1n\left(
\frac {\E Y_1^2}{(\E Y_1)^2}-1\right)}
}
\Eq(V.88)
$$
Now using Lemma \ver.7 one verifies easily that
$$
\E Y_1^2 \leq C N^\k + T_I(x) e^{-NK_N}
\Eq(V.89)
$$
So that \eqv(V.88) gives
$$
\P\left[\sum_{i=1}^n Y_i>t\right]
\geq  \left(1-\frac t{n\E Y_1}\right)^2 \frac 1{1+\frac 1n\left( C N^\k +
 T_I(x) e^{-NK_N}\right)}
\Eq(V.90)
$$
Thus the second factor is essentially equal to one if $n\gg \max(N^\k,T_I)$.
We now choose $n$ as the integer part of $n(t)$ where  
$$
n(t)=\min\left(\frac t{\E Y_1 (1-\a_t)},\frac 1{h(N)(C N^\k+T_I)}
\right)
\Eq(V.91)
$$
This yields
$$
\eqalign{
\P\left[\t^x_{I}>t\right]& \geq \left(1-\P\left[\t^x_{I}\leq \t^x_x\right]
\right)^{n(t)} \a_t^2  \frac 1{1+\frac 1{n(t)}\left( C N^\k +
 T_I(x) e^{-NK_N}\right)}\cr
&\geq e^{-t \frac{\P[\t^x_{I}\leq \t^x_x]}{\E Y_1(1-\a_t)} -t 
\OO\left(\P[\t^x_{I}\leq \t^x_x]^2\right)}
\a_t^2 \left(1- h(N)\right)
}
\Eq(V.92)
$$
if $t$ is such that the $n(t)$ is given by the second term in the minimum
in \eqv(V.91). This yields  the first case in \eqv(V.90).  
If $t$ is smaller than that, one sees easily that 
$ n(t) \P[\t^x_{I}\leq \t^x_x]$ tends to zero uniformly in $t$, as 
$N\uparrow\infty$ (in fact exponentially fast!) This implies the second case.
\endproof

\vskip1cm

\chap{6. Miscellaneous consequences for the processes}6

In this section we collect some soft consequences of the 
preceding analysis. We begin by substantiating the claim made in Section 1
that the process spends most of the time in the immediate vicinity of the 
minima. 
We formulate this in the following form.

\proposition {\ver.1} {\it There exists finite positive  constants $C,k$ 
such that for any $\rho>0$, 
 $x\in \MM_N$ and  $t>0$, 
$$
\P\left[|X_t-x|>\rho\big|\t^{x}_{\MM_N\ba x}>t,X_0=x\right]
\leq CN^{k}
\inf_{\rho'<\rho}\sup_{y\in\G_N:\rho'\leq|x-y|\leq3\rho'/2}
e^{-N[F_N(y)-F_N(x)]}
\Eq(7.1)
$$
}

\proof We start to decompose the event $\{\t^{x}_{\MM_N\ba x}>t\}$ as follows:
$$
\{\t^{x}_{\MM_N\ba x}>t\}
=\bigcup_{0<s<t}
\{\t^{x}_{\MM_N\ba x}>s\}\cap\left\{X_s=x, \forall_{s<s'<t} X_{s'}
\not\in\MM_N\right\}
\Eq(7.2)
$$
Then 
$$
\eqalign{
&\P\left[|X_t-x|>\rho\big|\t^{x}_{\MM_N\ba x}>t,X_0=x\right]
\cr
&=\sum_{0<s<t} \frac{\P\left[\t^{x}_{\MM_N\ba x}>s,X_s=x\right]}
{\P\left[\t^{x}_{\MM_N\ba x}>t\right]}
\P\left[|X_{t-s}-x|>\rho,\t^x_{\MM_N}>t-s|X_0=x\right]\cr
&\leq\sum_{0<s<t} \frac{\P\left[\t^{x}_{\MM_N\ba x}>s,X_s=x\right]}
{\P\left[\t^{x}_{\MM_N\ba x}>s,X_s=x\right]\P\left[\t^{x}_{\MM_N\ba x}>t-s
\right]}
\P\left[|X_{t-s}-x|>\rho,\t^x_{\MM_N}>t-s|X_0=x\right]\cr
&\inf_{\rho'<\rho}\leq   \sum_{y:\rho'\leq|x-y|\leq3\rho'/2}
\sum_{0<s<t}\frac{\P\left[\t^x_y<\t^x_{\MM_N},\t^x_{\MM_N}>t-s\right]}
{\P \left[\t^{x}_{\MM_N\ba x}>t-s
\right]}
\cr
&
\leq  \inf_{\rho'<\rho} \sum_{y:\rho'\leq|x-y|\leq3\rho'/2}
\sum_{0<s<t}\frac{\min\left(\P\left[\t^x_y<\t^x_{\MM_N}\right],
\P\left[\t^x_{\MM_N}>t-s\right] \right)}{\P \left[\t^{x}_{\MM_N\ba x}>t-s
\right]}
}
\Eq(7.3)
$$
Now 
$$
\P\left[\t^x_y<\t^x_{\MM_N}\right]\leq e^{-N[F_N(y)-F_N(x)]}
\Eq(7.4)
$$
while 
$$
\eqalign{
&\P\left[\t^x_{\MM_N}>t-s\right] =
\sum_{y\in \MM_N} \P\left[\t^x_y>t-s|\t^x_y<\t^x_{\MM_N\ba y}\right]
\P\left[\t^x_y<\t^x_{\MM_N\ba y}\right]
}
\Eq(7.41)
$$
Note that 
by Theorem 1.10 and the exponential Markov inequality,
for some $\k<\infty$
$$
\P\left[\t^x_y>t-s |\t^x_y<\t^x_{\MM_N\ba y}\right] \leq 
C N^{\k} e^{-(t-s)cN^{-\k}} 
\Eq(7.5)
$$
while the factors $ \P\left[\t^x_y<\t^x_{\MM_N\ba y}\right]$ are exponentially
small except if $y=x$. 
The denominator is bounded below by Proposition 5.9. Since it decays with 
$t-s$ at an exponentially smaller rate than the numerator (see \eqv(7.5)),
and is close to one for times up to the order $\exp(NC)$ (for some $C$),
it is completely irrelevant.

Thus we see that the second term in the minimum takes over for $t-s>
N^{\k+1} [F_N(y)-F_N(x)]$, a number small compared to the inverse of the
first term in the minimum. Thus using  that, for $ a,b \ll 1$,
$$
\sum_{t=0}^\infty \min\left(e^{-at},b\right)\leq \frac {b|\ln b|}a+\frac
b{1-e^{-a}}\approx \frac ba(|\ln b|+1)
\Eq(7.51)
$$
the result follows immediately. \endproof 

Based on this result, we will now show that during an admissible 
transition the process also stays  mostly close to its starting point, i.e. 
the lowest minimum of the valley concerned. The following proposition makes 
this precise.

\proposition {\ver.2}{\it Let $\FF(x,z,x')$ be an admissible transition.
Then there exists finite positive constants $C,k$  and $K_N$ s.t. $\lim N^{1-\a}K_N\uparrow\infty$, for some $\a>0$,  such that
for any $t$ and $\rho>0$, 
$$
\eqalign{
&\P\left[|X_t-x|>\rho\big|\t^{x}_{\TT_{z,x}^c}>t,X_0=x\right]\cr
&\leq CN^{\k}\inf_{\rho'<\rho}
\sup_{y\in\G_N:\rho'\leq|x-y|\leq3\rho'/2}e^{-N[F_N(y)-F_N(x)]}+Ce^{-NK_N}
}\Eq(7.6)
$$
}

\proof The proof of this proposition is in principle similar to that of 
Proposition \ver.1. 
We begin by applying the same decomposition as before to get
$$
\eqalign{
&\P\left[|X_t-x|>\rho\big|\t^{x}_{\TT^c_{z,x}}>t,X_0=x\right]
\cr
&=\sum_{0<s<t} \frac{\P\left[\t^{x}_{\TT^c_{z,x}}>s,X_s=x\right]}
{\P\left[\t^{x}_{\TT^c_{z,x}}>t\right]}
\P\left[|X_{t-s}-x|>\rho,\t^x_{\TT^c_{z,x}\cup x}>t-s|X_0=x\right]\cr
&\inf_{\rho'<\rho}\leq   \sum_{y:\rho'\leq|x-y|\leq3\rho'/2}
\sum_{0<s<t}
\frac{\P\left[\t^x_y<\t^x_{\TT^c_{z,x}\cup x},\t^x_{\TT^c_{z,x}\cup x}
>t-s\right]}{\P\left[\t^{x}_{\TT^c_{z,x}}>t-s\right]} 
\cr
&
\leq  \inf_{\rho'<\rho} \sum_{y:\rho'\leq|x-y|\leq3\rho'/2}
\sum_{0<s<t}\frac{\min\left(\P\left[\t^x_y<\t^x_{\TT^c_{z,x}\cup x}\right], 
\P\left[\t^x_{\TT^c_{z,x}\cup x}
>t-s\right] \right)}{\P\left[\t^{x}_{\TT^c_{z,x}}>t-s\right]} 
}
\Eq(7.7)
$$
As in the proof of Proposition \ver.1, the denominator is bounded by 
Proposition 5.9 and is seen to be insignificant. We concentrate on the
 estimates of the numerator.
Again we have the obvious bound 
$$
\P\left[\t^x_y<\t^x_{\TT^c_{z,x}\cup x}\right]\leq e^{-N[F_N(y)-F_N(x)]}
\Eq(7.71)
$$
but to deal with the second probability in the minimum will be a little more 
complicated.
Note first that as in \eqv(7.41) we can write
$$
\eqalign{
&\P\left[\t^x_{\TT^c_{z,x}\cup x}>t-s\right]=
\P\left[\t^x_x>t-s|\t^x_x<\t^x_{\TT^c_{z,x}}\right]
\P\left[\t^x_x<\t^x_{\TT^c_{z,x}}\right]\cr
&+\sum_{y\in\TT^c_{z,x}} 
\P\left[\t^x_y>t-s|\t^x_y<\t^x_{\TT^c_{z,x}\cup x\ba y}\right]
\P\left[\t^x_y<\t^x_{\TT^c_{z,x}\cup x\ba y}\right]\cr
}
\Eq(7.72)
$$
Now the terms in the second sum are all harmless, since by the estimates
of Lemma 5.7 and the geometry of our setting, using the exponential 
Chebyshev inequality as if Corollary 5.8, for any $\d>0$,
$$
\P\left[\t^x_y>t-s|\t^x_y<\t^x_{\TT^c_{z,x}\cup x\ba y}\right]
\P\left[\t^x_y<\t^x_{\TT^c_{z,x}\cup x\ba y}\right]
\leq e^{-(t-s)(1-\d)/T_{\TT^c_{z,x}\cup x}(y)} e^{-N[F_N(z)-F_N(x)]}
\Eq(7.73)
$$
with $T_{\TT^c_{z,x}\cup x}(y)$ much smaller than $ e^{N[F_N(z)-F_N(x)]}$.
The remaining term is potentially dangerous. 
To deal with  this efficiently, we need to 
classify the trajectories according to the deepest minimum they 
have visited before returning to $x$. 
In the present situation the relevant effective depth of a minimum 
$y\in \TT_{z,x}$ is (recall \eqv(V.24))
$$
d(y)\equiv d_{\TT_{z,x}^c}(y,x)=F_N(z^*(y,x))-F_N(y)
\Eq(7.61)
$$
We will  enumerate the minima in $\TT_{z,x}$ according to increasing
depth by $x=y_0,\dots, y_k$ (we assume for simplicity that no degeneracies 
occur). We set
$L(y)\equiv \{y'\in\TT_{z,x}:d(y')\geq d(y)\}$. 
Then the family of disjoint events $\{\t^x_x\geq \t^x_{y_i}\}\cap
\{\t^x_x< \t^x_{L(y_{i+1})}\}$ can serve as a partition of unity, 
i.e. we have that
$$
\eqalign{
&\P\left[\t^x_x>t-s|\t^x_x<\t^x_{\TT^c_{z,x}}\right]
=\sum_{i=0}^k \P\left[\t^x_x>t-s,\t^x_x\geq \t^x_{y_i},\t^x_x<\t^x_{L(y_{i+1})}
|\t^x_x<\t^x_{\TT^c_{z,x}}\right]\cr
&\leq \sum_{i=0}^k\min\left(
 \P\left[\t^x_x\geq \t^x_{y_i}
|\t^x_x<\t^x_{\TT^c_{z,x}}\right],
 \P\left[\t^x_x>t-s,\t^x_x< \t^x_{L(y_{i+1})}
|\t^x_x<\t^x_{\TT^c_{z,x}}\right]\right)
\cr
&\leq \sum_{i=0}^k\min\left(
 \P\left[\t^x_x\geq \t^x_{y_i}
|\t^x_x<\t^x_{\TT^c_{z,x}}\right],
 \P\left[\t^x_x>t-s|\t^x_x< \t^x_{L(y_{i+1})},
\t^x_x<\t^x_{\TT^c_{z,x}}\right]\right)
}
\Eq(7.62)
$$
Now again 
$$
\P\left[\t^x_x>t-s|\t^x_x< \t^x_{L(y_{i+1})},
\t^x_x<\t^x_{\TT^c_{z,x}}\right]
\leq e^{-(t-s)e^{-Nd(y_i)}}
\Eq(7.63)
$$
while 
$$
 \P\left[\t^x_x\geq \t^x_{y_i}
|\t^x_x<\t^x_{\TT^c_{z,x}}\right]
\leq e^{-N[F_N(z^*(y_i,x))-F_N(x)]}
\Eq(7.64)
$$
which is much smaller than $e^{-Nd(y_i)}$ (except in the case $i=0$ where 
we are back in the situation of Proposition \ver.1). 
Combining all these estimates and using again \eqv(7.51)
yields the claim of the proposition.
\endproof

\remark 
Note that Proposition \ver.2 again exhibits the special r\^ole played by 
admissible transitions. It justifies the idea that the behaviour of the process
during an 
admissible transition can
be described, on the time scale of the expected transition time\note{A finer
resolution will of course exhibit rare and rapid excursions to other minima
during the time of the admissible transition, and we have all the tools 
to investigate these interior cycles.},
as waiting in the immediate vicinity of the starting minimum for 
an exponential 
time until jumping quasi-instantaneously to the destination point. This idea
can also be expressed by passing to a measure valued description (as 
was done  in [GOSV]) which will exhibit that 
the empirical measure of
the process on any time scale small compared to the 
expected transition time but long compared to the next-smallest transition time
within the admissible transition, is close to the Dirac mass at the minimum;
since this, in turn, is asymptotically the invariant measure of the process 
conditioned to stay in the valley associated to the admissible transition, 
it can thus  justly be seen as a metastable state associated with this 
time scale. The corresponding measure-valued process is than close to a 
jump process on the Dirac measures centered at these points. These results 
can be derived easily from the preceding Propositions, and we will not go into
the details.

Let us also mention that from the preceding results and Corollary 5.8 (ii)
one can easily extract  statements concerning ``exponential convergence to 
equilibrium''. E.g., one has the following.

\corollary {\ver.3} {\it Let $N_k\uparrow \infty$ be a subsequence 
 such that for all $k$ the topological structure of the
tree from Section 5 is the same  
and such that along the subsequence, $F_{N_k}$ is generic. Let $m_0$ denote 
the lowest minimum of $F_{N_k}$.
Let $f\in C(\L,\R)$ be any continuous function
on the state space. Consider the process starting in some point $x\in\G_N$.
Then there is a unique minimum   $m(x)$ of $F_{N_k}$, 
converging to a minium $m(x)$ of $F$,
  such that, setting  $\bar\t^x(k)\equiv
\E\left[\t^{m_{N_k}(x)}_{m_0}\right]$
$$
\lim_{k\uparrow\infty} \E f(X_{t/\bar\t^x(k)})
=e^{-t}f(m(x))+(1-e^{-t})f(m_0)
\Eq(7.65)
$$
(where on the right hand side $m(x),m_0$  denote the corresponding minima of
the limiting function $F$).  
The point $m(x)$ is the lowest minimum of the deepest valley 
visited by the process in the canonical decomposition of the transition 
$x,m_0$ given in Theorem 4.4.
}

We leave the proof of the corollary to the reader. 
In a way such statements
that involve convergence on a single time-scale are rather poor reflections
of the complex structure of the behaviour of the process that is encoded in
the description given in Section 4. 


\medskip
\line{\bf Relation to spectral theory.\hfill}
\medskip
Contrary  to much of the work 
on the dynamics of spin systems we have not 
used the notion of ``spectral gap'' in this paper, and in fact the 
analysis of spectra has been limited in general to the rather 
auxiliary estimates in Section 2. Of course these approaches
 are closely related 
and our results could be re-interpreted in terms of spectral theory.

Most evidently, the  estimate given in Theorem 5.1 can  also be seen 
as precise estimates on the largest eigenvalue of the 
Dirichlet operator associated with the admissible transition 
$\FF(x,z,y)$. Moreover,  these Dirichlet eigenvalues are closely 
related to the low-lying spectrum of the stochastic matrix $P_N$. 
Sharp estimates on this relation require however some work, and
we will 
not pursue this analysis in this paper but relegate it to 
forthcoming work in which the relation between the metastability problem and 
associated quantum mechanical tunneling problem will be further elucidated.

\vskip1cm

\chap{7. Mean field models and mean field dynamics}7

Our main motivation is to study the properties of stochastic dynamics
for a class of models called ``generalized random mean field models''
that were introduced in [BG1]. We recall that such models require the
following ingredients:

\item {(i)} A single spin space $\SS$ that we will always take to be a 
           subset of some linear space,  equipped with 
            some a priori probability measure $q$. 
\item {(ii)} A state space $\SS^N$ whose elements we denote by $\s$ and call
             {\it spin configurations}, equipped with the product measure
             $\prod_i q(d\s_i)$.  
\item {(iii)} The dual space ${(\SS^N)^*}^M$ of linear maps 
             $\xi_{N,M}^T:\SS^N\rightarrow \R^M$.
\item {(iv)} A mean field potential  which is some real valued function
$E_M:\R^M\rightarrow \R$.
\item{(v)} An abstract probability space $(\O,\FF, \PP)$ and measurable maps
           $\xi^T: \O\rightarrow {(\SS^\N)^*}^\N$. Note that 
           if $\Pi_N$ is the canonical projection $\R^\N\rightarrow \R^N$, 
then $\xi_{M,N}^T[\o]\equiv\Pi_M\xi^T[\o]\circ\Pi_N^{-1}$ are random elements 
of   $ {(\SS^N)^*}^M$. 
\item {(vi)} The random order parameter 
$$
m_{N,M}[\o](\s)\equiv \frac 1N   \xi_{M,N}^T[\o]\s \in \R^M
\Eq(M.1)
$$
\item{(vii)} A random Hamiltonian 
$$
H_{N,M}[\o](\s) \equiv -N E_M\left(m_{N,M}[\o](\s)\right)   
\Eq(M.2)
$$

In [BG1] the equilibrium properties of such models were studied in the 
case where $M=M(N)$ grows with  $N$. Our aim in the long run is to be able to 
study dynamics in this situation, but in the present paper we restrict us to 
the case of fixed $M=d$. Also, we will only consider the case where $\SS$ is
a  finite set.

Typical dynamics studied for such models are Glauber dynamics, i.e.
(random) 
Markov chains $\s(t)$, 
defined on the configuration space $\SS^N$ that are reversible 
with respect to the (random) Gibbs measures 
$$
\mu_{\b,N}(\s) [\o] \equiv \frac {e^{-\b H_N[\o](\s)}\prod_{i=1}^N q(\s_i)}
{Z_{\b,N}[\o]}
\Eq(M.6)
$$
and in which the transition rates are non-zero only 
if the final configuration can be obtained from the initial one 
by changing the value of one spin only. 
To simplify notation we will henceforth drop  the reference to the random 
parameter $\o$. 

As always the final goal will be to understand the macroscopic dynamics, 
i.e. the behaviour of $m_N(\s(t))$ as a function of time. 
It would be very convenient in this situation if $m_N(\s(t))$ were itself a 
Markov chain with state space $\R^d$. Such a Markov chain would be reversible 
with 
respect to the measure induced by the Gibbs measure on $\R^d$ through the 
map $\frac 1N\xi^T$, and this measure has nice large deviation properties.
Unfortunately, $m_N(\s(t))$ is almost never a Markov chain. A notable
exception  is the (non-random) Curie-Weiss model (see the next section). 
There are special situations
in which it is possible to introduce a larger number of 
macroscopic order parameters in such a way that the corresponding induced 
process will be Markovian;  in general this will not be 
possible. However, there is a canonical construction of a new Markov 
process on $\R^d$ that can be expected to 
be a good approximation to the induced process.  This construction and the 
following results are all adapted from Ligget [Li], Section II.6. 

Let $r_N(\s,\s')$ be transition rates of a Glauber 
dynamics reversible with respect 
to  the measure $\mu_{\b,N}$, i.e. for $\s\neq\s'$,
 $r_N(\s,\s')=\sqrt{\frac{\mu_N(\s')}
{\mu_N(\s)}}g_N(\s,\s')$ where $g_N(\s,\s')=g_N(\s',\s)$. 
We denote by $\RR_N$ the law of this Markov chain and by $\s(t)$
the coordinate variables.
Define the induced 
measure 
$$
\Q_{\b,N}=\mu_{\b,N}\circ m_{N,d}^{-1}
\Eq(M.7)
$$
and the new transition rates for a Markov chain with state space 
the $\G_N=m_{N,d}(\SS^N)$  (we drop the indices of $m_{N,d}$ in the sequel) by
$$
p_N(x,y)\equiv \frac 1{\Q_{\b,N}(x)} \sum_{\s:{m(\s)=x}}\sum_{\s':{m(\s')=y}}
\mu_{\b_N}(\s)r_N(\s,\s')
\Eq(M.8)
$$

\theo{\ver.1} {\it Let $\P_N$ be the law of the Markov chain $x(t)$ with state 
space $\G_N$ and transitions rates  $p_N(x,y)$ given by \eqv(M.8).
 Then $\QQ_{\b,N}$ is the unique reversible invariant measure 
for the chain $x(t)$.  
Moreover, for any $\s\in \SS_N$ and $D\subset\SS_N$, one has
$$
\mu_{\b,N}(\s)\RR_N\left[\t^\s_D\leq \t^\s_\s\right]\leq 
\Q_{\b,N}(m(\s))\P_N\left[\t^{m(\s)}_{m(D)}\leq \t^{m(\s)}_{m(\s)}\right]
\Eq(M.9)
$$
Finally, the image process $m(\s(t))$  is Markovian and  has 
law $\P_N$ if 
for all $\s,\s''$ such that $m(\s)=m(\s'')$, $r(\s,\cdot)=r(\s'',\cdot)$.
If the initial measure $\pi_0$ is such that for all $\s$, $\pi_0(\s)>0$,
then this condition is also necessary.
}

\remark Notice that by the ergodic theorem, we can rewrite \eqv(M.9)
in the less disturbing form 
$$
\frac
{\E\t^\s_\s}{\RR_N\left[\t^\s_D\leq \t^\s_\s\right]}\geq
\frac {\E\t^{m(\s)}_{m(\s)}}
{\P\left[\t^{m(\s)}_{m(D)}\leq \t^{m(\s)}_{m(\s)}\right]}
\Eq(M.9bis)
$$
from which we see that the theorem really implies an ineqality for the 
arrival times in $D$.

\proof Note that we can write
$$
p_N(x,y)= \sqrt{\frac {\Q_{\b,N}(y)}{\Q_{\b,N}(x)}} 
 \sum_{\s:{m(\s)=x}}\sum_{\s':{m(\s')=y}}
\frac{\sqrt{\mu_{\b_N}(\s)\mu_{\b,N}(\s')}}
{\sqrt{\Q_{\b,N}(x)\Q_{\b,N}(y)}}
 g_N(\s,\s')
\Eq(M.10)
$$
which makes the reversibility of the new chain obvious. 
Note also that if $r_N(\s,\s')$ is constant on the sets 
$m^{-1}(x)$, then
$$
p_N(x,y)=\sum_{\s':{m(\s')=y}}r_N(\s,\s')
=\RR_N\left[m(\s(t+1))=y|\s(t)=\s\right] 
\Eq(M.11)
$$ 
which is only a function of 
$m(\s)$. From this one sees easily that in this case, the law of
$m(\s(t))$ is $\P$.  The proof of the converse statement is a little more
involved and can be found in [BR].

Finally, the inequality \eqv(M.10) is proven in [Li] (Theorem 6.10). However,
the proof given there is rather cumbersome, and meant to illustrate
coupling techniques, while the result follows 
in a much simpler way from Theorem 6.1 in the same book. It may be worthwhile
to outline the argument. 
Theorem 6.1 in [Li] states that
$$
 \mu_{\b,N}(\s)\RR_N\left[\t^\s_D<\t^\s_\s\right]
=\frac 12\inf_{h\in \HH^\s_D}\Phi_N(h)
\Eq(M.12)
$$
where $ \HH^\s_D$ is the set of functions
$$
\HH^\s_D\equiv\left\{h:\SS^N\rightarrow[0,1]:h(\s)=0,\forall_{\s'\in D}h(\s')=1
\right\}
\Eq(M.13)
$$
and $\Phi_N$ is the Dirichlet form associated to the chain $\RR_N$,
$$
\Phi_N(h)\equiv \sum_{\s,\s'\in\S^N}\mu_{\b,N}(\s)\RR_N(\s,\s')
\left[h(\s)-h(\s')\right]^2
\Eq(M.14)
$$
Now we clearly majorize the infimum by restricting it to functions
that are constant on the level sets of the map $m$, that is if we define the 
set
$$
\wt \HH^x_{\wt D}\equiv\left\{\tilde h:\G_N\rightarrow[0,1]:\tilde h(x)=0,
\forall_{y\in \wt D}\tilde h(y)=1
\right\}
\Eq(M.15)
$$
we have that
$$
\eqalign{
\inf_{h\in \HH_\s}\Phi_N(h)&\leq
\inf_{\tilde h\in \HH^{m(s)}_{m(D)}}\Phi_N(\tilde h\circ m)\cr
}
\Eq(M.16)
$$
But 
$$
\eqalign{
\Phi_N(\tilde h\circ m)=
&\sum_{x,x'\in \G_N} \left[\tilde h(x)-\tilde h(x')\right]^2
 \sum_{\s: m(\s)=(x),\s':m(\s')=x'}\mu_{\b,N}(\s)\RR_N(\s,\s')\cr
&=\sum_{x,x'\in \G_N} \left[\tilde h(x)-\tilde h(x')\right]^2
\QQ_{\b,N}(x)p_N(x,x')\equiv \wt\Phi_N(\tilde h)
}
\Eq(M.17)
$$
where $\wt\Phi_N$ is the Dirichlet form of the chain $\P$. Using the analog
of \eqv(M.12) for this new chain we arrive at the inequality 
\eqv(M.9).\endproof

We certainly expect that in many situations the Markov chain  $x(t)$ under the 
law $\P_N$ has essentially the same long-time behaviour than the 
non-Markovian image process $m(\s(t))$. However, we have no general results and 
there are clearly situations imaginable in which this would not be true. In 
the next section we will apply our general results to a specific model where
this issue in particular can be studied nicely.

\newpage
\def\Pr{\PP}
\def\mm{\underline{m}}

\chap{8. The random field Curie-Weiss model}8

The simplest example of disordered mean field  models is the random 
field Curie-Weiss model.
Here $\SS=\{-1,1\}$, $q$ is the uniform distribution on this set. 
Its Hamiltonian is 
$$
H_N[\o](\s) \equiv -N \frac{\left(M_N^1(\s)\right)^2}2
-\sum_{i=1}^N \th_i[\o]\s_i
\Eq(M.5)
$$
where 
$$
M_N(\s)\equiv \frac 1N\sum_{i=1}^N\s_i
\Eq(M.36000)
$$
is called the {\it magnetization}. Here $\th_i$, $i\in \N$ are i.i.d.
random variables. 
The dynamics of this model has been studied before:
dai Pra and den Hollander studied the short-time dynamics using large 
deviation results and obtained the analog of the McKeane-Vlasov equations
[dPdH]. Matthieu and Picco [MP1]  
considered convergence to equilibrium in a 
particularly simple 
case where the random field takes only the two  values 
$\pm \e$ (with further restrictions on the parameters that exclude the presence
 of more than two minima). 

In this section we take up this simple model in the more general situation 
where the random field is allowed to take values in an arbitrary finite set.
The main  idea here is that in this case we are, as we will see, in the
position to construct an image of the Glauber dynamic in a finite dimensional
space that is Markovian, while it will be possible to compare this to the 
Markovian dynamics defined on the single parameter $M_N$ in the manner 
described in the previous section.

We consider the Hamiltonian \eqv(M.5) where $\th_i$ 
take values in the set
$$
\HH\equiv \{h_1,\dots,h_{K-1},h_K\}
\Eq(9.1)
$$
Each realization of the random field $\{\th[\o]\}_{i\in\N}$ induces a random 
partition of the  set $\L\equiv \{1,\dots,N\}$ into subsets
$$
\L_k[\o]\equiv \{i\in\L:\th_i[\o]=h_k\}
\Eq(9.2)
$$
We may 
introduce $k$ order parameters
$$
m_k[\o](\s)\equiv \frac 1N\sum_{i\in\L_k[\o]}\s_i
\Eq(9.3)
$$
We denote by $\mm[\o]$ the $K$-dimensional vector $(m_1[\o],\dots,m_K[\o])$. 
Note that these take values in the set
$$
\G_N[\o] \equiv\times_{k=1}^K \left\{-\rho_{N,k}[\o],-\rho_{N,k}[\o]+\sfrac 2N,\dots,
\rho_{N,k}[\o]-\sfrac 2N,\rho_{N,k}[\o]\right\}
\Eq(9.4)
$$
where
$$
\rho_{N,k}[\o]\equiv \frac {|\L_k[\o]|}N
\Eq(9.5)
$$
Note that the random variables $\rho_{N,k}$ concentrate exponentially (in $N$)
around their mean values $\E_h\rho_{N,k}=\Pr[\th_i=h_k]\equiv p_k$.  
Obviously $m^1[\o](\s)=\sum_{k=1}^Km_k[\o](\s)$ and
$m^2[\o](\s)=\sum_{\ell=1}^k h_k m_k(\s)$, so that the Hamiltonian 
can be written as a function of the variables $\mm[\o](\s)$, via
$$
H_N[\o](\s)= -N E(\mm[\o](\s))
\Eq(9.6)
$$
where $E:\R^K\rightarrow \R$ is the deterministic function
$$
E(x)\equiv \frac 12\left(\sum_{k=1}^K x_k\right)^2+\sum_{k=1}^K h_kx_k
\Eq(9.7)
$$
The point is
now that the image of the 
Glauber dynamics  under the family of 
functions $m_\ell$ is again Markovian. This follows easily by verifying the
criterion given in Theorem 7.1.  

On the other hand, it is easy to compute the equilibrium distribution of the 
variables $\mm[\o]$. Obviously,
$$
\mu_{\b,N}[\o](\mm[\o](\s)=x)\equiv
\Q_{\b,N}[\o](x)= \frac 1{Z_N[\o]} e^{\b NE(x)} \prod_{k=1}^K 2^{-N\rho_{N,k}[\o]}
{{N\rho_{N,k}[\o]}\choose {N (1+x_k)/2}}
\Eq(9.8)
$$
where $Z_N[\o]$ is the normalizing partition function. Stirling's formula
yields the well know asymptotic expansion for the binomial coefficients
$$
2^{-N\rho_{N,k}[\o]}{{N\rho_{N,k}[\o]}\choose {N (1+x_k)/2}}
=e^{- N\rho_{N,k}[\o] [I(x_k/\rho_{N,k}[\o]) + J_{N}(x_k,\rho_{N,k}[\o])]}
\Eq(9.9)
$$
where
$$
I(x)\equiv \frac {1+x}2\ln (1+x)+\frac {1-x}2\ln(1-x)
\Eq(9.10)
$$ 
is the usual Cram\`er entropy and
$$
J_N(x,\rho) =-\frac 1{\rho N}\ln\left(\frac {1-\left(x/\rho\right)^2}4+\frac
{2x/\rho}{1-\left(x/\rho\right)^2} \right)
+O\left(\sfrac 1{(\rho N)^2}\right)+\frac 1{N^2}C(\rho N)
\Eq(9.11)
$$
with $C(\rho N)$ a constant independent of $x_k$  
(and thus irrelevant) that satisfies
$C(\rho N)=O \left(\ln(\rho N)\right)$. 
Thus 
$$
F_N[\o](x)\equiv -\frac {1}{\b N} \ln \Q_{\b,N}[\o](x)
=F_{0,N}[\o](x) +F_{1,N}[\o](x) +C_N
\Eq(9.12)
$$
with 
$$
F_{0,N}(x)=-E(x) +\frac 1\b\sum_{k=1}^K{\rho_{N,k}}I(x_k/\rho_{N,k})
\Eq(9.13)
$$
$C_N=\b^{-1}\sum_{k=1}^K \rho_{N,k}C(\rho_{N,k}N)$ is constant and of order 
$\ln N$,
and $F_{1,N}$ of order $1/N$, uniformly on compact subsets of 
$\G\equiv \times_{k=1}^K (-p_k,p_k)$. 
Moreover,
$F_N(x)$ converges almost surely to the deterministic function
$$
F_0(x)=-E(x) +\frac 1\b \sum_{k=1}^K {p_k}I(x_k/p_k)
\Eq(9.15)
$$
uniformly on compact subsets of $\G$.
The dominant contribution to the finite volume corrections thus comes from the 
fluctuations part of the function $F_{0,N}$, $F_{0,N}(x)- F_{0}(x)$.    
One easily verifies that all conditions imposed on the functions $F_N$
in Section 1 are verified in this example.

\medskip
\line{\bf The landscape given by $F$. The deterministic picture.\hfill}
\medskip

To see how the landscape of the function $F_N$ looks like, we begin by 
studying the deterministic limiting function $F_0$. 
Let us first look at the critical points. They are solutions of the equation
$\nabla  F_0(x)=0$, which reads explicitly
$$
0=\frac {\del}{\del x_k}F_0(x)=-\sum_{\ell=0}^K x_\ell -h_kx_k +\frac 1\b 
{p_k}I'(x_k/p_k),
\,\,\,k=1,\dots, K
\Eq(9.16)
$$
or equivalently
$$
\eqalign{
&x_k=p_k\tanh(\b( m+h_k)),\,\,\,k=1,\dots, K\cr
&m=\sum_{k=1}^Kx_k
}
\Eq(9.17)
$$
These equations have a particularly pleasant structure. Their 
solutions are generated by   
solutions of the transcendental equation
$$
m=\sum_{k=1}^Kp_k\tanh(\b ( m+h_k))=\E_h\tanh\b(m+h)
\Eq(9.18)
$$
Thus if $m^{(1)},\dots, m^{(r)}$ are the solutions of \eqv(9.18), then the 
full set of solutions of the equations \eqv(9.16) is given by the vectors
$x^{(1)},\dots, x^{(r)}$ defined by 
$$
x^{(\ell)}_k\equiv p_k\tanh\b(m^{(\ell)}+h_k)
\Eq(9.19)
$$
Next we analyze the structure of the critical points. Using that 
$I''(x)=\frac 1{1-x^2}$, we see that
$$
\frac{\del^2}{\del x_k\del x_{k'} }F_0(x)=-1+\frac {\d_{k,k'}}{\b p_k\left(
1-x^2_k/p^2_k\right)}
\Eq(9.20)
$$
Thus at a critical point $x^{(\ell)}$, 
$$
\frac{\del^2}{\del x_k\del x_{k'} }F_0\left(x^{(\ell)}\right)=-1+\d_{k,k'}\l_{k}(m^{(\ell)})
\Eq(9.21)
$$
where
$$
\l_{k}(m) \equiv\frac 1{\b p_k(1-\tanh^2(\b(m+h_k)))}
\Eq(9.22)
$$

\lemma {\ver.1} {\it The Hessian  of $F_0$ at $(x^{(\ell)})$ 
 has at most one negative
eigenvalue. A negative eigenvalue exists if and only if
$$
\b\E_h\left(1-\tanh^2(\b(x^{(\ell)}+h))\right)>1
\Eq(9.221)
$$
}

\proof Consider any matrix of the form 
$A_{kk'}=-1+\d_{k,k'}\l_{k}$ with $\l_k\geq 0$. To see this, let 
$\{\z_1,\dots,\z_L\}$ denote the set of distinct values that
are taken by $\l_1,\dots, \l_K$. Put
$k_\ell=\{k:\l_k=\z_\ell\}$ and denote by $|\k_\ell|$ the cardinalities of 
these sets. 
Now the eigenvalue equations read
$$
-\left(\sum_{k=1}^K u_k\right)+(\l_k-\g)u_k=0
\Eq(9.23)
$$
Let $\z_\ell$ be such that $|\k_\ell|>1$, if such a $\z_\ell$ exists.
Then we will construct $|\k_\ell|-1$ orthogonal solutions to
\eqv(9.23) with eigenvalue $\g=\z_\ell$. Namely, we
set $u_k=0$ for all $k\not\in\k_\ell$. The remaining components must satisfy
$\sum_{k\in\k_\ell}u_k=0$. But obviously, this equation has $|\k_\ell|-1$
orthonormal solutions. Doing this for every $\z_\ell$, we construct
altogether $K-L$  eigenvectors corresponding to the eigenvalues
$\z_\ell$. Note that for all these solutions, $\sum_ku_k=0$.
We are left with finding the remaining $L$ eigenfunctions.
Now take $\g\not \in \{\z_1,\dots,\z_L\}$. 
Then \eqv(9.23) can be rewritten as
$$
u_k=\frac {\sum_{k=1}^K u_k}{\l_k-\g}
\Eq(9.24)
$$
Summing equation \eqv(9.24) over $k$, we get 
$$
\sum_{k=1}^K u_k =\sum_{k=1}^K u_k\sum_{k=1}^K \frac {1}{\l_k-\g}
\Eq(9.25)
$$
Since we have already exhausted the solutions with $\sum_{k=1}^Ku_k=0$,
we get for the remaining ones the condition
$$
1=\sum_{k=1}^K \frac {1}{\l_k-\g} =\sum_{\ell=1}^L\frac {|\k_\ell|}{\z_k-\g}
\Eq(9.26)
$$
Inspecting the right-hand side of \eqv(9.26)
one sees immediately that this equation has precisely $L$ solutions
$\g_i$ that satisfy
$$
\g_1<\z_1<\g_2<\z_2<\g_3<\dots<\g_L<\z_L
\Eq(9.27)
$$
of which at most $\g_1$ can be negative.
Moreover, a negative solution $\g$ implies that
$$
1=\sum_{\ell=1}^L\frac{|\k_\ell|}{\l_k-\g} <\sum_{\ell=1}^L
\frac{|\k_\ell|}{\l_k}
=\sum_{k=1}^K\frac 1{\l_k}
\Eq(9.28)
$$
which upon inserting the specific form of $\l_\k$ yields \eqv(9.221). On the 
other hand, if $\sum_{k=1}^K\frac 1{\l_k}>1$, then by monotonicity 
there exists a negative solution to \eqv(9.26).
 This proves the lemma.\endproof

The following general features are now easily verified due to the fact that
the analysis of the critical points is reduced to equations of one single 
variable. The following facts hold:

\item{(i)} For any distribution of the field, there exists $\b_c$
such that: If $\b<\b_c$, there exists a single critical point and $F_0$ is
strictly convex. If $\b>\b_c$, there exist at least $3$ critical points,
the first and the last of which (according to the value of $m$)
 are local minima, and each minimum is followed by a saddle with 
one negative eigenvalue, and vice versa, with possibly 
intermediate saddles with one zero eigenvalue interspersed.

\item{(ii)} Assume $\b>\b_c$. Then each pair of consecutive critical points of
$F_0$ can be joined by a unique integral curve of the the vector field
$\nabla F_0(x)$.

The exact picture of the landscape depends of course on the particular 
distribution of the magnetic field chosen. In particular, the exact number of
critical points, and in particular of minima, depends on the distribution
(and on the temperature). The reader is invited to use e.g. mathematica and 
produce diverse pictures for her favorite choices. We see that a major effect 
of the disorder enters into the form of the deterministic function $F_0(x)$.
Only a secondary r\^ole is played by the remnant disorder whose effect will be 
most notable in symmetric situations where it can break symmetries present on
the level of $F_0$.

\medskip
\line{\bf Fluctuations\hfill}
\medskip

In the present simple situation it turns out that  the fluctuations of the 
function $F_{0,N}$ can also be controlled in a precise way.
We will show the following result.

\proposition {\ver.3} {\it Let $g_k,   k=1,\dots, K$ be a family of
independent 
  Gaussian random variables with mean zero and variance $p_k(1-p_k)$. Then
the function $\sqrt N[F_N(x)- F_0(x)]$
converges in distribution, uniformly on compact subsets of $\G$ to 
the random function
$$
\frac 1\b\sum_{k=1}^K g_k\left(\frac {x_k}{p_k^2}I'(x_k/p_k)-I(x_k/p_k)\right)
\Eq(9.31)
$$
}

\proof Since $F_N-F_{0,N}$ converges to zero uniformly, it is enough to 
consider 
$$
\eqalign{
F_{0,N}(x)-F_{0}(x)
&=\frac 1\b\sum_k\left(\rho_{N,k} I(x_k/\rho_{N,k})-p_kI(x_k/p_k)\right)\cr
&=\frac 1\b
\sum_k\Bigl((\rho_{N,k}-p_k) I(x_k/p_k) +p_k(I(x_k/\rho_{N,k})-I(x_k/p_k))\cr
&
+(\rho_{N,k}-p_k)(I(x_k/\rho_{N,k})-I(x_k/p_k))\Bigr)
}
\Eq(9.32)
$$
Now in the interior of $\G$ we may develop 
$$
I(x_k/\rho_{N,k})-I(x_k/p_k)=(\rho_{N,k}-p_k)\frac 1{p_k^2} I'(x_k/p_k)+
O((\rho_{N,k}-p_k)^2) 
\Eq(9.33)
$$
Now the $\rho_{N,k}$ are actually sums of
independent Bernoulli random variables with mean $p_k$, namely
$\rho_{N,k}=\frac 1N\sum_{i=1}^N\d_{h_k,\th_i}$.
Thus,  by the exponential Chebyshev inequality,
$$
\Pr\left[|\rho_{N,k}-p_k|>\e\right]\leq 2 \exp\left(-NI_{p_k}(\e)\right)
\Eq(9.34)
$$
where $I_p(\e)\geq 0$ is a strictly convex function that takes its minimum 
value $0$ at $\e=0$. Thus with probability tending to one rapidly, 
we have that e.g. $(\rho_{N,k}-p_k)^2\leq N^{-3/4}$ which allows us to neglect all 
second order remainders. Finally, by the central limit theorem
the family of random variables
$\sqrt N(\rho_{N,k}-p_k)$ converges to a family of 
independent Gaussian random variables with variances $p_k(1-p_k)$. This yields
the proposition.\endproof

\medskip
\line{\bf Relation to a one-dimensional problem.\hfill}
\medskip

We note that the structure of the landscape in this case is quasi 
one-dimensional. This is no coincidence. 
In fact, it is governed by the rate function of the total 
magnetization, $-\frac 1{\b N} \mu_{\b,N} (m^1(\s)=m)$
 which to leading orders is computed, using standard techniques,
as
$$
G_{0,N}(m)=-\frac {m^2}2+\sup_{t\in \R}\left(mt-\frac 1\b\sum_{k=1}^K\rho_{N,k}
\ln\cosh\b(h_k+t)\right)
\Eq(9.29)
$$
The most important facts for us are collected in the following Lemma.

\lemma{\ver.2}{\it The functions $G_{0,N}$ and $F_{0,N}$ are related in the 
following ways.
\item{(i)} For any $m\in [-1,1]$,
$$
G_{0,N}(m)=\inf_{x\in\R^K:\sum_k x_k=m} F_{0,N}(x)
\Eq(9.30)
$$
\item{(ii)} If $x^*$ is a critical point of $F_{0,N}$, then
$m^*\equiv \sum_kx^*_k$ is a critical point of $G_{0,N}$.
\item{(iii)} If $m^*$ is a critical point of $G_{0,N}$,
then $x^*(m^*)$, with components $x^*_k(m)\equiv \rho_{N,k}\tanh\b(m^*+h_k)$
is a critical point of $F_{0,N}$.
\item{(iv)} At any critical point $m^*$, $G_{0,N}(m^*)=F_{0,N}(x^*(m^*))$.
}

The prove of this lemma is based on elementary analysis and will be left to 
the reader. 

The point we want to make here is that while the 
dynamics induced by the Glauber dynamics
on the total magnetization is not Markovian, if we define a Markov chain
$m(t)$ that is reversible with respect to the  distribution of
the magnetization in the spirit of Section 7 and compare its behaviour
to that of the Markov
 chain $\mm(t)=\mm(\s(t))$, the preceding result assures that
their long-time dynamics are identical  since all that matters are the 
precise values of the respective free-energies at its critical points, and
 these coincide according to the preceding lemma (up to 
terms of order $1/N$, and the asymptotics
given in \eqv(9.10), \eqv(9.11), up to ($K$-dependent) constants). 
In other words, the two 
dynamics, when observed on the set of minima of their respective free energies,
are identical on the level of our precision.

\vskip1cm

\chap{References}0
\item{[Az]} R. Azencott, ``Petites perturbations al\'eatoires des syst\`emes 
dynamiques: d\'eveloppements asymptotiques'', Bull. Sc. Math. {\bf 109},
253-308 (1985).
\item{[BG1]} A. Bovier and V. Gayrard, ``Hopfield models as generalized 
             random mean field models'',  in 
            ``Mathematical aspects of spin glasses and neural networks'',  
             A. Bovier and P. Picco (eds.), Progress in Probability {\bf 41},
             (Birkh\"auser, Boston 1998).
\item{[BG2]} A. Bovier and V. Gayrard, ``Sample path large deviations
for a class of Markov chains related to disordered mean field models'',
WIAS-preprint 487 (1999).
\item{[BR]} C.J. Burke and M. Rosenblatt, ``A Markovian function of a 
Markov Chain'',
Ann. Math. Statist. {\bf  29}, 1112-1122, (1958).
\item{[CC]} O. Catoni and R. Cerf, ``The exit path of a Markov chain with rare
transitions'', ESAIM Probab. Statist. {\bf 1}, 95-144 (1995/97).
\item{[CGOV]} M. Cassandro, A. Galves, E. Olivieri, and M.E. Var\'es,
``Metastable behaviour of stochastic dynamics: a pathwise approach, J. Stat. 
Phys. {\bf 35}, 603-634 (1988).
\item{[FJ]} W.H. Fleming and M.R. James, 
``Asymptotic series and exit time probabilities'',
Ann. Probab. {\bf 20}, 1369-1384 (1992). 
\item{[FMP]}  R.L. Fontes, P. Mathieu and P. Picco, ``On the averaged dynamics
of the 
 random field Curie-Weiss model'', private communication.
\item{[FW]} M.I. Freidlin and A.D. Wentzell, ``Random perturbations of 
dynamical systems'', Springer, Berlin-Heidelberg-New York, 1984.
\item{[GT]} I. Gaudron and A. Trouv\'e, ``Fluctuations of empirical means at 
low temperature for finite Markov chains with rare transitions in the 
general case''. Probab. Theor. Rel. Fields {\bf 111}, 215-251 (1998).
\item{[Ho]} J.J. Hopfield, ``Neural networks and physical systems
with emergent collective computational abilities'', Proc. Natl.
Acad. Sci. USA {\bf 79}, 2554-2558 (1982).
\item{[vK]} N.G. van Kampen, ``Stochastic processes in physics and
chemistry'', North-Holland, Amsterdam, 1981 (reprinted in 1990).
\item{[KMST]} C. Knessl, B.J. Matkowsky, Z. Schuss, and C. Tier,
``An asymptotic theory of large deviations for Markov jump processes'',
SIAM J. Appl. Math. {\bf 46}, 1006-1028 (1985).
\item{[Ki1]} Y. Kifer, ``On the principle eignevalue of a singular 
perturbation problem with hyperbolic limit points and circles'', 
J. Diff. Eq. {\bf 37}, 108-139 (1980).
\item{[Ki1]} Y. Kifer, ``The exit problem for small random perturbations
 of 
dynamical systems  with a hyperbolic fixed point'', Israel J. Math. {\bf 40}, 
74-94 (1981).
\item{[Ki3]} Y. Kifer, ``Random perturbations of 
dynamical systems'', Progress in Probability and Statistics 16, Birkh\"auser,
Boston-Basel, 1988.
\item{[Ki4]} Y. Kifer, ``A discrete time version of the Wentzell-Freidlin 
theory'', Ann. Probab. {\bf 18}, 1676-1692 (1990).
\item{[Ku]}  T. G. Kurtz, ``Solutions of ordinary differential equations 
as limits of pure jump Markov processes'', J. Appl. Probab. {\bf 7},
49--58 (1970).
\item{[Li]} T.M. Liggett, ``Interacting particle systems'', Springer, Berlin, 
1985.
\item{[MP]} P. Mathieu and P. Picco, ``Metastability and convergence to 
equilibrium for the random field Curie-Weiss model'',
J. Stat. Phys. {\bf 91}, 679-732 (1998).
\item{[OS1]} E. Olivieri, E. Scoppola, ``Markov chains with exponentially small
transition probabilities: First exit problem from a general domain -I. The
reversible case'' J. Stat. Phys.  {\bf 79}, 613 (1995).
\item{[OS2]} E. Olivieri, E. Scoppola, `` Markov chains with exponentially 
small
transition probabilities: First exit problem from a general domain -II. The
general case''.  J. Stat. Phys.  {\bf 84}, 987-1041 (1996).
\item{[SS]} R. Schonmann and S. Shlosman, ``Wulff droplets and the 
metastable relaxation of kinetic Ising models'', Commun. Math. Phys.
{\bf 194}  389-462 (1998).
\item{[Sc]} E.  Scoppola, ``Renormalization group for Markov chains and
application to metastability'', J. Statist. Phys. {\bf 73}, 83-121 (1993). 
\item{[Va]}M.E. Vares, ``Large deviations and metastability'', in 
 {\it Disordered systems},
1-62, Travaux en Cours Vol.{\bf 53}, Hermann, Paris, 1996.
\item{[W1]} A.D. Wentzell, ``Rough limit theorems on large deviations for 
Markov stochastic processes I.'', 
Theor. Probab. Appl. {\bf 21}, 227-242 (1976). 
\item{[W2]} A.D. Wentzell, ``Rough limit theorems on large deviations for 
Markov stochastic processes II.'', 
Theor. Probab. Appl. {\bf 21}, 499-512 (1976).
\item{[W3]} A.D. Wentzell, ``Rough limit theorems on large deviations for 
Markov stochastic processes III.'', 
Theor. Probab. Appl. {\bf 24}, 675-692 (1979).
\item{[W4]} A.D. Wentzell, ``Rough limit theorems on large deviations for 
Markov stochastic processes IV.'', 
Theor. Probab. Appl. {\bf 27}, 215-234 (1982).

\end